\pdfoutput=1
\documentclass[usenatbib]{mn2e}
\usepackage{apjfonts}
\usepackage{amssymb}
\usepackage{amsmath}
\usepackage{ctable}
\usepackage{verbatim}
\usepackage{url}
\usepackage{fixltx2e} 
\usepackage[implicit=false,breaklinks,colorlinks,citecolor=blue]{hyperref}

\newcommand{\be}{\begin{equation}}
\newcommand{\ee}{\end{equation}}

\newcommand{\etal}{et al.}

\newcommand{\msun}{M_{\sun}}

\newcommand{\paperone}{Paper {\small I}}
\newcommand{\papertwo}{Paper {\small II}}

\newcommand{\demofigrestart}{Fig.~\ref{fig:restart.sfmodel}}

\newcommand{\LEBRON}{Locally-Extincted Background Radiation in Optically-thin Networks}
\newcommand{\tauavg}{\langle \tau \rangle_{b}}

\newcommand{\ICsurl}{\href{http://www.tapir.caltech.edu/~phopkins/publicICs}{\url{http://www.tapir.caltech.edu/~phopkins/publicICs}}}
\newcommand{\FIREurl}{\href{http://fire.northwestern.edu}{\url{http://fire.northwestern.edu}}}
\newcommand{\gizmourl}{\href{http://www.tapir.caltech.edu/~phopkins/Site/GIZMO.html}{\url{http://www.tapir.caltech.edu/~phopkins/Site/GIZMO.html}}}
\newcommand{\movieurl}{\href{http://www.tapir.caltech.edu/~phopkins/Site/animations/}{\url{http://www.tapir.caltech.edu/~phopkins/Site/animations/}}}

\newcommand{\apjl}{ApJL}
\newcommand{\apjs}{ApJS}

\newcommand{\aap}{A\&A}

\newcommand\plotone[1]
 {\centering \leavevmode \includegraphics[width={0.99\columnwidth}]{#1}}

\newcommand\plotonesize[2]
 {\centering \leavevmode \includegraphics[width={#2\columnwidth}]{#1}}
\newcommand{\plotsidesize}[2]
 {\centering \leavevmode \includegraphics[width={#2\textwidth}]{#1}}
\newcommand{\acknowledgments}{\begin{small}\section*{Acknowledgments}\end{small}}
\newcommand\altaffilmark[1]{$^{#1}$}
\newcommand\altaffiltext[1]{$^{#1}$}
\title[Radiative Feedback in Galaxies]{Radiative Stellar Feedback in Galaxy Formation: Methods and Physics
\vspace{-0.5cm}}

\vspace{-0.2cm}
\author[Hopkins \etal]{
\parbox[t]{\textwidth}{ 
Philip F.~Hopkins\thanks{E-mail:phopkins@caltech.edu}\altaffilmark{1},
Michael Y.~Grudi\'{c}\altaffilmark{1}, 
Andrew Wetzel\altaffilmark{2}, 
Du\v{s}an Kere\v{s}\altaffilmark{3}, 
Claude-Andr{\'e} Faucher-Gigu{\`e}re\altaffilmark{4}, 
Xiangcheng Ma\altaffilmark{5}, 
Norman Murray\altaffilmark{6},
Nathan Butcher\altaffilmark{3}
} 
\vspace*{6pt} \\
\altaffiltext{1}{TAPIR, Mailcode 350-17, California Institute of Technology, Pasadena, CA 91125, USA} \\
\altaffiltext{2}{Department of Physics, University of California, Davis, CA 95616, USA} \\
\altaffiltext{3}{Department of Physics, Center for Astrophysics and Space Science, University of California at San Diego, 9500 Gilman Drive, La Jolla, CA 92093} \\ 
\altaffiltext{4}{Department of Physics and Astronomy and CIERA, Northwestern University, 2145 Sheridan Road, Evanston, IL 60208, USA} \\ 
\altaffiltext{5}{Department of Astronomy and Theoretical Astrophysics Center, University of California Berkeley, Berkeley, CA 94720} \\ \altaffiltext{6}{Canadian Institute for Theoretical Astrophysics, 60 St. George Street, University of Toronto, ON M5S 3H8, Canada} \vspace{-0.5cm}
}

\date{Working Document\vspace{-0.6cm}}
\begin{document}
\maketitle
\label{firstpage}

\vspace{-0.2cm}
\begin{abstract}
Radiative feedback (RFB) from stars plays a key role in galaxies, but remains poorly-understood. We explore this using high-resolution, multi-frequency radiation-hydrodynamics (RHD) simulations from the Feedback In Realistic Environments (FIRE) project. We study ultra-faint dwarf through Milky Way mass scales, including H+He photo-ionization; photo-electric, Lyman Werner, Compton, and dust heating; and single+multiple scattering radiation pressure (RP). We compare distinct numerical algorithms: ray-based LEBRON (exact when optically-thin) and moments-based M1 (exact when optically-thick). The most important RFB channels on galaxy scales are photo-ionization heating and single-scattering RP: in all galaxies, most ionizing/far-UV luminosity ($\sim 1/2$ of lifetime-integrated bolometric) is absorbed. In dwarfs, the most important effect is photo-ionization heating from the UV background suppressing accretion. In MW-mass galaxies, meta-galactic backgrounds have negligible effects; but local photo-ionization and single-scattering RP contribute to regulating the galactic star formation efficiency and lowering central densities. Without some RFB (or other ``rapid'' FB), resolved GMCs convert too-efficiently into stars, making galaxies dominated by hyper-dense, bound star clusters. This makes star formation more violent and ``bursty'' when SNe explode in these hyper-clustered objects: thus, including RFB ``smoothes'' SFHs. These conclusions are robust to RHD methods, but M1 produces somewhat stronger effects. Like in previous FIRE simulations, IR multiple-scattering is rare (negligible in dwarfs, $\sim10\%$ of RP in massive galaxies): absorption occurs primarily in ``normal'' GMCs with $A_{V}\sim1$.
\end{abstract}

\begin{keywords}
galaxies: formation --- galaxies: evolution --- galaxies: active --- 
stars: formation --- cosmology: theory\vspace{-0.5cm}
\end{keywords}

\section{Introduction}
\label{sec:intro}

Stars are not passive gravitational ``sinks.'' Rather, once formed, the radiation, winds, and explosions of (especially massive) stars dramatically alter subsequent star and even galaxy formation. If these ``feedback'' processes are not included, gas in galaxies or giant molecular clouds (GMCs) collapses, fragments and turns entirely into stars within a couple free-fall times \citep{bournaud:2010.grav.turbulence.lmc,hopkins:rad.pressure.sf.fb,tasker:2011.photoion.heating.gmc.evol,dobbs:2011.why.gmcs.unbound,harper-clark:2011.gmc.sims}, eventually turning most of the baryons in the Universe into stars \citep{katz:treesph,somerville99:sam,cole:durham.sam.initial,springel:lcdm.sfh,keres:fb.constraints.from.cosmo.sims,faucher-giguere:2011.halo.inflow.properties}. In reality, observed GMCs appear to convert just a few percent of their mass into stars before being disrupted via feedback \citep{zuckerman:1974.gmc.constraints,williams:1997.gmc.prop,evans:1999.sf.gmc.review,evans:2009.sf.efficiencies.lifetimes}, only a percent or so of the gas on a galaxy scale turns into stars per (galactic) free-fall time \citep{kennicutt98}, and only a few percent of the baryons remain in galaxies \citep{conroy:monotonic.hod,behroozi:mgal.mhalo.uncertainties,moster:stellar.vs.halo.mass.to.z1} while the rest are expelled into the circum-galactic and inter-galactic medium in outflows \citep{martin99:outflow.vs.m,heckman:superwind.abs.kinematics,pettini:2003.igm.metal.evol,songaila:2005.igm.metal.evol,martin:2010.metal.enriched.regions,sato:2009.ulirg.outflows,steidel:2010.outflow.kinematics}. 

``Feedback'' is an umbrella term incorporating many processes including proto-stellar jets, photo-heating, stellar mass loss, radiation pressure, supernovae (Types Ia \&\ II), cosmic ray acceleration, and more \citep[e.g.][]{evans:2009.sf.efficiencies.lifetimes,lopez:2010.stellar.fb.30.dor}. Until recently, in simulations of galaxies, the ISM on $\sim$kpc scales, or large GMC complexes ($\gtrsim 10^{6}\,\msun$), it was not possible to model these processes directly and so simplified ``sub-grid'' prescriptions were used to model the ultimate effects of feedback (e.g.\ directly launching winds from clouds or galaxies); however, in recent years simulations have begun to directly resolve the multi-phase structure in the ISM and therefore have attempted to treat these feedback channels explicitly \citep[e.g.][]{tasker:2011.photoion.heating.gmc.evol,hopkins:rad.pressure.sf.fb,hopkins:fb.ism.prop,wise:2012.rad.pressure.effects,kannan:2013.early.fb.gives.good.highz.mgal.mhalo,agertz:2013.new.stellar.fb.model,roskar:2014.stellar.rad.fx.approx.model}. For one example, studies of galaxy formation and star cluster formation from the Feedback In Realistic Environments (FIRE)\footnote{\label{foot:movie}See the {\small FIRE} project website:\\
\FIREurl \\
For additional movies and images of FIRE simulations, see:\\
\movieurl} project \citet{hopkins:2013.fire} explicitly treat the multi-phase structure of gas from $\sim 10- 10^{10}$\,K with star formation only in self-gravitating, self-shielding dense gas and resolution reaching $\sim 30\,M_{\sun}$ in cosmological galaxy-formation simulations \citep{ma:fire2.reion.gal.lfs,wheeler:ultra.highres.dwarfs} or $\sim 0.01\,M_{\sun}$ in massive GMC complex/star cluster simulations \citep{grudic:sfe.cluster.form.surface.density,grudic:cluster.properties}. At this resolution, the simulations attempt to explicitly follow different stellar feedback channels with the spectral, energy, momentum, mass, and metal fluxes in stellar mass-loss, SNe, and radiation taken directly from stellar evolution models. 

On these scales, it is widely-agreed that radiation from stars -- ``radiative feedback'' -- plays a role in galaxy and star formation. Photo-ionization by starlight is necessary to sustain HII regions (which can in turn expand and destroy GMCs), the warm interstellar medium (WIM), and the meta-galactic UV background \citep[e.g.][]{tielens:2005.book}, which in turn can suppress dwarf galaxy formation \citep{barkana:reionization.review}. Photo-electric heating is critical to the structure of the cold/warm neutral medium (C/WNM; \citealt{wolfire:1995.neutral.ism.phases}). Radiation pressure inputs a single-scattering momentum flux into the ISM of $\dot{p} \sim L/c$, comparable (at least initially) to the momentum injection from stellar winds and SNe \citep{starburst99}, and in very dense regions which are optically-thick to infrared (IR) radiation, multiple-scattering can increase this by a factor up to a maximum of $\sim 1 + \tau_{\rm IR}$ (the IR optical depth; \citealt{murray:momentum.winds}). At this point, essentially all observational and theoretical studies agree that radiation has an important impact on star formation at the scales of individual stars \citep{rosen:massive.sf.rhd} and cores \citep{bate:2014.core.collapse.rad.mhd.sph,guszejnov:universal.scalings}, clumps \citep{hopkins:rhd.momentum.optically.thick.issues,grudic:sfe.cluster.form.surface.density}, GMCs \citep{murray:molcloud.disrupt.by.rad.pressure,harper-clark:2011.gmc.sims,hopkins:fb.ism.prop,howard:gmc.rad.fx}, and clusters \citep{lopez:2010.stellar.fb.30.dor,colin:2013.star.cluster.rhd.cloud.destruction,grudic:cluster.properties,grudic:max.surface.density}. However, controversy about the ultimate impact of these processes for bulk {\em galaxy} properties (e.g.\ SFRs, galaxy masses) has abounded, much of it centered on questions of the numerical methods used to treat the radiation \citep{krumholz:2012.rad.pressure.rt.instab,sales:2013.phototion.fb.strong,davis:2014.rad.pressure.outflows,jiang:RT.RHD.instabilities,tsang:monte.carlo.rhd.dusty.wind}. Most (though not all) of this work has focused on idealized studies, which (while critical for understanding the micro-physics involved; see e.g. \citealt{skinner:2015.cloud.sf.frag,takeuchi:rhd.mhd.dusty.wind.instability,raskutti:2016.m1.cloud.sims,zhang:2017.rhd.dusty.winds,zhang:dusty.cloud.acceleration}) do not clearly map to consequences for global galaxy properties; even galactic studies have largely been focused on high-resolution simulations of idealized, non-cosmological, single galaxies \citep{bieri:qso.rhd.fb,costa:dusty.wind.driving,kannan:2018.arepo.rhd,emerick:rad.fb.important.stromgren.ok}. But this misses potentially key regimes in galaxy mass and/or redshift, where the physics may change in important ways. On the other hand, obviously, large-scale simulations must capture at least {\em some} of the key scales (e.g.\ the existence of GMCs and phase structure in the ISM around massive stars) needed to compute the effects of radiative feedback, or else they cannot predict its consequences.

In a companion paper \citet[][hereafter \paperone]{hopkins:fire2.methods}, we presented an updated version of the FIRE code, ``FIRE-2,'' and considered a wide range of numerical effects (e.g.\ resolution, hydrodynamic solvers), as well as the effects of various ``non-feedback'' physics (e.g.\ cooling, star formation) on galaxy formation, in fully-cosmological simulations which follow the physics described above. These can begin to explore these critical questions for feedback. A follow-up paper \citep[][hereafter \papertwo]{hopkins:sne.methods} specifically explored the importance of numerical methods and resolution-scale physics in coupling mechanical feedback from stars (e.g.\ SNe and stellar mass-loss). This paper therefore continues these studies by exploring the role of radiative feedback in galaxy formation. 

Specifically, we will use these simulations to explore the importance of both numerical methods and different aspects of radiative feedback physics, for {\em global} galaxy properties. This includes e.g.\ star formation rates and histories (SFRs/SFHs); stellar masses; metallicities/abundances;  stellar, baryonic, and dark matter mass profiles and content within the halo; circular velocity profiles; visual morphologies; and the distribution of gas in different phases. In order to explore how the effects depend on mass, we will consider a range of galaxy masses from ultra-faint dwarfs to Milky Way and Andromeda-mass systems -- however, we will exclusively focus on radiative feedback from {\em stars} in this manuscript, meaning we will not consider AGN feedback, so we cannot consider galaxies much more massive than $\sim L_{\ast}$. We will survey a wide range of different radiative feedback channels (e.g.\ radiation pressure in UV vs.\ IR, H and He photo-ionization, photo-electric heating, Compton heating, etc.) as described in \S~\ref{sec:rad.overview} below. Finally, because the RHD methods which can be solved efficiently in large cosmological simulations are necessarily approximate, we will compare all of the above using two fundamentally distinct numerical methods for approximating and discretizing the RHD equations. 

In \S~\ref{sec:rad.overview}, we briefly review the various processes collectively referred to as ``radiative feedback'' explored here. In \S~\ref{sec:methods} we discuss our simulation methods, and in \S~\ref{sec:results} we systematically explore the effects of each of these radiative feedback processes in our galaxy formation simulations. We summarize the major effects of radiative feedback as a function of galaxy mass (\S~\ref{sec:ov}) and specifically discuss degeneracies with other ``early'' feedback processes (\S~\ref{sec:early}) before systematically exploring each radiative feedback process in turn (\S~\ref{sec:mechanisms}), as well as discussing where and how photons actually couple in the simulations (\S~\ref{sec:where}) and the effects of numerical methods (\S~\ref{sec:numerics}). We summarize and conclude in \S~\ref{sec:discussion}.

\begin{footnotesize}
\ctable[
  caption={{\normalsize FIRE-2 simulations run to $z=0$ used for our detailed study of stellar radiative feedback here.}\label{tbl:sims}},center,star
  ]{lcccccccr}{
\tnote[ ]{Parameters describing the FIRE-2 simulations from \citet{hopkins:fire2.methods} that we use for our case studies. Halo and stellar properties listed refer only to the original ``target'' halo around which the high-resolution region is centered. All properties listed refer to our highest-resolution simulation using the standard, default FIRE-2 physics and numerical methods. All units are physical. 
{\bf (1)} Simulation Name: Designation used throughout this paper. 
{\bf (2)} $M_{\rm halo}^{\rm vir}$: Virial mass \citep[following][]{bryan.norman:1998.mvir.definition} of the ``target'' halo at $z=0$.
{\bf (3)} $R_{\rm vir}$: Virial radius at $z=0$. 
{\bf (4)} $M_{\ast}$: Stellar mass of the central galaxy at $z=0$. 
{\bf (5)} $R_{1/2}$: Half-mass radius of the stars in the central $M_{\ast}$ at $z=0$. 
{\bf (6)} $m_{i,\,1000}$: Mass resolution: the baryonic (gas or star) particle/element mass, in units of $1000\,\msun$. The DM particle mass is always larger by the universal ratio, a factor $\approx 5$. 
{\bf (7)} $\epsilon_{\rm gas}^{\rm MIN}$: Minimum gravitational force softening reached by the gas in the simulation (gas softenings are adaptive so always exactly match the hydrodynamic resolution or inter-particle spacing); the Plummer-equivalent softening is $\approx 0.7\,\epsilon_{\rm gas}$.
{\bf (8)} $r_{\rm DM}^{\rm conv}$: Radius of convergence in the dark matter (DM) properties, in DM-only simulations. This is based on the \citet{power:2003.nfw.models.convergence} criterion using the best estimate from \citet{hopkins:fire2.methods} as to where the DM density profile is converged to within $<10\%$. The DM force softening is much less important and has no appreciable effects on any results shown here, so is simply fixed to $40\,$pc for all runs here. The initial conditions are all publicly available at \ICsurl
}
}{
\hline\hline
Simulation & $M_{\rm halo}^{\rm vir}$ & $R_{\rm vir}$ & $M_{\ast}$ & $R_{1/2}$ & $m_{i,\,1000}$ & $\epsilon_{\rm gas}^{\rm MIN}$ & $r_{\rm DM}^{\rm conv}$ & Notes \\
Name \, & $[\msun]$ & $[{\rm kpc}]$ & $[\msun]$  & $[{\rm kpc}]$ & $[1000\,\msun]$ & $[{\rm pc}]$ & $[{\rm pc}]$ & \, \\ 
\hline 
{\bf m09} & 2.4e9 & 35.6 & 9.4e3 & 0.29 & 0.25 & 1.1 & 65 & Early-forming, ultra-faint field dwarf \\
{\bf m10q} & 8.0e9 & 52.4 & 1.8e6 & 0.63 & 0.25 & 0.52 & 73 & Early-forming, dwarf spheroidal, small core  \\
{\bf m11b} & 4.3e10 & 92.2 & 1.1e8 & 2.4 & 2.1 & 2.9 & 250 & Intermediate-forming, disky, gas-rich dwarf \\
{\bf m11q} & 1.4e11 & 136 & 9.8e8 & 4.1 & 7.0 & 1.5 & 300 & Early-forming, LMC-mass dIrr with large core \\
{\bf m12i} & 1.2e12 & 278 & 1.0e11 & 2.3 & 56 & 1.4 & 290 & Milky-Way mass, compact disk (at low resolution) \\
{\bf m12m} & 1.5e12 & 302 & 1.4e11 & 5.0 & 56 & 1.4 & 360 & Milky-Way mass, extended disk (at all resolutions) \\
\hline\hline
}
\end{footnotesize}

\vspace{-0.5cm}
\section{What Is Radiative Feedback?}
\label{sec:rad.overview}

``Radiative feedback'' is itself an umbrella term referring to a huge range of processes. We attempt to enumerate some of these here, because although we will explore many of them, it is impossible to be exhaustive and we wish to be clear about which processes we do {\em not} address in this paper. We will focus exclusively on galaxy scales.

Broadly-speaking, ``radiative feedback'' can be divided into three categories: ``radiative heating,'' ``indirect feedback  (ionization/dissociation effects),'' and ``radiation pressure.''

\vspace{-0.5cm}
\subsection{Radiative Heating}

Here consider processes that directly transfer thermal energy to gas, probably the best-studied form of radiative feedback on large scales. This takes many forms, including: 

\begin{enumerate}

\item{\bf Photo-ionization heating from local sources}: Photo-ionization heating (to $\sim 10^{4}-10^{5}\,$K) by ionizing photons around massive stars (with flux dominated by nearby stars in a galaxy/cluster) is critical to the WIM and HII regions. This will be considered throughout this paper.

\item{\bf Photo-ionization heating from ``collective'' effects}: Although ultimately the same sources, we distinguish the collective photo-ionization heating by many galaxies (via ``escaped'' photons) in the form of the meta-galactic UV background (UVB), critical for the phase structure of the circum-galactic medium (CGM) and inter-galactic medium (IGM), Ly-$\alpha$ forest, and believed to suppress SF in small dwarf galaxies. This will also be considered throughout, with the approximation of a (spatially) uniform background plus local self-shielding.

\item{\bf Photo-electric heating}: Local UV (non-ionizing) luminosity absorbed by dust generates photo-electric heating (photo-adsorption/desorption can also be included), important for thermal balance/chemistry in the WNM/CNM ($T\sim 100-8000\,$K). It has been controversial whether this has any {\em dynamical} effects on SF on galaxy scales; it is included in our default RHD treatment, but we will explore removing it as well.

\item{\bf Compton heating}: Hard photons can Compton-heat diffuse gas to high temperatures; this is potentially important around AGN ($T_{\rm Compton} \gtrsim 10^{7}\,$K). From stars alone (where it depends on the much-less-luminous X-ray binary [XRB] population) it is not likely to be dominant \citep{sazonov04:qso.radiative.heating,oppenheimer:2018.flickering.heating.cgm}, so we will only briefly consider it (in a limited subset of tests with our ``extended'' RHD network) for completeness.

\item{\bf Thermal dust/collisional heating}: At high densities ($n \gg 10^{6}\,{\rm cm^{-3}}$) and low temperatures ($T\ll 100\,$K) dust is collisionally tightly-coupled to gas and thermal heating of dust by IR radiation can transfer heat to gas. This is critical to protostellar accretion and may explain the Universality of the IMF \citep{offner:2013.imf.review,bate:2014.core.collapse.rad.mhd.sph,guszejnov.2015:feedback.imf.invariance,guszejnov:protostellar.feedback.stellar.clustering.multiplicity,guszejnov:imf.var.mw}, but is negligible at the much lower density scales (and mass scales $\gg 0.1\,M_{\sun}$) resolved here. We will only briefly consider it for completeness (although we may be effectively modeling it implicitly, by assuming an IMF).

\end{enumerate}

\vspace{-0.5cm}
\subsection{Indirect Radiative Feedback (Ionization/Dissociation Effects)}

This refers to processes which alter subsequent gas cooling rates or star formation via, e.g.\ ionization or dissociation. 

\begin{enumerate}

\item{\bf H \&\ He Ionization}: Photo-ionization of H and He, as described above, not only directly contributes a heating term but also alters the cooling rates non-linearly (changing e.g.\ the number of free electrons, recombination rates, etc.). This is always included in our simulations alongside the self-consistent ionization calculations (see \paperone\ for details).

\item{\bf Metal-Ionization}: In the CGM/IGM, the UVB partially ionizes metals, which alters their line cooling properties. But local, hard sources can (under special circumstances) dominate over the collective UVB and ``over-ionize'' those metals, further suppressing line cooling (lowering the cooling rates), although this likely requires non-equilibrium chemistry to treat properly. Whenever we include the effects of the UVB, the effect on metal ionization of the UVB is also self-consistently included (in the tabulated cooling rates for metal-line cooling). However, we will not explore the effects of non-equilibrium over-ionization from {\em local} sources in this paper \citep[see][for more discussion]{richings:2016.chemistry.uvb.photoelec.fx,oppenheimer:2018.flickering.heating.cgm}.

\item{\bf Lyman-Werner Feedback}: Dissociation of H$_{2}$ by Lyman-Werner radiation can suppress molecular-hydrogen cooling in low-temperature gas, potentially important in extremely metal-poor first-star environments (by $Z\gtrsim 10^{-5} - 10^{-3}\,Z_{\odot}$, metal-line and dust cooling dominates H$_{2}$ cooling at low temperatures). Because we do not follow explicit molecular chemistry, and our feedback and yield physics does not include any separate model for Pop III stars, we will not consider this in detail in this paper, although we will briefly discuss approximate treatments \citep[but see e.g.][]{wise:2008.first.star.fb,wise:2012.rad.pressure.effects}.

\end{enumerate}

\vspace{-0.5cm}
\subsection{Radiation Pressure}

Here consider processes that transfer momentum or kinetic energy to gas.

\begin{enumerate}

\item{\bf Single-Scattering (``Direct'')}: Photons carry momentum $h\,\nu/c$; if gas or dust absorption ``destroys'' the photon (no re-emission, or isotropic re-emission at longer wavelengths), then this is transferred to dust and gas as a momentum flux $\sim L_{\rm absorbed}/c$. This can be comparable to or larger than gravitational forces in HII regions, massive GMCs, and even galaxy scales. We will consider this throughout for all photons at frequencies above infrared.

\item{\bf Multiple-Scattering (Continuum)}: If photons are repeatedly scattered (e.g.\ Thompson scattering) or re-emitted at wavelengths with similar opacities (e.g.\ IR re-emission and re-absorption) the repeated scattering can transfer additional momentum to gas, up to a maximum momentum flux $\sim \tau\,L/c$. Our default RHD treatment accounts for this for infrared photons, in the grey approximation (frequency-independent IR opacity).

\item{\bf Line-Driving}: This is the same concept as multiple-scattering above, but we distinguish multiple scattering in resonance lines because it requires line transfer, and the ``escape'' by scattering out-of-resonance is distinct. This is believed to be critical for mass-loss from massive stars, and certain types of accretion-disk winds in AGN, but on galactic scales is not believed to be critical except, potentially, for resonant Ly-$\alpha$ line scattering. However properly treating resonant Ly-$\alpha$ scattering is physically challenging (usually requiring custom algorithms even to post-process results; \citealt{faucher-giguere:2010.lya.cooling.selfshield,smith:lyalpha.rt.fire.sim.escape.fraction}) and  extremely computationally demanding, so we will not consider it here \citep[but see][for recent studies]{2017MNRAS.464.2963S,2018MNRAS.479.2065S,kimm:lyman.alpha.rad.pressure}.

\end{enumerate}

\begin{figure*}
    \plotsidesize{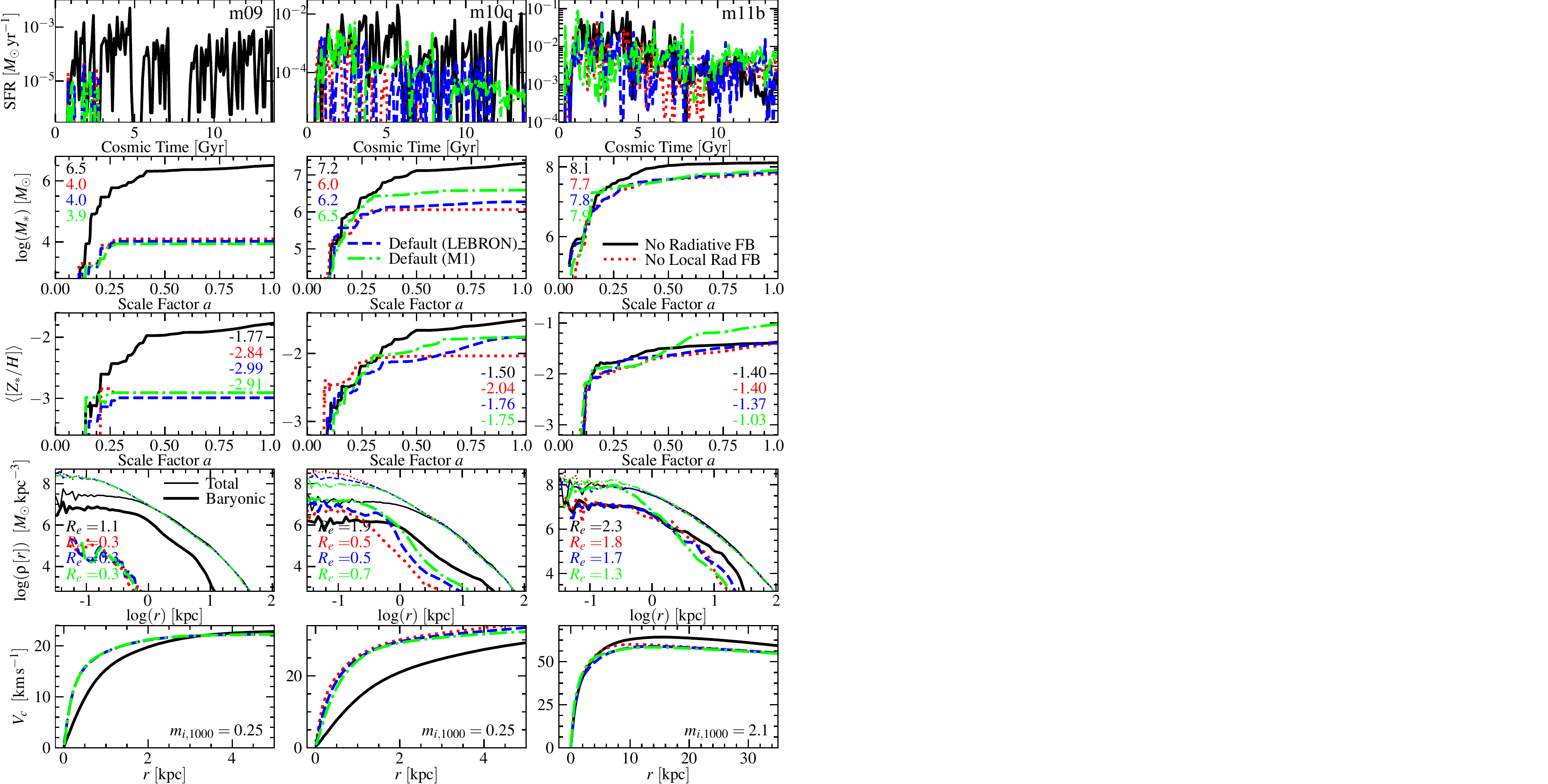}{0.99}
    \vspace{-0.25cm}
    \caption{Effects of radiative feedback on gross galaxy properties in cosmological simulations (dwarfs; continued in Fig.~\ref{fig:ov.2}). 
    {\em Top:} Star formation history (averaged in $100\,$Myr intervals) of the primary ($z=0$) galaxy 
     from Table~\ref{tbl:sims}. 
     {\em Second:} Total stellar mass in box (dominated by primary) vs.\ scale factor ($a=1/(1+z)$). 
     The value at $z=0$ for each run is shown as the number in the panel. 
     {\em Middle:} Stellar mass-weighted average metallicity vs.\ scale factor ($z=0$ value shown). 
     {\em Third:} Baryonic ({\em thick}) and total ({\em thin}) mass density profiles (averaged in spherical shells) as a function of radius around the primary galaxy at $z=0$. Number is the stellar effective ($1/2$-mass) radius at $z=0$. 
     {\em Bottom:} Rotation curves (circular velocity $V_{c}$ versus radius) in the primary galaxy. 
     Value $m_{i,\,1000}$ of the mass resolution is shown. 
     In each, we compare variations from \S~\ref{sec:ov}: 
     {\bf (1)} {\em Default (LEBRON)}: The standard FIRE-2 radiative FB implementation including photo-ionization, photo-electric heating, near UV/optical/IR single-scattering and re-emission with multiple-scattering in the IR, using the LEBRON radiative transport algorithm.
     {\bf (2)} {\em Default (M1)}: This uses the same default source functions, opacities, etc., for all radiation quantities, but replaces the photon transport with the moments-based M1 RHD algorithm. 
     {\bf (3)} {\em No Local Rad FB}: Removes all radiative FB from stars in the simulation, but keeps the (uniform) meta-galactic UVB.
     {\bf (4)} {\em No Radiative FB}: Removes all radiative FB (including the UVB). 
     For dwarfs, the UVB has a large effect; this remains significant even up to LMC-like masses. 
     Removing local radiative FB leads to more ``violent'' SF ({\bf m10q} ``overshoots'' and ``self-quenches'' at $z\sim 2$ and has almost no gas, and no SF, at later times, without radiative FB, producing a lower total stellar mass) in dwarfs, and somewhat more dense centrally-concentrated SF in massive galaxies (Fig.~\ref{fig:ov.2}). M1 and LEBRON algorithms produce qualitatively similar effects, though quantitatively effects appear somewhat stronger in the M1 runs.\vspace{-0.5cm}
    \label{fig:ov}}
\end{figure*}

\begin{figure*}
    \plotsidesize{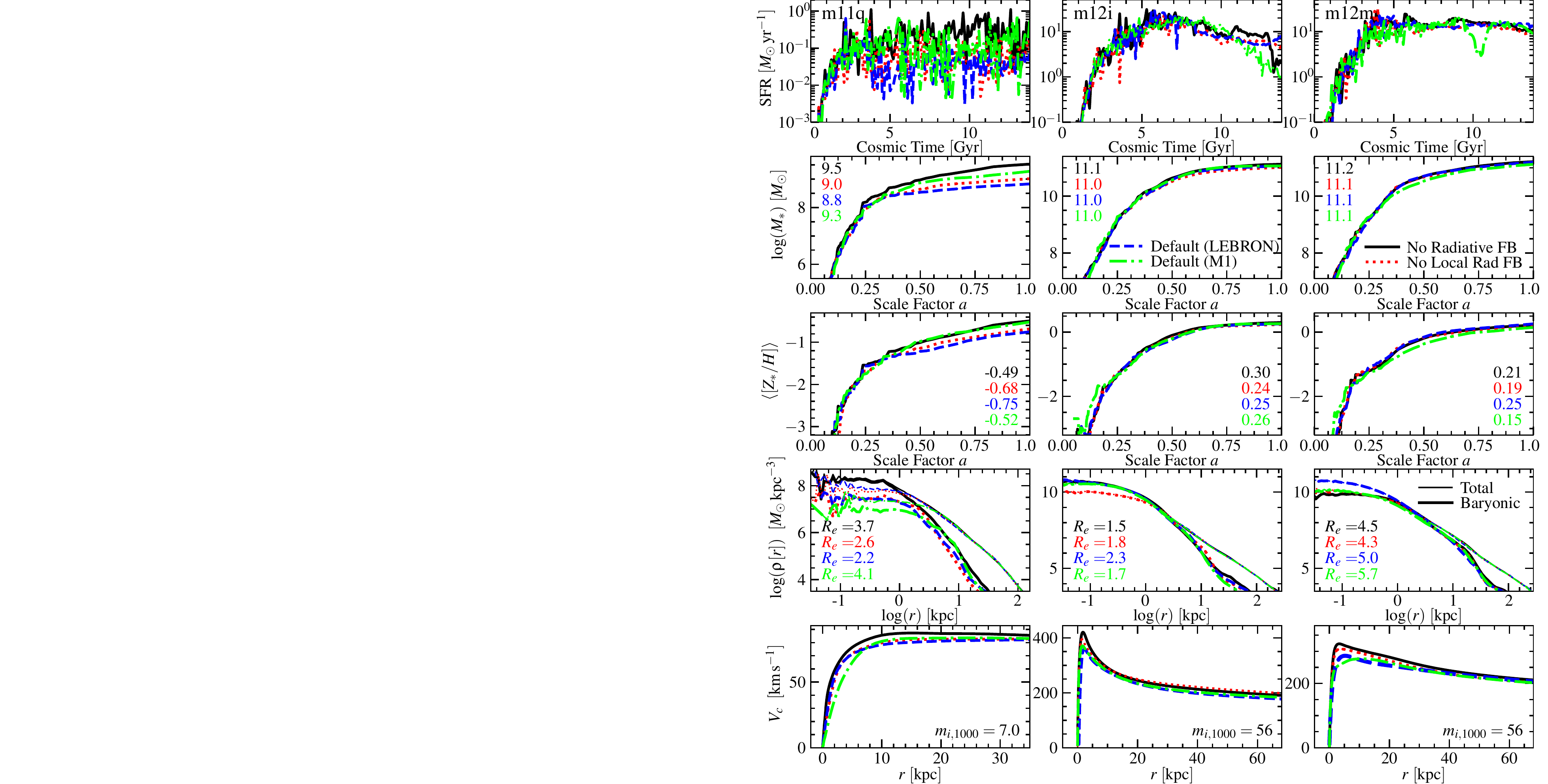}{0.99}
    \vspace{-0.25cm}
    \caption{Fig.~\ref{fig:ov}, continued for more massive galaxies. Removing local radiative FB leads to somewhat more dense centrally-concentrated SF in massive galaxies. Again, M1 and LEBRON algorithms produce qualitatively similar effects, though quantitatively effects appear slightly stronger in the M1 runs.\vspace{-0.5cm}
    \label{fig:ov.2}}
\end{figure*}

\vspace{-0.5cm}
\section{Simulation Methods}
\label{sec:methods}

\subsection{Overview}

The simulations in this paper were run as part of the Feedback in Realistic Environment project (FIRE); specifically using the ``FIRE-2'' version of the code from \paperone. In this paper we will systematically vary the treatment of radiative feedback, but all other simulation physics, initial conditions, and numerical parameters are held fixed. \paperone\ contains all details of these aspects of the method, so we only briefly summarize them here. 

The simulations are run using {\small GIZMO}, a radiation-magnetohydrodynamics code\footnote{A public version of {\small GIZMO} is available at \gizmourl} in its Lagrangian Godunov ``meshless finite mass'' (MFM) mode (for extensive test problems see \citealt{hopkins:gizmo,hopkins:mhd.gizmo,hopkins:cg.mhd.gizmo,hopkins:gizmo.diffusion}). They are fully-cosmological ``zoom-in'' simulations which embed a high-resolution Lagrangian region, that will surround a single $z=0$ galaxy, in a large cosmological box initialized at $z\sim100$. Gravity is treated with adaptive softenings so hydrodynamic and force softenings are always matched, with no artificial minimum enforced. Gas cooling is followed over $T=10-10^{10}\,$K including free-free, Compton, metal-line, molecular, fine-structure, dust collisional, cosmic ray, photo-electric and photo-ionization processes and self-shielding, accounting for both a meta-galactic background and local stellar sources (see details below). Gas turns into stars according to a sink-particle prescription if it is self-gravitating at the resolution scale \citep{hopkins:virial.sf} as well as self-shielding \citep{krumholz:2011.molecular.prescription}, thermally Jeans unstable, and denser than $n_{\rm crit}>1000\,{\rm cm^{-3}}$. Star particles are then considered single-age stellar populations with IMF-averaged feedback properties calculated following standard stellar evolution models \citep{starburst99}: we explicitly treat mechanical feedback from stellar mass loss (O/B and AGB winds), SNe Ia and II, as described in \papertwo, and radiative feedback as described below.

For consistency and brevity, our physics study here will focus on a small sub-set of galaxies from \paperone, with properties in Table~\ref{tbl:sims} which span a range of mass and are typical of other simulated galaxies in the same mass range. For parameter surveys we will particularly focus on two representative galaxies: a dwarf ({\bf m10q}) and Milky Way (MW) mass system ({\bf m12i}), which were studied in detail in \papertwo.

Note that, in \paperone\ and \papertwo, the highest resolution simulations for {\bf m10q} ($30\,M_{\sun}$), {\bf m11q} ($880\,M_{\sun}$), {\bf m12i} and {\bf m12m} ($7000\,M_{\sun}$), run using our ``Default (LEBRON)'' model, were a factor $\sim 8$ better mass resolution than the versions studied here. This owes to computational cost: especially with the M1 RHD solver, which requires a Courant factor limited by the (reduced) speed-of-light, it was not feasible to simulate a large parameter survey of the sort here at these extremely high resolution levels. As shown in \paperone\ and \papertwo, most properties here are insensitive to resolution over this range; the exception is the central ``spike'' in the rotation curve of {\bf m12i}, which appears ubiquitously here, but is substantially reduced in our ``Default'' method at higher resolution. We discuss explicit resolution tests below.

\vspace{-0.5cm}
\subsection{Radiation Hydrodynamics}
\label{sec:feedback:radiation}

\subsubsection{Sources \&\ Frequencies}
\label{sec:feedback:radiation:sources}

Each star particle is a unique source, and is treated as a single stellar population with a known age ($t_{\ast}$) and metallicity ($Z$). We directly tabulate the IMF-averaged luminosity $L_{\nu}(t_{\ast},\,Z)$ as a function of frequency $\nu$, age $t_{\ast}$, and metallicity $Z$ from the same stellar evolution models ({\small STARBURST99} with a \citet{kroupa:imf} IMF) used for SNe and stellar mass loss. Appendix~A of \paperone\ gives approximate expressions for $L_{\nu}(t_{\ast},\,Z)$. 

For the physics of interest, to good approximation it is not necessary to follow a finely-resolved spectrum $L_{\nu}$, so in our ``default'' simulations we integrate into five broad bands:
\begin{enumerate}
\item{Hydrogen ionizing ($L_{\rm ion}$, $\lambda<912\,$\AA), used in computing photo-ionization (dominated by young, massive stars).}
\item{Far-UV ($L_{\rm FUV}$, $912\,$\AA$<\lambda<1550\,$\AA), used for photo-electric heating (also dominated by young stellar populations).}
\item{Near-UV ($L_{\rm UV}$, $1550<\lambda<3600\,$\AA), primarily relevant as continuum single-scattering photons (dominated by young stellar populations).}
\item{Optical/near-IR ($L_{\rm Opt}$, $3600\,$\AA$<\lambda < 3\,\mu$), primarily relevant as continuum single-scattering photons (dominated by older stellar populations).}
\item{Mid/far-IR ($L_{\rm IR}$, $\lambda > 3\,\mu$), representing radiation absorbed and re-radiated by dust.} 
\end{enumerate}
The spectrum $L_{\nu}$ of each star particle is integrated over these wavelengths to give the broad-band $(L_{\rm ion},\,L_{\rm FUV},\,L_{\rm UV},\,L_{\rm Opt},\,L_{\rm IR})$. In \citet{hopkins:fb.ism.prop}, we compare full radiative transfer calculations in galaxy simulations using a detailed full spectrum (with $\sim 10^{7}$ frequency bins) to our simple broad-band approach, and found that discretizing the spectrum into these bands introduces $\lesssim10\%$-level changes in the energy and/or momentum coupled and radiative transfer solutions.

Appendix~\ref{sec:luminosity.opacity.descriptions} gives details. There we also describe a more extensive 10-band frequency network which we use for some additional tests here, which includes: hard X-ray ($L_{\rm HX}$), soft X-ray ($L_{\rm SX}$), He-II ionizing ($L_{\rm HeII}$), He-I ionizing ($L_{\rm HeI}$), H-ionizing ($L_{\rm HI}$), Lyman-Werner ($L_{\rm LW}$), photo-electric ($L_{\rm PE}$), near-UV ($L_{\rm NUV}$), optical/near-IR ($L_{\rm opt}$), and multi-temperature mid/far-IR ($L_{\rm FIR}$, which tracks an effective blackbody of dynamically-evolving radiation temperature). 

Opacities within in each narrow band (e.g.\ separate $\kappa_{\rm ion}$, $\kappa_{\rm FUV}$, $\kappa_{\rm UV}$, $\kappa_{\rm Opt}$, $\kappa_{\rm IR}$) are calculated as flux-weighted means based on the {\small STARBURST99} mean spectra, as a function of the gas neutral fractions in the relevant states, and metallicity (assuming a constant dust-to-metals ratio); see Appendix~\ref{sec:luminosity.opacity.descriptions}.

\vspace{-0.5cm}
\subsubsection{Photon Transport}
\label{sec:photon.transport}

{\small GIZMO} includes several different RHD solvers: the direct intensity solver from \citet{jiang:2014.rhd.solver.local}, the ray-based ``LEBRON'' (Locally Extincted Background Radiation in Optically Thin Networks; \citealt{hopkins:fb.ism.prop}), and moments-based flux-limited diffusion (FLD), first-moment (M1; \citealt{levermore:1984.FLD.M1}) and optically-thin variable Eddington-tensor (OTVET; \citealt{gnedin.abel.2001:otvet}) methods. Because exact or Monte Carlo solutions of the general eight-dimensional RT equation are simply not tractable in ``real-time'' in our simulations, we study two approximate methods (both of which are computationally tractable) in this paper: (1) LEBRON, and (2) M1. 

\begin{enumerate}

\item{LEBRON} is an approximate ray-tracing method, which assumes (1) negligible light-travel times (as most ray methods and our gravity solver also assume), and (2) local extinction in the vicinity of sources and absorbers dominates over absorption ``in between,'' so the intervening transport can be approximated as optically thin. This means it trivially reduces to the exact ray-tracing RT solution in the optically thin limit, even for arbitrary numbers of sources. It also does not require a ``reduced speed of light'' (RSOL) and properly treats photons as collisionless (so rays can intersect/cross one another). However it fails to capture shadowing, anisotropic photon diffusion (e.g.\ diffusion along the ``path of least resistance'' in optically thick media), and the optically-thin assumption means the long-range flux ${\bf F}_{\nu}$ is not strictly photon-conserving (if there is, for example, a shadowing clump along a line-of-sight).\footnote{We will show below that although the LEBRON scheme is not exactly photon-conserving, the net sense of its errors are to slightly {\em under}-estimate the total photon number/momentum/energy coupled to gas (at the $\sim 10\%$ level). For more explicit tests see \citet{hopkins:fb.ism.prop}.} Our LEBRON implementation is described in detail in \citet{hopkins:fire2.methods} (see Appendix~E therein).

\item{M1} is a moments-based method which reduces to exact solutions in the infinitely optically-thick regime, can capture certain shadowing effects and anisotropic photon propagation, and is manifestly photon-conserving. However it imposes strict timestep requirements which necessitate a RSOL approximation. More important, like {\em any} moments-based approximation, the closure imposed on the Eddington tensor prevents photons from behaving collisionlessly (e.g.\ intersecting rays ``shock'' and merge, and ``new'' rays isotropically diffuse out from their new location, like in FLD), so it cannot converge to correct solutions (at any resolution) in the optically-thin limit for $>1$ source. This can be especially problematic in systems with many sources, like galaxies. We adopt the ``face-integrated'' formulation of M1 \citep{hopkins:rhd.momentum.optically.thick.issues}, with the gradient treatment\footnote{Specifically, as shown in \citet{rosdahl:2015.galaxies.shine.rad.hydro}, one obtains more accurate results with M1 in the limit where UV/ionizing photon mean-free-paths are un-resolved in neutral gas (always the case here) if we replace the explicit flux ${\bf F}_{\nu}$ with $e_{\nu}\,c\,\hat{\bf F}_{\nu}$ (the ``incident free-streaming flux'') in the calculation of the RHD momentum transfer (``radiation pressure'') term.} described in \citet{rosdahl:2015.galaxies.shine.rad.hydro} -- this is critical for correctly capturing the RP forces, as described below. Additional details of our M1 implementation (including e.g.\ how photons are isotropically ``injected'' onto the grid each timestep) are given in \citet{hopkins:rhd.momentum.optically.thick.issues}, Appendix~A.

\end{enumerate}

Clearly, both methods have (serious) limitations. However they form a particularly useful ``pair'' because their advantages/disadvantages, and regimes where they correctly converge to exact solutions, are almost exactly opposite/complementary. Thus where they give similar results, those results are likely to be robust, and where they differ, they will tend to bracket the allowed range of solutions.

\vspace{-0.5cm}
\subsubsection{Radiative Acceleration (Radiation Pressure)}
\label{sec:methods:rad.pressure}

The (non-relativistic) radiative acceleration in a differential volume $d^{3}{\bf x}$ is just $\kappa\,{\bf F}/c$; we couple this to the gas using the ``face-integrated'' formulation from \citet{hopkins:rhd.momentum.optically.thick.issues} where this is integrated over a cell domain ``towards'' each effective face:
\begin{align}
\label{eqn:rad.accel} \frac{\partial {\bf v}}{\partial t}{\Bigr |}_{\nu} =&\, \frac{\kappa_{\nu}\,{\bf F}_{\nu}}{c} \\  
\nonumber (\dot{\bf p}_{\nu})_{ab} \equiv &\, \int_{\Delta {\rm Vol}_{a}} d^{3}{\bf x}\,\frac{\rho\,\kappa\,{\bf F}_{\nu}({\bf x})}{c}\,\Theta({\bf x},\,\hat{\bf F}_{\nu},{c}\,{\bf A}_{ab})
\end{align}
where $\Theta = 1$ if the flux vector $\hat{\bf F}_{\nu}$ at point ${\bf x}$ (within the domain of cell $a$) points ``towards'' face ${\bf A}_{ab}$ (of the faces surrounding $a$, it is the first one intercepted by the ray $\hat{\bf F}_{\nu}$), and $\Theta=0$ otherwise. 

As shown in \citet{hopkins:rhd.momentum.optically.thick.issues}, older ``cell-integrated'' or ``cell-centered'' methods -- where Eq.~\ref{eqn:rad.accel} is simply integrated over the whole cell volume (or evaluated at the cell center) instead of at cell faces -- artificially suppress the radiation pressure force by at least an order of magnitude if the mean-free-path of photons around sources is un-resolved. For ionizing photons, resolving the mean free path in neutral gas at the location of a star particle would require an un-achievable mass resolution $m_{i} \lesssim 10^{-10}\,M_{\sun}\,(n/100\,{\rm cm^{-3}})^{-2}$.

\vspace{-0.5cm}
\subsubsection{Radiative Heating \&\ Indirect Feedback}
\label{sec:feedback:radiation:rad.pressure}

Radiative heating/cooling and photo-ionization rates follow standard expressions (all given explicitly in \paperone; App.~B). These include (among other processes): photo-ionization (HI, HeI, HeII), photo-electric heating, dust collisional heating/cooling, Compton heating/cooling. Each of these depends on the radiation energy density $e_{\nu}$ in some band[s]: these are taken from the RT solution and used directly in the appropriate heating/cooling functions. If our ``default'' runs do not include RT in some band (e.g.\ X-rays, only followed in our extended set, or if we ``turn off'' radiative transfer in a given band), then we assume a universal Milky Way background for the term in the heating/cooling routine.

\vspace{-0.5cm}
\subsubsection{The Meta-Galactic UV Background}
\label{sec:UVB}

For the sake of consistency with our previous FIRE simulations and considerable historical work, we do not explicitly solve for the UV background from a set of ``sources'' -- this means we are not self-consistently solving the RT equations for the photons in the meta-galactic UV background. Doing so is highly non-trivial in our zoom-in simulations, since there are only a small number of galaxies inside the high-resolution region (vastly smaller than the $\gg 100\,$Mpc scales needed to correctly capture the collective generation of the background), and the ``boundaries'' of the hydrodynamic grid on which the M1 equations are solved for the RHD are constantly changing and irregular (so even a ``photon inflow boundary condition'' is not well-defined). Instead, we follow most previous galaxy-formation simulations and assume a spatially-uniform but redshift-dependent UV background tabulated from \citet{faucher-giguere:2009.ion.background}, with self-shielding accounted for via local attenuation with a Sobolev approximation (see \citealt{hopkins:fire2.methods} for details). After this self-shielding correction, the remaining UVB spectrum is added to the explicitly-followed relevant ionizing RT band intensities, for use in computing e.g.\ photo-heating and ionization states above.

\vspace{-0.5cm}
\section{Results}
\label{sec:results}

\begin{figure*}
\begin{tabular}{cc}
\includegraphics[width=0.54\textwidth]{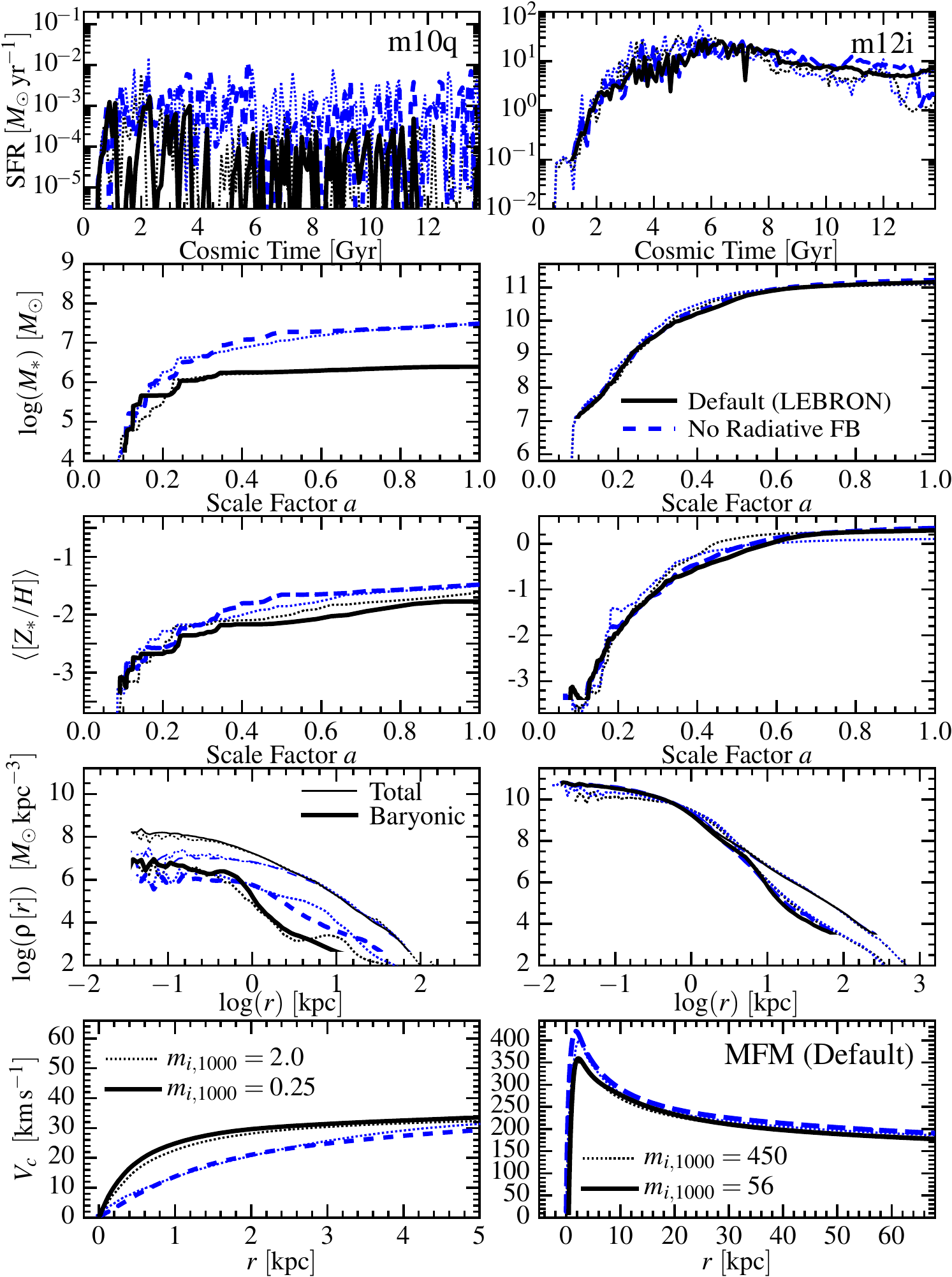} &
\includegraphics[width=0.30\textwidth]{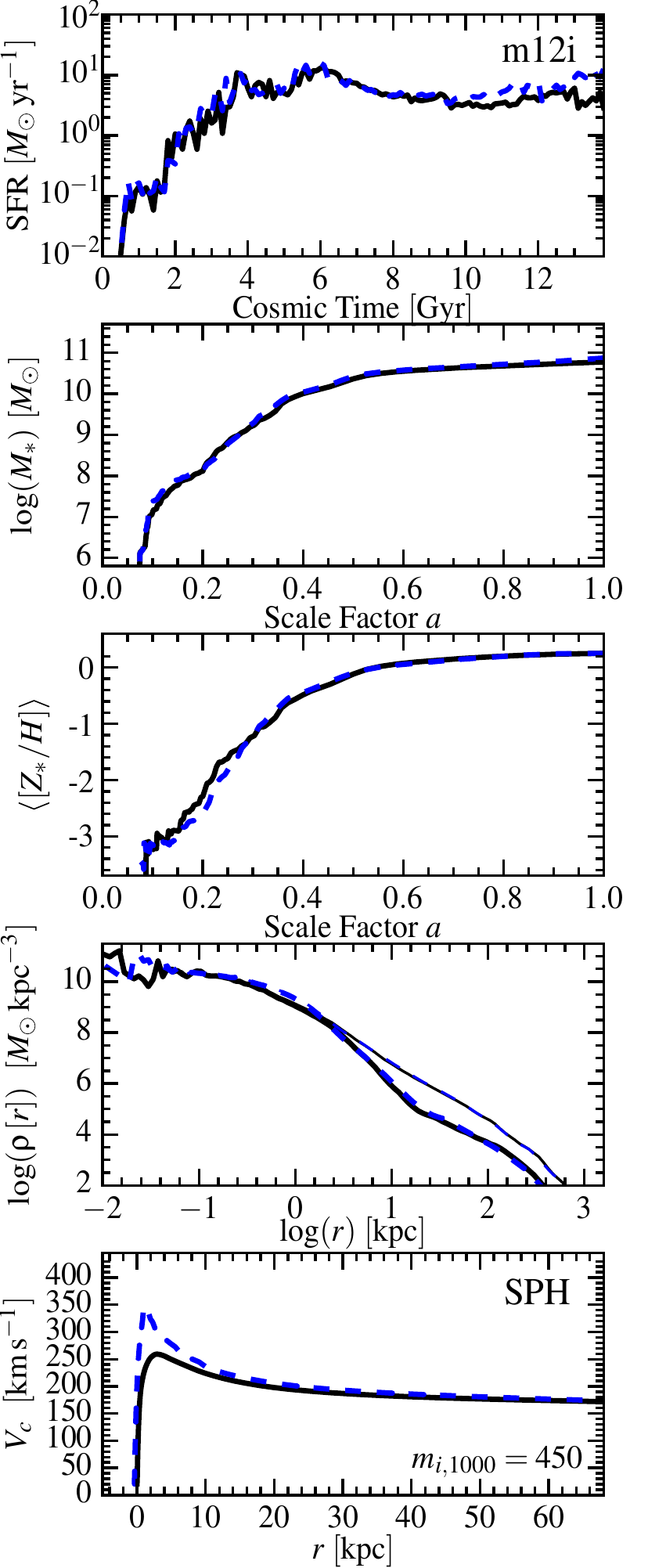}
\end{tabular}
    \vspace{-0.25cm}
    \caption{Effects of radiative FB on {\bf m10q} \&\ {\bf m12i} as Figs.~\ref{fig:ov}-\ref{fig:ov.2}. {\em Left:} We show {\bf m10q} and {\bf m12i} at lower and higher resolution, as in Fig.~\ref{fig:ov}. {\em Right:} We show {\bf m12i} re-run at the lower resolution using the smoothed-particle hydrodynamics implementation used in the original FIRE-1 simulations \citep{hopkins:2013.fire}, as described and compared in detail in \citet{hopkins:fire2.methods}. The qualitative effects of ``turning off'' RFB are similar across resolution levels and hydro solvers. However SPH, at low resolution, produces a slightly less-compact and lower-SFR {\bf m12i} (the difference decreases at higher resolution) -- discussed in detail in \paperone\ -- so the {\em relative} effect of radiation suppressing the steep central rotation curve excess is more obvious. 
    \vspace{-0.5cm}
    \label{fig:ov.sph}}
\end{figure*}

\begin{figure*}
\includegraphics[width={0.99\textwidth}]{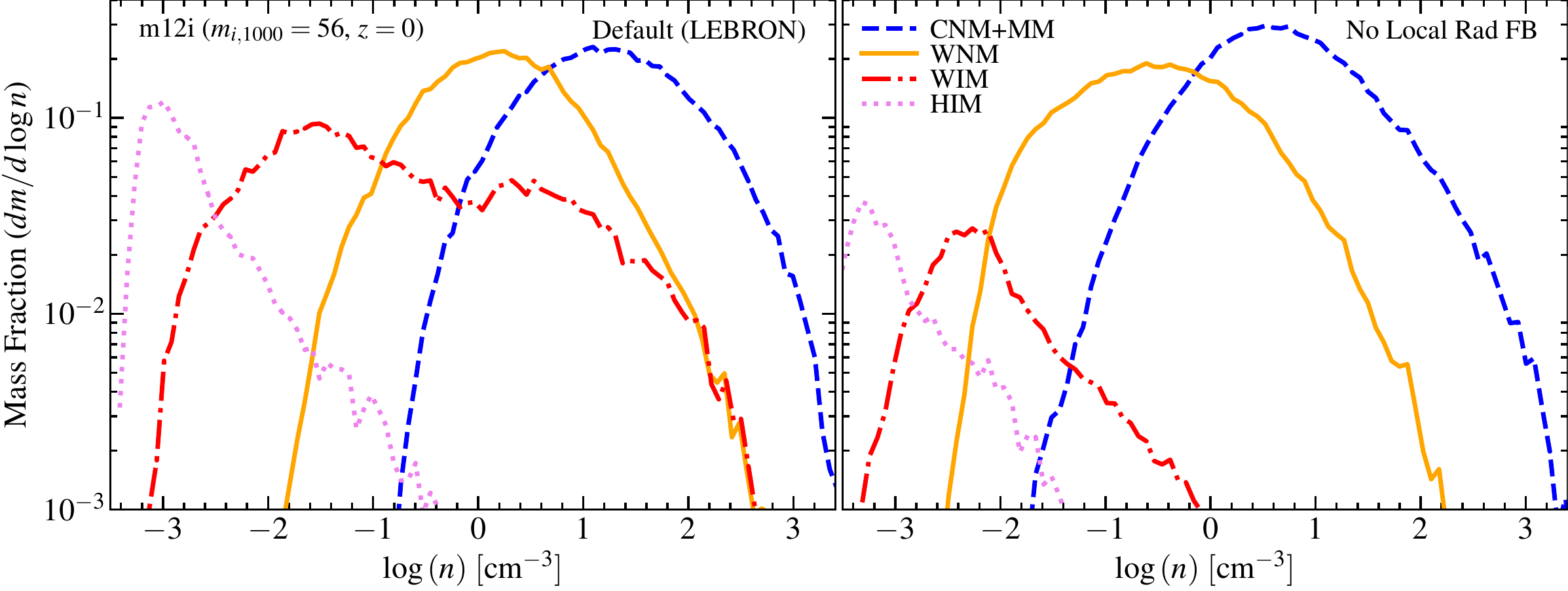}\\
%\hspace{0.5cm}\includegraphics[width={0.96\textwidth}]{figs_fb_physics/phase_diagram_2D_RHD.pdf}
    \vspace{-0.2cm}
    %{\em Top:}
    \caption{Effect of radiative FB on the distribution of ISM gas phases. Differential mass fraction ($dm/d\log{n}$, normalized to the {\em total} gas mass) of gas within $<20\,$kpc of the galactic center of {\bf m12i} at $z=0$, in our ``Default (LEBRON)'' ({\em left}) and ``No Local Rad FB'' ({\em right}) runs . We separately show the cold ($T<1000\,$K) neutral and molecular medium (CNM+MM), warm ($T>1000\,$K) neutral medium (WNM), warm ($T<10^{6}$\,K) ionized (WIM) medium, and hot ($T>10^{6}$\,K) ionized medium (HIM). Ionization states are taken directly from the self-consistent values in-code. In ``Default,'' the  distribution of gas in different phases and densities broadly agrees with canonical Milky Way values \citep[e.g.][]{draine:ism.book}. In ``No Local Rad FB,'' the mass of {\em ionized} gas in the ISM decreases by a factor $\sim 10$; there are no HII regions (no WIM at high $n \gg 1 \,{\rm cm^{-3}}$); lack of (local) photo-ionization and photo-electric heating means the overall mass fraction of CNM is larger (and it dominates the mass budget at lower densities $n\gtrsim 1\,{\rm cm^{-3}}$); the WNM persists (because cooling is inefficient below $\sim 8000\,$K) but dominates even at extremely low densities $n\sim 10^{-2}\,{\rm cm^{-3}}$ (approximately the density where the ISM becomes self-shielding against the meta-galactic UV background). Hot gas is also suppressed owing to the less efficient pre-processing of GMCs (reducing their densities, hence increasing cooling times) before SNe explode within them. 
%{\em Bottom:} Temperature-density diagram (no ionization information) at the same time, weighted by the log of the total gas mass in each pixel (darker is higher). The same trends are evident, but the decrease in the WIM at intermediate densities is less obvious, because this is mostly a change in ionization state as WIM becomes WNM without local radiative FB.
    \label{fig:phase}}
\end{figure*}

\begin{figure*}
%\includegraphics[width={0.99\textwidth}]{figs_fb_physics/phase_diagram_RHD.pdf}\\
%\hspace{0.5cm}
\includegraphics[width={0.96\textwidth}]{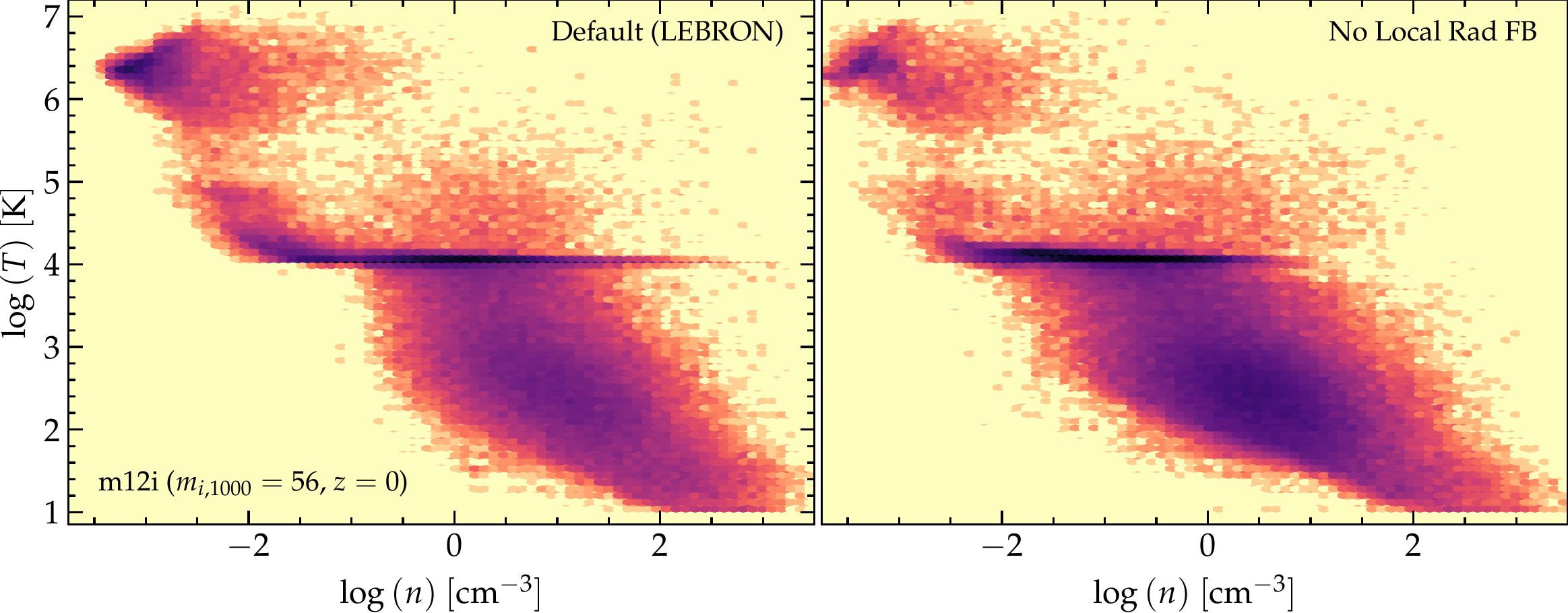}
    \vspace{-0.2cm}
    %{\em Top:}
    \caption{Effect of radiative FB on the distribution of ISM gas phases, as Fig.~\ref{fig:phase}, continued. Here we show a 2D histogram of temperature and density (no ionization information) at the same time, weighted by the log of the total gas mass in each pixel (darker is larger gas mass per unit density-temperature). The same trends are evident, but the decrease in the WIM at intermediate densities from Fig.~\ref{fig:phase} is less obvious, because this is mostly a change in ionization state as WIM becomes WNM without local radiative FB.
    \label{fig:phase.2D}}
\end{figure*}

\subsection{Overview: Net Effects of Radiative Feedback on Galaxy Properties}
\label{sec:ov}

Figs.~\ref{fig:ov}-\ref{fig:ov.2} compares the effects of radiative FB as a function of galaxy mass. For each galaxy IC in our sample (ranging from ultra-faint to MW mass), we compare four simulations: (1) ``Default (LEBRON)'' -- this uses the ``standard'' FIRE-2 radiative FB implementation from \paperone, including five spectral bands accounting for photo-ionization (by both local sources and the UVB), photo-electric heating, continuum absorption (and associated radiation pressure/momentum transfer) by dust in near-UV, optical/NIR, and mid/far IR, and re-emission in the IR. (2) ``Default (M1)'' -- this uses the same source functions, rates, etc., but replaces the photon transport step with the M1 algorithm. (3) ``No Local RHD'' -- this disables all local sources, i.e.\ star particles emit no radiation, but the UVB remains. (4) ``No Radiative FB'' -- this disables local sources and the UVB. 

We run each simulation to $z=0$, then show the resulting star formation history (for all stars which reside within the $z=0$ galaxy), stellar mass growth history,\footnote{We plot the archaeological mass growth, i.e.\ the stellar mass formed at each time, which at $z=0$ resides within the virialized halo.} mean stellar metallicity as a function of time, and $z=0$ (spherically-averaged) mass density profile of baryons and dark matter (and stellar effective radius), and the $z=0$ circular velocity curve. The resolution of each simulation is labeled. Additional details about how each quantity are computed are in \paperone. 

Removing radiative FB entirely, we see order-of-magnitude larger SFRs and stellar masses in dwarfs.\footnote{Note that, for the lowest-mass dwarfs {\bf m09} and {\bf m10q}, the higher stellar and baryonic mass associated with removing all radiative FB actually correlates with {\em lower} central circular velocity. This effect owes to the formation of a ``cored'' dark matter profile via stellar feedback, the efficiency of which is a strong function of the stellar-to-dark matter mass ratio \citep[see detailed studies in][]{onorbe:2015.fire.cores,chan:fire.dwarf.cusps,chan:fire.udgs}.} Not surprisingly, most of this effect comes from the ``external'' UVB. With this fixed, ultra-faint dwarfs (e.g.\ {\bf m09}) are quenched or ``starved'' of new gas by the UVB; as a result (also given their very low stellar masses) the ``local'' radiative FB generated by stars is sub-dominant. In more massive dwarfs, the ``local'' radiative FB has an important effect ``smoothing out'' feedback. Because it provides a ``gentle'' feedback mechanism (e.g.\ keeping gas warm, which can maintain $Q>1$ and prevent runaway gravitational instability in a dwarf galaxy; see \citealt{shetty:2008.sf.feedback.model,kannan:photoion.feedback.sims,rosdahl:2015.galaxies.shine.rad.hydro}) and helps disrupt GMCs before they turn most of their mass into stars \citep{lopez:2010.stellar.fb.30.dor,murray:molcloud.disrupt.by.rad.pressure,harper-clark:2011.gmc.sims,hopkins:fb.ism.prop,colin:2013.star.cluster.rhd.cloud.destruction,grudic:cluster.properties}, it makes star formation less ``violent.'' Without local radiative FB, {\bf m10q}, for example, undergoes complete ``self-quenching'' (gas collapses and ``overshoots,'' forming too many SNe, which then blow out all the gas, and no stars form in the last $\sim 10\,$Gyr, the galaxy has no gas, and has a highly-suppressed metallicity). At still higher masses the potential becomes deeper, so these effects are progressively less prominent.\footnote{The detailed SFRs versus time of the dwarfs, in particular, are quite stochastic and subject to large run-to-run fluctuations in e.g.\ burst timing: as shown in \paperone, systematic differences in mass or trends versus time are much more robust. For example, {\bf m10q} at some resolutions in Fig.~\ref{fig:ov.sph} briefly ceases forming stars for $\sim 1$\,Gyr before ``rejuvenating'' (this occurs at high resolution near $z\sim 0$, with rejuvenation at $z\sim 0.04$). But other runs of {\bf m10q} with seeded minor perturbations in \citet{su:discrete.imf.fx.fire} show the timing and depth of this specific burst-quench cycle are stochastic, while e.g.\ the total stellar mass is robust to $\lesssim 0.1$\,dex.}

In detail: we show explicitly below that star formation is more strongly clustered without ``early'' feedback to disrupt clouds, but this is already well-established from much higher-resolution simulations of individual GMCs, which show they collapse more efficiently and turn much more of their mass into stars in a single free-fall time, without early radiative feedback from HII regions and radiation pressure \citep[see e.g.][]{harper-clark:2011.gmc.sims,colin:2013.star.cluster.rhd.cloud.destruction,grudic:sfe.cluster.form.surface.density,howard:gmc.rad.fx,kim:2017.art.uv.starclusters}.\footnote{Worth noting here, simulations using the same physics as those here but simulating single clouds with resolution reaching $\sim 0.01\,M_{\odot}$ actually find cloud-lifetime integrated star formation efficiencies and radiative feedback efficiencies in good agreement with our galaxy-scale simulations \citep[compare][]{hopkins:fb.ism.prop,oklopcic:clumpy.highz.gals.fire.case.study.clumps.not.long.lived,grudic:sfe.cluster.form.surface.density,grudic:sfe.gmcs.vs.obs}.} We see this below directly in the galaxy morphologies (Fig.~\ref{fig:morphology}, \S~\ref{sec:early}), amplitude of the bursts (Fig.~\ref{fig:ov}), and in our previous studies showing the efficiency of star formation in individual GMCs \citep{hopkins:fb.ism.prop,kim:gc.form.FIRE,moran:2018.metallicity.cooling.direct.collapse}. In turn, many idealized studies of SNe-driven outflows have shown that enhancing the clustering (in time or space) of star formation (hence young stars and SNe explosions) leads to stronger outflows as the clustered explosions more easily produce super-bubbles and chimneys, and lose less energy to radiation \citep[see e.g.][and references therein]{2015MNRAS.454..238W,gentry:clustered.sne.momentum.enhancement,2018MNRAS.481.3325F}. We see this reflected indirectly in the suppressed stellar masses and total baryonic masses (by definition inversely proportional to the outflow ``mass loading factor'' averaged over cosmic time) within $\sim 1-10\,$kpc  in our dwarfs without early or local radiative FB (see e.g.\ {\bf m10q} and {\bf m11b} in Figs.~\ref{fig:ov} \&\ \ref{fig:early.fb}). Most dramatic are the ``self-quenching'' events discussed above. A more detailed study of the wind mass-loading factors and how they vary in-and-out of ``bursts'' (with different feedback physics included or excluded) is in preparation, but for some results, see \citet{muratov:2015.fire.winds}.

By MW-mass, the UVB has only weak effects (on the {\em primary} galaxy -- a more detailed study of how the UVB alters e.g.\ ultra-faint galaxies in the Local Group, as satellites of the MW, will be the subject of future work). And overall, the effects of local radiative FB on gross properties of the galaxy (SFRs, stellar masses, metallicities, baryonic mass profiles, in Fig.~\ref{fig:ov.2}-\ref{fig:ov.sph}) are generally weaker in all respects (though we show below this is not true for their detailed small-scale morphology/structure). This is expected: the galaxies have deep potential wells with escape velocities $\gg 100\,{\rm km\,s^{-1}}$, so photo-ionization heating does not suppress accretion or appreciably ``thicken'' the disk, they have higher gas densities so cooling times are relatively short (warm gas can still radiate efficiently and collapse to form stars), and winds (driven by radiation as well as SNe) simply become much less efficient (the baryonic mass of the galaxy at MW masses is an appreciable fraction of the total supply $\sim f_{\rm baryon}\,M_{\rm halo}$, see \paperone) so it is simply the case that {\em all} feedback effects are much weaker. 

However, in our MW-mass runs some modest ``excessive burstiness'' without local radiative FB is still evident at early times (when the galaxy is a dwarf; see e.g.\ SFRs for {\bf m12i} at $t<6\,$Gyr in Fig.~\ref{fig:ov.2}-Fig.~\ref{fig:ov.sph}), but we also see that the runs without local radiative FB have slightly higher central rotation curve peaks -- this is more obvious in Fig.~\ref{fig:ov.sph} (and in the tests presented in \citealt{hopkins:2013.fire}). Recall, {\bf m12i} in particular exhibits a sharp central rotation curve ``spike'' at this (relatively low) resolution, so the relative effect is small. As discussed in detail in \paperone, this feature in $V_{c}$ is sensitive to both resolution and hydrodynamic methods; which is why the effects of radiation on this feature are more obvious in Fig.~\ref{fig:ov.sph}. These galaxies have high central densities, so failure to destroy GMCs before SNe explode means those explosions would occur in dense environments, suppressing SNe bubble overlap and therefore expulsion of material from dense galaxy centers in galactic super-winds -- we see this below in a suppressed hot gas content within the ISM (Figs.~\ref{fig:phase}-\ref{fig:phase.2D}). Note that the lack of local/early FB still produces more strongly-clustered SF (see \S~\ref{sec:early} below); but unlike in dwarfs (1) the densities are much higher in the galactic nuclei, so SNe become less efficient and bubble overlap is more challenging (as the SNe cooling radii scale $\sim {\rm pc}\,(n/10^{4}\,{\rm cm^{-3}})^{-1/3}$; see \papertwo), (2) optical depths to radiation pressure are much larger (given higher densities and metallicities, compared to dwarfs), and (3) the dynamical times in galaxy centers $\sim 2\,{\rm Myr}\,(R/{\rm kpc})\,(400\,{\rm km\,s^{-1}}/V_{c})$ become shorter than the time over which most SNe explode ($\sim 30\,$Myr), so early feedback becomes more important in regulating against runaway SF \citep{torrey.2016:fire.galactic.nuclei.star.formation.instability,grudic:sfe.cluster.form.surface.density}. In short, in both dwarfs and MW-mass systems, local radiative FB is most important in suppressing SF efficiencies on small spatial/time scales (and/or high density scales). As we show below, for {\em fixed} gas properties, the radiation also has an important role regulating how fast the gas turns into stars in massive systems on the disk scale, but that is hidden here because faster star formation at early times exhausts the gas, leading to lower supply at late times. 

Fig.~\ref{fig:ov.sph} compares ``No Radiative FB'' and ``Default (LEBRON)'' models at different resolution, and using a different hydrodynamic solver (the ``pressure'' formulation of smoothed-particle hydrodynamics, i.e. ``P-SPH'' from \citealt{hopkins:lagrangian.pressure.sph} as used for the FIRE-1 simulations, instead of our default meshless-finite mass or MFM solver). As noted above, \paperone\ contains extensive discussion and comparison of how the hydrodynamic solver and resolution influence our results; our only purpose here is to illustrate that the qualitative effects of radiation are independent of both resolution and hydro solver, even if the quantitative details differ. It is particularly worth noting in this context that \paperone, \citet{guszejnov:imf.var.mw}, and \citet{guszejnov:fire.gmc.props.vs.z} all consider explicit resolution studies of ISM phase structure including GMC mass functions, the cloud linewidth-size relation, and size-mass relations (hence typical cloud surface densities and opacities), and show that these are actually quite robust over factor $>100$ improvements in resolution. Although there is certainty un-resolved sub-structure in the cold gas at any resolution level, quantities like the GMC mass function and size-mass relation are not modified at the resolved, largest masses (which contain most of the star formation, feedback, and cold gas mass in galaxies; see \citealt{solomon:gmc.scalings,blitz:gmc.properties,murray:2010.sfe.mw.gmc,rice:2016.gmc.mw.catalogue}) -- rather increasing the resolution simply extends these to smaller and smaller clouds (which do not contribute much to the total mass or star formation budget). 

\begin{figure}
\begin{center}
\includegraphics[width=0.8\columnwidth]{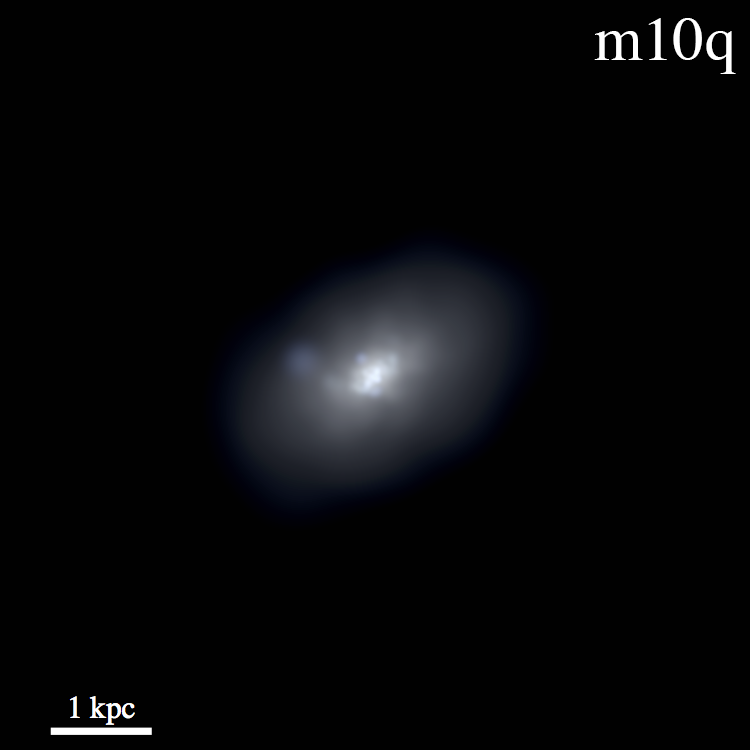} \\
\includegraphics[width=0.8\columnwidth]{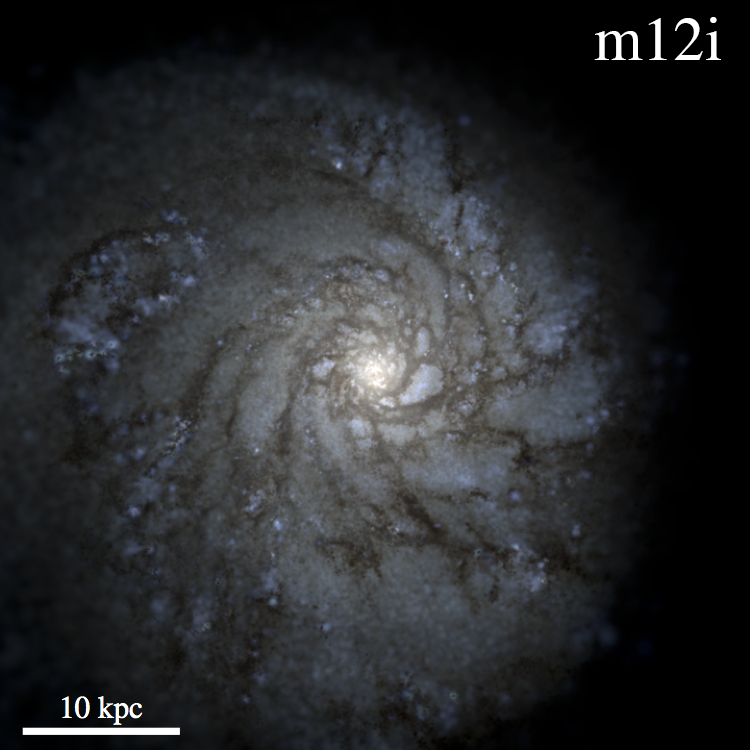} 
\end{center}
    \vspace{-0.25cm}
    \caption{Mock image of our highest-resolution ``Default (LEBRON)'' {\bf m10q} run and {\bf m12i} run at $z=0$. These are compared in Figs.~\ref{fig:ov}-\ref{fig:rad.pressure.tau.distribution}. {\bf m10q} is a dSph with no coherent morphological structure -- as a result, there is no discernable difference in its visual morphology in our different runs. Moreover, we see that there is essentially no dust obscuration, owing to low gas densities and (more importantly) very low metallicities $\sim 0.01\,Z_{\odot}$, so only ionizing photon absorption by neutral gas produces large effects. 
     {\bf m12i} exhibits substantial differences (see e.g.\ Fig.~\ref{fig:morphology}) owing to its thin-disk morphology; the run here shows the young stars are mostly, at late times, forming in spiral arms which are highly dust-obscured, where most of the absorption studied in Figs.~\ref{fig:rad.pressure.coupling}-\ref{fig:rad.pressure.tau.distribution} occurs (these lanes/clouds have gas surface densities $\sim 50-100\,M_{\odot}/{\rm pc}^{-2}$). \vspace{-0.5cm}
    \label{fig:morphology.m10q.m12i.z0}}
\end{figure}

\begin{figure*}
\begin{tabular}{cc}
\hspace{-0.27cm}
\includegraphics[width=0.5\textwidth]{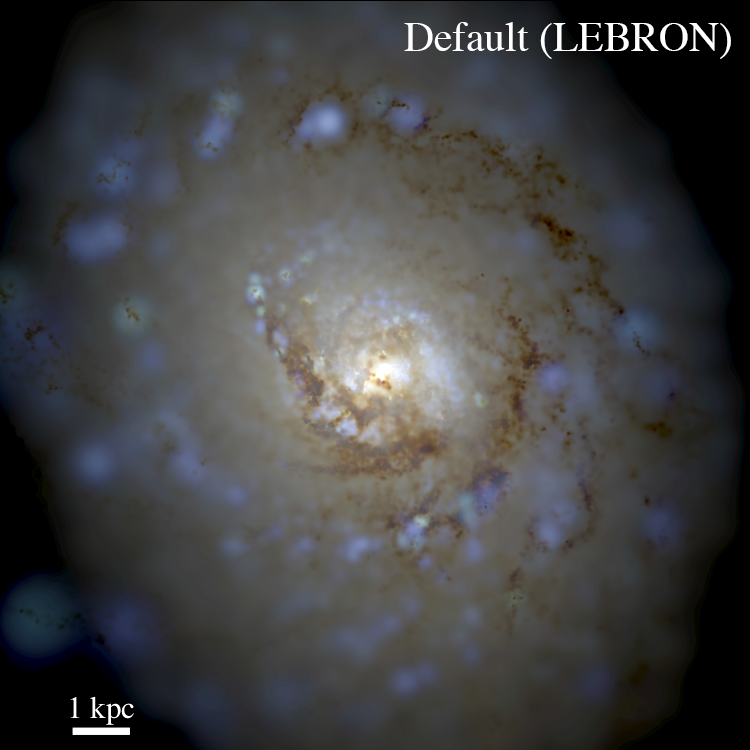} &
\hspace{-0.45cm}
\includegraphics[width=0.5\textwidth]{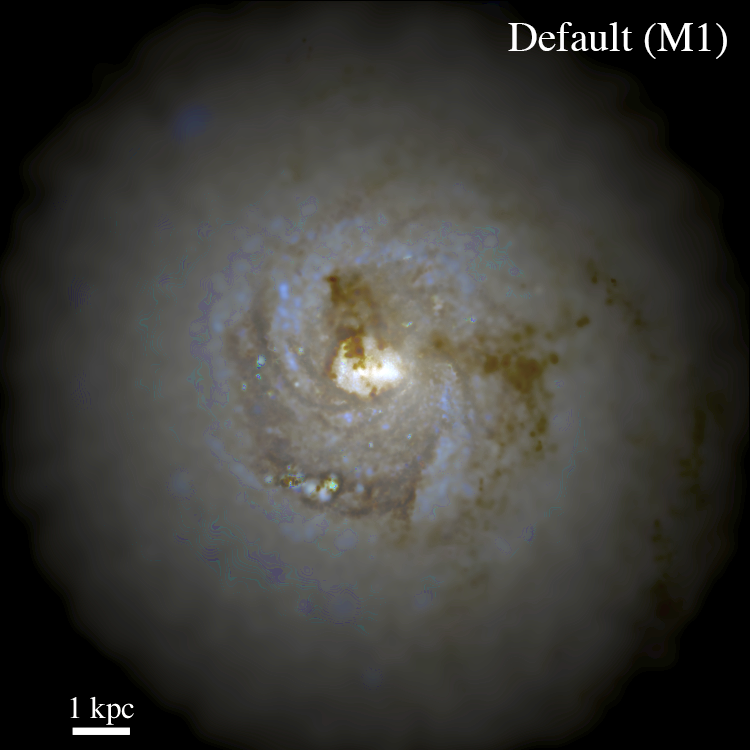} \\
\hspace{-0.27cm}
\includegraphics[width=0.5\textwidth]{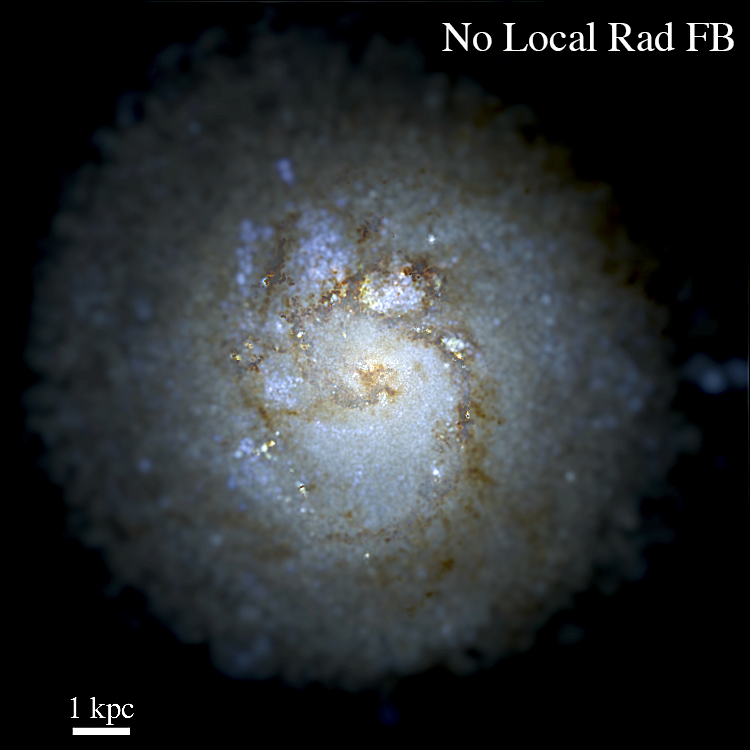} &
\hspace{-0.45cm}
\includegraphics[width=0.5\textwidth]{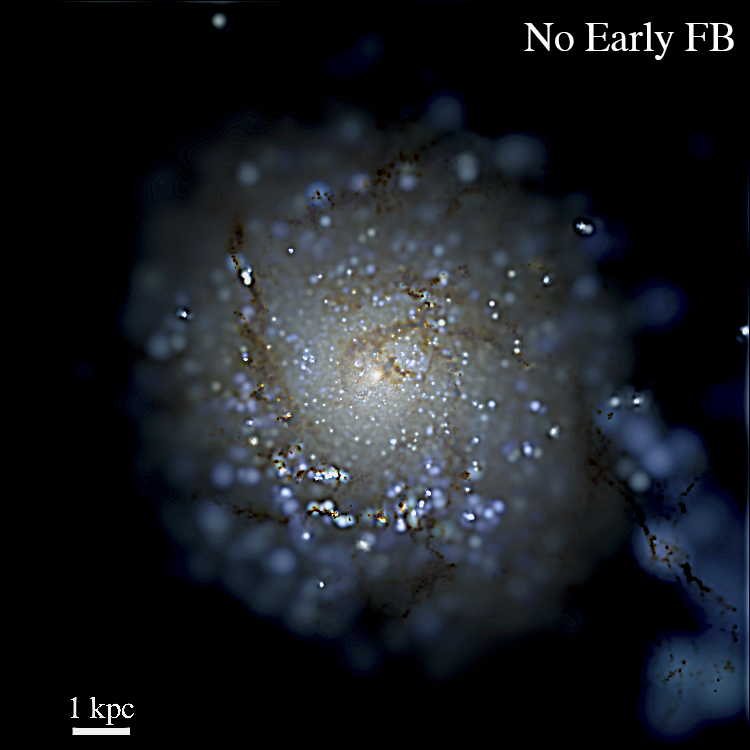} 
\end{tabular}
    \vspace{-0.25cm}
    \caption{Mock images of our Milky Way-mass {\bf m12i} galaxy ($m_{i,\,1000}=56$) at $z\approx 0.9$ (the lowest redshift to which all runs were run). We compare the ``Default (LEBRON)'' and ``Default (M1)'' ({\em top}) and ``No Local Rad FB'' and ``No Early FB'' runs ({\em bottom}) from Fig.~\ref{fig:early.fb}. While the details of e.g.\ disk thickness and spiral arm structure differ between LEBRON and M1, these are highly time-variable and subject to stochastic run-to-run variations. Without local radiative FB, the galaxy size and mass and integrated color are similar, but the spiral structure is significantly less obvious owing to the more-bursty episodes blowing out gas, and there are more small, discrete star clusters. Without any early FB, the galaxy morphology and stellar mass is entirely dominated by hyper-compact star clusters. Although some vague spiral-like structure appears here it is an artifact of a recent merger -- the star clusters which dominate the stellar mass are mostly on nearly-radial orbits.  \vspace{-0.5cm}
    \label{fig:morphology}}
\end{figure*}

Figs.~\ref{fig:phase}, \ref{fig:phase.2D}, \ref{fig:morphology}, \ref{fig:morphology.m10q.m12i.z0} compare the phase distributions and visual morphologies of a subset of these runs. We specifically examine how radiative feedback alters the temperature distribution of dense, cool gas in the halo, and how the combination of radiative and other ``early'' FB channels alters the visual morphology of the galaxies. Note that although the visualizations of morphology shown are mock images (i.e.\ light-weighted), the differences in morphology persist if we make a simple stellar mass-weighted map. Though the effect is subtle, we see the MW-mass runs without local radiative FB produce somewhat less-disky morphologies. Runs without radiation have a substantially different gas temperature distribution, as expected (e.g.\ HII regions do not exist).\footnote{In Fig.~\ref{fig:phase}, we include all ionized gas with $T<10^{6}\,$K in the ``warm ionized'' medium, but because it is weighted by the ionized gas mass, material with $T\ll 10^{4}\,$K contributes negligibly.} The lack of ``pre-processing'' of GMCs by HII regions and winds also means that the clouds have higher densities when SNe explode, which means they, in turn, have shorter cooling times, producing less hot gas (see e.g.\ the results in non-cosmological simulations in \citealt{hopkins:stellar.fb.winds} or in idealized experiments in \citealt{haid:snr.in.clumpy.ism}). Note that for these runs we focus on MW-mass systems, where the differences are most pronounced. The low-mass dwarfs all have irregular/spheroidal morphologies -- the fact that small dwarfs ($M_{\rm halo} \ll 10^{11}\,M_{\sun}$) tend to be spheroidal or irregular in morphology, and that the stellar orbits are dispersion (as compared to rotation)-dominated, is not surprising (this is expected for many reasons in systems with $V_{c} \ll 100\,{\rm km\,s^{-1}}$ for many reasons; see e.g.\ \citealt{wheeler.2015:dwarfs.isolated.not.rotating}). Indeed previous studies have shown these morphological traits remain even if we remove {\em all} stellar feedback \citep{hopkins:2013.fire,hopkins:fire2.methods,hopkins:sne.methods}, or add additional physics such as magnetic fields and cosmic rays \citep{su:2016.weak.mhd.cond.visc.turbdiff.fx,hopkins:cr.mhd.fire2}. The ``quantitative photometric morphologies,'' on the other hand, defined using e.g.\ colors or Sersic indices may vary, however, as we have shown in Fig.~\ref{fig:ov} that the central stellar mass profile and SFRs do vary.

\vspace{-0.5cm}
\subsection{The Role of ``Early'' Feedback}
\label{sec:early}

\begin{figure*}
    \plotsidesize{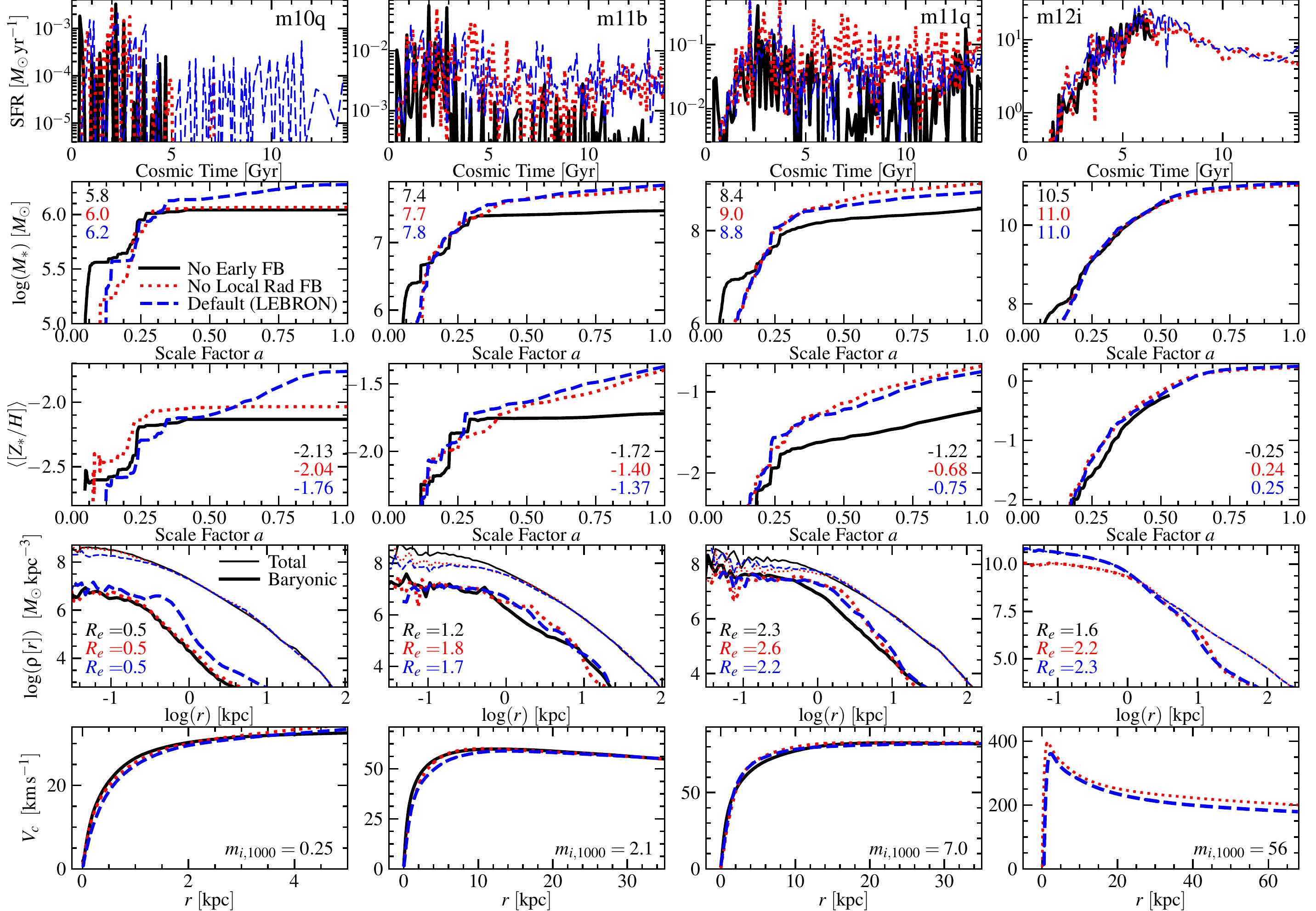}{0.99}
    \vspace{-0.25cm}
    \caption{As Fig.~\ref{fig:ov}, but comparing the effects of removing {\em all} ``early'' FB (local radiative FB and ``early'' fast O/B winds from massive stars), keeping the UVB in place. 
    Some of the effects of local radiative FB can be ``made up for'' by fast stellar winds (e.g.\ the small differences between ``Default'' and ``No Local Rad FB'' in {\bf m11q}), as implemented in FIRE-2 (see text for discussion). Removing both produces strong ``self-quenching'' in both {\bf m10q} and {\bf m11b}, and nearly in {\bf m11q} (up to LMC mass scales) -- these galaxies overshoot (form many more stars early, at high redshifts), then blow out much of their baryonic gas mass. This leads to them having almost no late-time SF (despite being isolated dwarfs), and having strongly suppressed metallicities ($\sim 0.3 -0.5$\,dex below the observed mass-metallicity relation). In {\bf m12i} the differences in formation history are less obvious, but as shown below the galaxy has a wildly different morphology and is dominated by tiny, dense star clusters (the run has to be stopped at $z\sim1$). \paperone\ \&\ \papertwo\ show that removing {\em just} stellar O/B winds, while keeping radiative FB, leads to much smaller effects than shown here -- the important thing is that {\em some} early FB is present.
     \vspace{-0.5cm}
    \label{fig:early.fb}}
\end{figure*}

Fig.~\ref{fig:early.fb} compares runs where we remove all local radiative FB and all other ``early FB'' (FB from massive stars before they explode). In our implementation, this includes radiation from massive stars, as well as stellar mass loss in O/B winds from massive stars before they explode (see \paperone\ for details). These runs are also compared in Figs.~\ref{fig:phase}, \ref{fig:phase.2D}, \ref{fig:morphology}, \ref{fig:morphology.m10q.m12i.z0}.

The effects described above in \S~\ref{sec:ov} become much more dramatic without {\em any} ``early FB.'' Every galaxy forms many more stars early, at redshifts $z\gtrsim 6-7$ (while a small fraction of their $z=0$ mass, this makes them order-of-magnitude more massive at these times). The SFR ``spikes'' at much higher values and these over-violent bursts produce ``self-quenching'' in {\bf m10q} and {\bf m11b}. Even {\bf m11q}, an LMC-mass system, essentially self-quenches for $\sim 5-10\,$Gyr, although it recovers below redshift $z \lesssim 0.5$. In all cases these ``blowouts'' strongly suppress the metallicity (by $\sim 0.3-0.4$\,dex), pushing the galaxies significantly below the observed stellar mass-metallicity relation (compare \citealt{ma:2015.fire.mass.metallicity}). We also see in the baryonic mass profiles that the systems are significantly more baryon-poor out to $\gtrsim 10\,$kpc -- i.e.\ they have ejected most of their gas (see e.g.\ the suppression of $\rho_{\rm baryon}$ from $\sim 1-10\,$kpc with ``No Early FB'' compared to ``Default'' in the second-from-bottom panels in Fig.~\ref{fig:early.fb}). These effects are also evident in Fig.~\ref{fig:morphology}-\ref{fig:morphology.m10q.m12i.z0}. 

In {\bf m12i}, these effects are proportionally smaller, but in fact we have to stop the run at $z\approx1$ as the timesteps become extremely small ($\sim 1\,$yr). The reason is obvious in Fig.~\ref{fig:morphology}: without {\em any} early FB, dense GMCs collapse on a timescale faster than their stellar evolution timescale (at densities $n\gtrsim 1000\,{\rm cm^{-3}}$ typical of dense GMCs, the free-fall time is $\lesssim 1\,$Myr). As shown in many previous, much higher-resolution studies of {\em individual} GMCs (see references in \S~\ref{sec:intro} or \citealt{grudic:sfe.cluster.form.surface.density}) or ``zoom-ins'' of GMCs in galaxy simulations \citep{kim:gc.form.FIRE}, this leads to the GMC turning most of its mass into stars, and leaves behind a very dense, bound remnant. As a result, {\em most} of the stellar mass is composed of extremely dense bound star clusters (for comparison, $\lesssim 1\%$ of the stellar mass in the MW is in such objects; see e.g.\ \citealt{harris96:mw.gcs,peng:2008.gc.specific.frequencies}). Because the galaxy is essentially assembling hierarchically from ``minor mergers'' of dense collisionless stellar clumps, its morphology has no recognizable disk and little angular momentum. 

This is consistent with previous studies, which have shown that without {\em some} form of ``early FB,'' galaxy-scale simulations disagree at the order-of-magnitude level with observations of quantities such GMC mass functions, size-mass and virial parameter scalings, and GMC lifetimes \citep{hopkins:fb.ism.prop,oklopcic:clumpy.highz.gals.fire.case.study.clumps.not.long.lived,grudic:sfe.cluster.form.surface.density}, as well as the ratio of various dense gas tracers in the ISM (e.g.\ CO vs.\ HCN; see \citealt{hopkins:dense.gas.tracers}). Similarly, other studies have shown that ``early FB'' has an order-of-magnitude effect on ionizing photon escape fractions (as, absent any such FB to create channels in GMCs before the most massive stars -- which produce almost all the ionizing photons -- explode, the escape fraction is negligibly small; see \citealt{ma:2015.fire.escape.fractions,ma.2016:binary.star.escape.fraction.effects}). 

We emphasize that, in \paperone\ \&\ \papertwo, as well as several of the references above, it was shown that disabling {\em only} O/B mass-loss (or all stellar mass-loss), while retaining radiative FB from both photo-ionization and radiation pressure, produces only minor effects (significantly smaller, in fact, than removing radiative FB while retaining stellar mass-loss). It is therefore {\em not} the case that the O/B winds ``dominate.''\footnote{Moreover, as briefly noted in \citet{grudic:sfe.cluster.form.surface.density}, the default FIRE scaling (used here) from an older version of {\small STARBURST99} for O/B mass loss rates extrapolates, at low metallicities, to higher mass-loss rates than given by other more recent stellar evolution models (particularly those favored by the massive black hole mergers in LIGO; see discussion in \citealt{lamberts:bh.mgr.progenitors.ligo,lamberts:lisa.bh.binary.pop.pred}). We have experimented (not shown here) with a more recent model, which has weak effects overall (see \paperone), but this does make the effects of removing radiation {\em as well} more dramatic in dwarfs, since the O/B winds can ``make up for'' less.} Rather, it seems that the different ``early FB'' channels: photo-ionization heating, single-scattering radiation pressure, and O/B mass-loss, can ``compensate'' to some extent for one another (\citet{2009MNRAS.396..377D,2017MNRAS.464.2963S,kimm:lyman.alpha.rad.pressure} argue that multiply-scattered Ly-$\alpha$ photons can also act in this manner, in metal-poor dwarfs). This should not be surprising: at solar metallicities the IMF-averaged momentum flux in massive stellar winds is $\sim L/c$, the same as that from single-scattering radiation pressure, and the momentum flux from warm gas pressure in a ``typical'' massive HII region is also similar (see discussion in \citealt{lopez:2010.stellar.fb.30.dor}). They all act on similar (short) time and (small) spatial scales, as they come from the same massive stars. At low metallicities (e.g.\ our {\bf m10q} dwarfs), stellar winds have a proportionally lower mass-loss rate and momentum flux, so they are less able to ``compensate'' for a lack of radiative FB, hence the stronger effects of removing local radiative FB in these runs.

Following the more detailed discussion in \citet{grudic:sfe.cluster.form.surface.density}, feedback from massive young stars (so $L \sim (1200\,L_{\sun}/M_{\sun})\,M_{\ast}$) imparts a momentum flux $\dot{p} \sim \eta\,L/c$ on gas in a typical GMC with surface density $M_{\rm GMC}/\pi\,R_{\rm GMC}^{2} \sim 100\,M_{\sun}\,{\rm pc^{-2}}$. Equating this to the gravitational force ($\sim G\,M_{\rm GMC}^{2}/R_{\rm GMC}^{2}$) implies that the cloud will be destroyed when $M_{\ast}/M_{\rm GMC} \sim 0.05/\eta$. So as long as a modest fraction of $L/c$ can couple, the cloud will self-regulate with $M_{\ast}/M_{\rm GMC} \ll 1$, producing an open-cluster type, unbound remnant, and re-cycling the mass, producing a low star formation efficiency locally. So while the difference between e.g.\ $\eta=1$ and $\eta=3$ (one versus all three mechanisms above acting in concert) might be detectable in individual cloud properties and star formation efficiencies \citep[see e.g.][]{grudic:sfe.gmcs.vs.obs}, all will produce effectively the same large-scale result. Given this, it is clear that strong constraints on which ``early FB'' mechanisms dominate (under various conditions) will not come from galaxy-scale properties, but from observations which can probe these small-scale phenomena \citep[consistent with many previous studies that have found galaxy-scale star formation efficiencies are de-coupled from cloud-scale star formation efficiencies; see e.g.][]{hopkins:rad.pressure.sf.fb,hopkins:virial.sf,hopkins:dense.gas.tracers,federrath:2012.sfr.vs.model.turb.boxes,agertz:2013.new.stellar.fb.model,orr:ks.law,orr:non.eqm.sf.model,semenov:local.vs.global.sfe}.

\begin{figure*}
    \plotsidesize{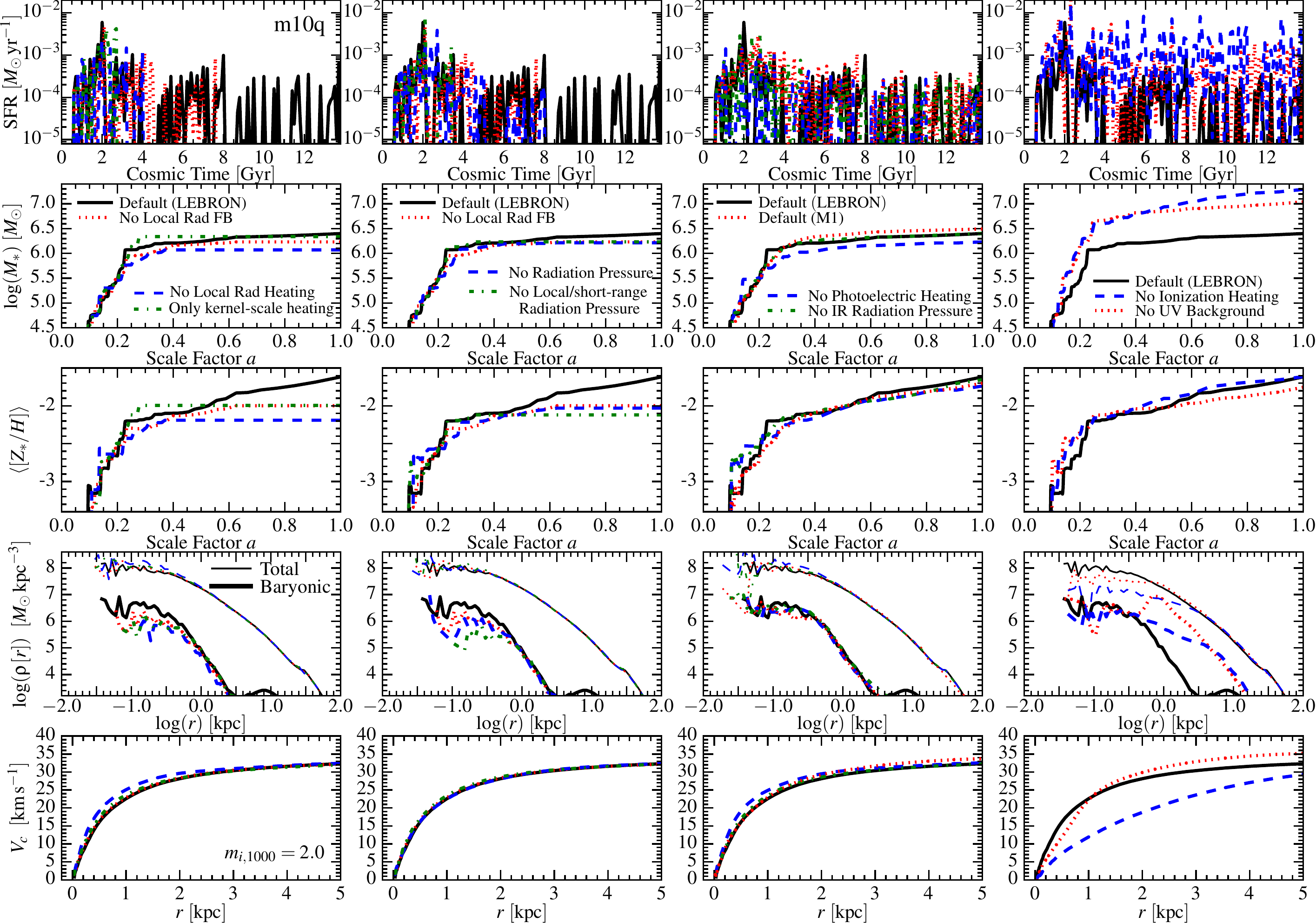}{0.99}
    \vspace{-0.25cm}
    \caption{Effects of individual radiative FB channels/algorithmic aspects, using our Default (LEBRON) scheme, in {\bf m10q}, as Fig.~\ref{fig:ov}, at one level lower resolution. All panels compare our Default (LEBRON) runs to variations.
    {\bf (1)} ``No Local Rad FB,'' we see without local radiative FB the early-time SFH is again overly-bursty (``gentler'' radiative FB fails to slow down SF before SNe), leading to more violent ``blowout'' of metals and gas, suppressing [Z/H] and the $z=0$ gas fraction. 
   {\bf (2)} ``No Local Rad Heating'': we turn off all local (non-UVB) radiative heating terms (Compton, photo-ionization, photo-electric, from simulation stars) but keep radiation pressure; results are similar to ``No Local Rad FB.'' 
   {\bf (3)} ``Only kernel-scale heating'' only allows local radiative heating in gas which is an immediate neighbor of a star particle (keeping UVB and radiation pressure); effects are similar to ``No Local Rad Heating.'' Long-range photo-heating appears to be sub-dominant.
   {\bf (4)} ``No Radiation Pressure'' turns off all RP; results are similar to ``No Local Rad FB.'' 
   {\bf (5)} ``No local/short-range radiation pressure'' turns off the ``short-range'' (kernel-scale) RP terms, but keeps long-range RP; the effects are similar to removing all RP (dynamical RP effects mostly occur on small scales).
   {\bf (6)} ``Default (M1)'' uses M1 instead of LEBRON, results are similar to Default (LEBRON).
   {\bf (7)} ``No Photoelectric Heating'' turns off just photo-electric heating; effects are weak. Photo-ionization is the most important heating term.   
   {\bf (8)} ``No IR Radiation Pressure'' ignores IR re-radiation and multiple-scattering. Effects are negligible in low-metallicity dwarfs.
   {\bf (9)} ``No Ionization Heating'' turns off local {\em and} UVB-based photo-ionization heating; effects are similar to ``No Radiative FB'' and slightly stronger than ``No UV Background.''
   {\bf (10)} ``No UV Background'' keeps all local radiative FB, but disables the UVB; effects are similar to ``No Ionization Heating.''
   Although the UVB clearly has the most dramatic effects on dwarfs, local warm gas pressure from photo-ionization heating and single-scattering RP are both significant.
    \label{fig:rad.fx.m10q}}
\end{figure*}

\begin{figure}
    \includegraphics[width=0.94\columnwidth]{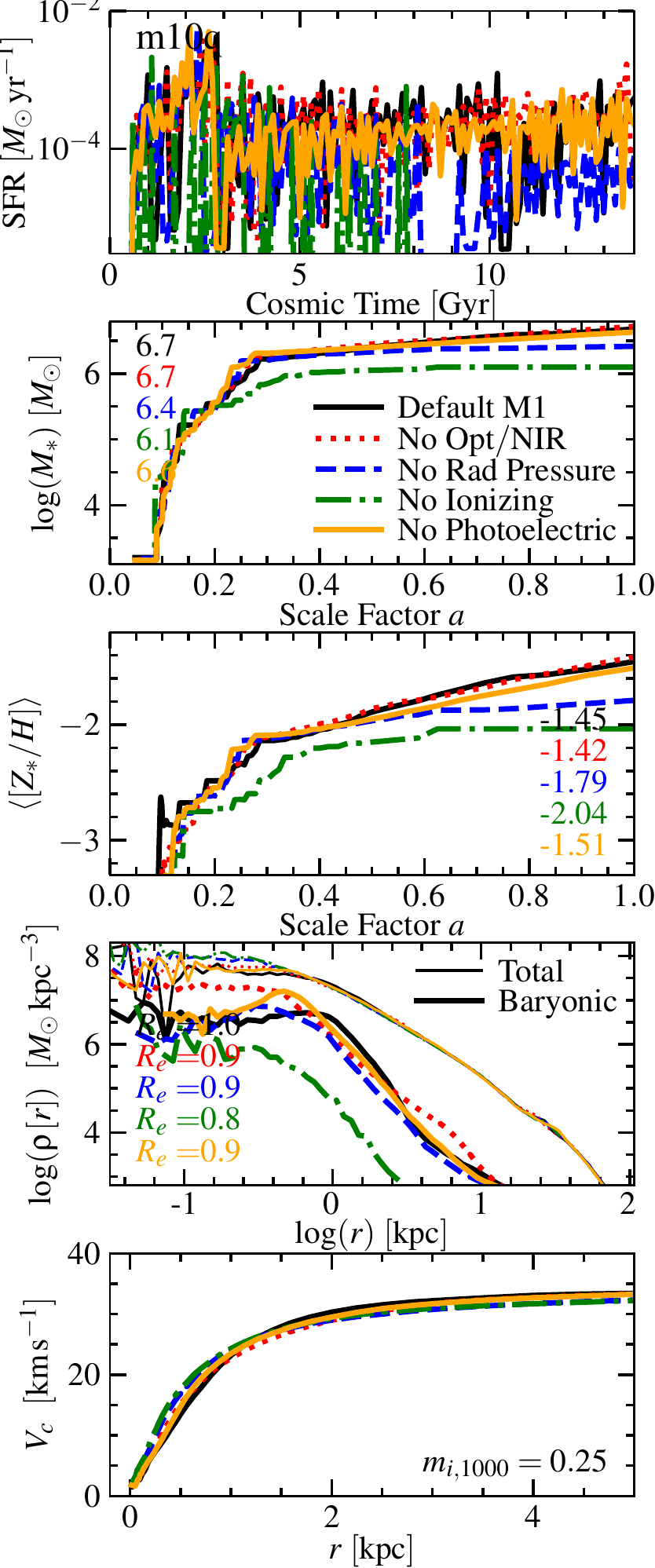}
    \vspace{-0.25cm}
    \caption{Effects of different local radiative FB channels, as Fig.~\ref{fig:rad.fx.m10q}, in our dwarf {\bf m10q}, but for runs using the M1 photon transport algorithm (turning off different wavebands in turn). Consistent with Fig.~\ref{fig:rad.fx.m10q}, single-scattering radiation pressure has a modest effect (with photo-heating present). Removing all local ionizing photons (both their heating and radiation pressure), but keeping the UVB, is similar to removing all local radiative FB. Other bands have weaker effects at dwarf masses.
    \vspace{-0.5cm}
    \label{fig:rad.fx.m10q.m1}}
\end{figure}

\begin{figure}
\includegraphics[width=0.94\columnwidth]{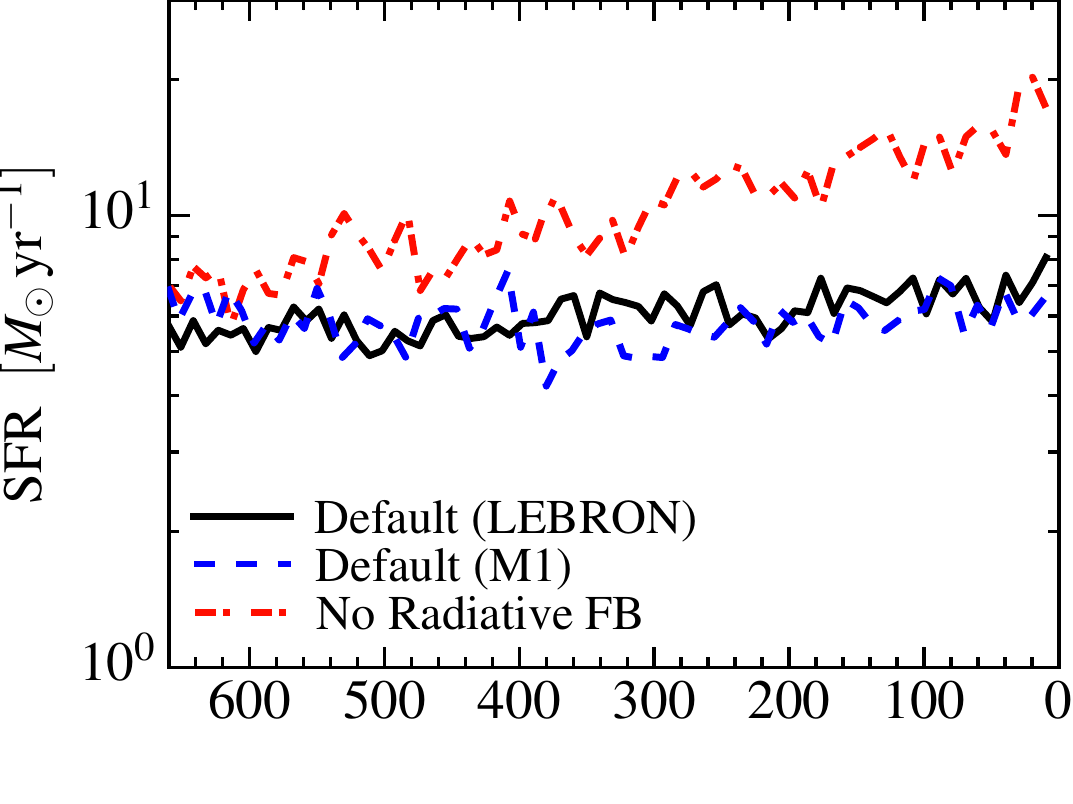} 
\vspace{-0.5cm}\\
\includegraphics[width=0.94\columnwidth]{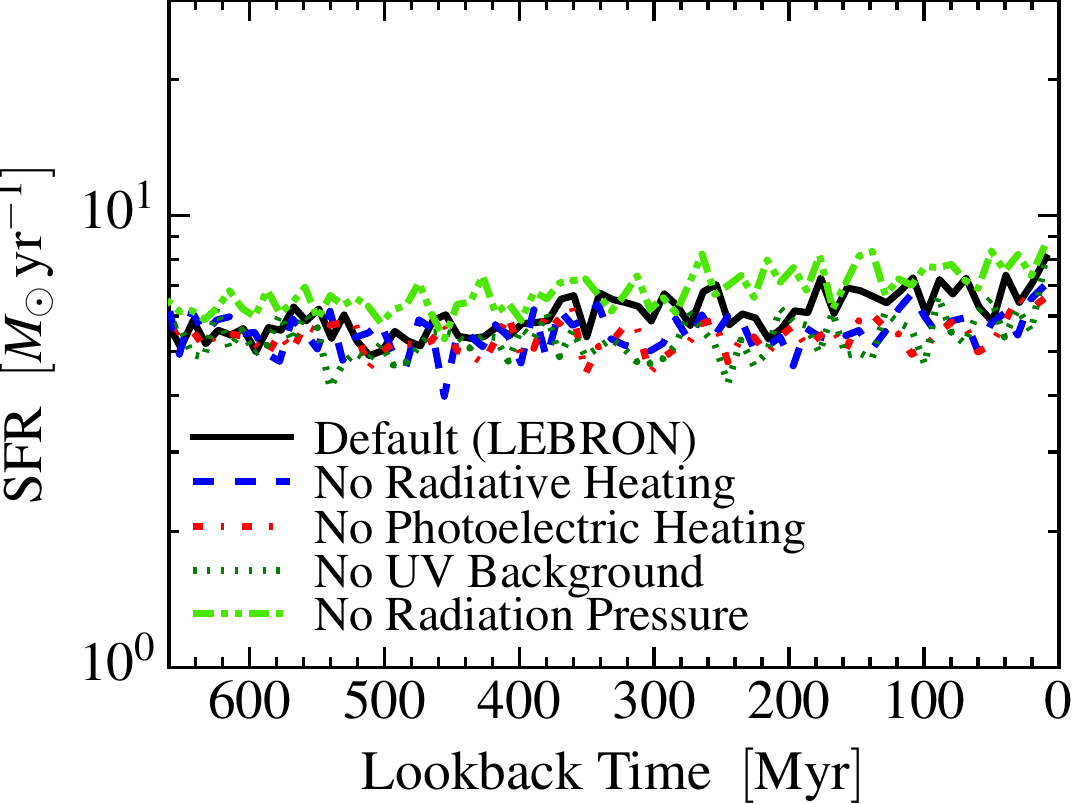} \\
    \vspace{-0.5cm}
    \caption{Variations of the radiative feedback, as Fig.~\ref{fig:rad.fx.m10q}, but re-starting a run of {\bf m12i} from identical ICs at $z\approx 0.05$, and running it to $z=0$ with varied physics. 
    Removing radiative FB leads to a significantly higher SFR at $z\sim0$, {\em for fixed initial conditions} (e.g.\ galaxy gas mass) -- the apparently smaller effect in Fig.~\ref{fig:ov.2} is because the ``No Radiative FB'' run exhausts somewhat more gas earlier, changing the late-time galaxy properties. 
    LEBRON \&\ M1 give similar results. Removing {\em either} RP or radiative heating alone produces little effect, but removing both together has a large effect -- one can ``make up for'' the other (along with stellar O/B mass-loss).
    \label{fig:rad.fx.z0.m12i.restart}}
\end{figure}

\vspace{-0.5cm}
\subsection{Effects of Different Radiative Feedback Channels}
\label{sec:mechanisms}

Figs.~\ref{fig:rad.fx.m10q}, \ref{fig:rad.fx.m10q.m1}, \ref{fig:rad.fx.z0.m12i.restart} break down the effects of individual radiative feedback channels, in turn. In Fig.~\ref{fig:rad.fx.m10q}, we take the ``Default (LEBRON)'' algorithm as a base, then turn off individual components of radiative feedback in turn to examine their separate effects. Owing to computational expense, we focus on comparison of one cosmological simulation ({\bf m10q}) where the total effect of radiative FB is significant, and run these tests at one level lower resolution than our initial comparison in Figs.~\ref{fig:ov}-\ref{fig:ov.2} (but note that the behaviors in all four variations ``Default (LEBRON),'' ``Default (M1),'' ``No Local Rad FB,'' and ``No Radiative FB'' are essentially identical at this and the higher resolution level). Fig.~\ref{fig:rad.fx.m10q.m1} begins from our ``Default (M1)'' model and similarly turns off, in turn, different wavebands evolved here. Fig.~\ref{fig:rad.fx.z0.m12i.restart} repeats the exercise from Fig.~\ref{fig:rad.fx.m10q} in a MW-mass system, re-starting our {\bf m12i} simulation at redshift $z\approx 0.05$ and re-running it to $z=0$ (approximately $\sim 1\,$Gyr) as in \paperone. The advantage of the ``controlled restart'' is that it allows us to see the effects of different feedback {\em specifically} in high-mass galaxies: since the MW-mass system ``begins'' as a dwarf (at high redshift), effects there resemble our {\bf m10q} run, and propagate forward (confusing the comparison). 

Fig.~\ref{fig:rad.fx.m10q.xnetwork} compares our dwarf simulations, with our ``Default (M1)'' implementation (five-band RHD including ionizing/EUV, photo-electric/FUV, NUV, optical/NIR, and MIR/FIR radiation) compared to the ``Extended Network (M1)'' described in \S~\ref{sec:feedback:radiation:sources} and Appendix~\ref{sec:luminosity.opacity.descriptions}, which expands this to a 10-band RHD treatment (dividing the ionizing band into 4 sub-bands with separate HeI and HeII ionizing bands, and adding Lyman-Werner, soft and hard X-rays, and dynamical radiation temperature-dependent FIR bands).

\vspace{-0.5cm}
\subsubsection{The Meta-Galactic UV Background}
\label{sec:feedback:radiation:tests:effects:UVB}

As shown above, for dwarfs, the most important form of radiative FB is the UVB: e.g.\ turning off {\em all} photo-heating in {\bf m10q} results in order-of-magnitude larger mass, while keeping just the local stellar radiation but removing the UVB produces factor of $\sim 5$ larger stellar mass (fourth column of Fig.~\ref{fig:rad.fx.m10q}). At $z=0$, assuming a continuous SFR $\dot{M}_{\ast}$, the same stellar SED templates used in-code, optically thin photon escape, and that SF is concentrated near a galaxy center ($r=0$, so flux scales $\propto 1/r^{2}$), the UVB (also taking the in-code values) should dominate the UV radiation energy density emitted by the young stars at distances $r \gtrsim 2\,{\rm kpc}\,(\dot{M}_{\ast}/10^{-3}\,M_{\sun}\,{\rm yr}^{-1})^{1/2}$, so this is not surprising. 

Somewhat less obviously, but consistent with previous studies \citep[e.g.][]{thoul.weinberg:uvb.effects.on.dwarfs}, we see significant effects from the UVB extending to $V_{\rm max}$ as large as $\sim 100\,{\rm km\,s^{-1}}$. In halos with $M_{\rm vir} \sim 10^{11}\,M_{\odot}$ and $V_{\rm max} \sim 50-100\,{\rm km\,s^{-1}}$ ({\bf m11b} and {\bf m11q}), the UVB suppresses the $z=0$ stellar mass by factors $\sim 2-3$, and in halos with $V_{\rm max}\sim 40-50\,{\rm km\,s^{-1}}$ ($M_{\rm vir}\sim 10^{10}\,M_{\odot}$, our {\bf m10q}) the difference is order-of-magnitude (Figs.~\ref{fig:ov}-\ref{fig:ov.2}).\footnote{Also as noted in \S~\ref{sec:ov}, because of the effects of stellar feedback on dark matter core creation, the circular velocity itself can be non-linearly sensitive to the change in star formation efficiency, while other quantities like metallicity change as expected with stellar mass.} These are well above the classical UVB ``quenching'' threshold ($V_{\rm max} \sim 10-20\,{\rm km\,s^{-1}}$), and indeed are not ``quenched'' with a UVB. The calculation above shows that the UV radiation energy density from local stars dominates inside the galaxy effective radii (compare $R_{e}$ and $\dot{M}_{\ast}$ to the equation above), so this is not where the UVB has an effect. However, at radii $\gtrsim R_{\rm vir}$ at $z\sim0$, the circular velocities are $\sim 20-50\,{\rm km\,s^{-1}}$ in these more massive halos, so the pressure support from the UVB contributes substantially, and suppresses the baryonic mass inside of $R_{\rm vir}$ by factors as large as $\sim 5-10$ (directly visible in the baryonic mass profiles; Figs.~\ref{fig:ov}-\ref{fig:ov.2}). Thus, we confirm that the UVB provides an important ``preventive'' or ``suppressive'' form of FB up to $V_{\rm max} \sim 50-100\,{\rm km\,s^{-1}}$. 

By MW-mass ($V_{\rm max} \gtrsim 200\,{\rm km\,s^{-1}}$), the UVB has almost no effect on the primary galaxy, as expected. Of course, from the arguments above, we expect it to have a large effect on the mass function of satellites (small dwarfs) around the local group. This will be investigated in more detail in future studies (Wheeler et al., in prep.) which examine more realistic allowed variations in the UVB and their effect on satellite properties.

\vspace{-0.5cm}
\subsubsection{Local Photo-Ionization Heating}
\label{sec:feedback:radiation:tests:effects:photoionization.heating}

Fig.~\ref{fig:rad.fx.m10q} shows that turning off {\em either} local-photo ionization heating, {\em or} radiation pressure, within the LEBRON scheme, produces a similar effect to turning off all local radiative feedback. We have also run a parallel set of M1 runs to Fig.~\ref{fig:rad.fx.m10q}, presented in Fig.~\ref{fig:rad.fx.m10q.m1}. With the M1 runs in Fig.~\ref{fig:rad.fx.m10q.m1} we see that removing all local ionizing/UV photons (i.e.\ both their heating and radiation pressure) produces effects similar to removing all local radiative FB, while removing {\em just} the radiation pressure (keeping photo-heating) produces a significant, but not-as dramatic effect. The two schemes (LEBRON and M1, in Figs.~\ref{fig:rad.fx.m10q} and \ref{fig:rad.fx.m10q.m1}, respectively) are therefore qualitatively consistent. In both schemes, {\em both} local photo-ionization heating and radiation pressure are important for the effects we described above. 

Interestingly, in Fig.~\ref{fig:rad.fx.z0.m12i.restart} at MW masses we see the opposite: turning off photo-ionization heating alone or radiation pressure alone produces almost no effect, but turning off both at the same time produces a large effect on the SFR. It appears that in {\bf m12i}, photo-heating and radiation pressure are able to more directly ``take over'' from one another (either one can pre-process large GMCs, such that the effects of SNe exploding in those clouds, for example, is similar). This is similar to our conclusions in FIRE-1 \citep{hopkins:2013.fire}, although the overall effect was stronger there owing to the different treatment of O/B mass-loss, as discussed above. This is also consistent with previous studies of non-cosmological, isolated galactic disks, including full RHD treatment of photo-ionizing stellar feedback \citep{kannan:photoion.feedback.sims,rosdahl:2015.galaxies.shine.rad.hydro,emerick:rad.fb.important.stromgren.ok}.

Why are radiation pressure and photo-ionization (as well as other ``early feedback'' mechanisms discussed above) able to ``take over'' or ``compensate'' for one another in MW-mass systems more efficiently than in dwarfs? This is at least partly a combination of metallicity and density effects. At solar metallicities (e.g.\ typical MW-like conditions), the optical depth to NIR/optical/NUV photons through GMCs is appreciable ($\tau \gtrsim 1$), so single-scattering radiation pressure carries $\sim L/c$ momentum flux, while at low metallicities, the dust opacity is small so only ionizing photons are efficiently absorbed -- thus the radiation pressure forces are suppressed by a factor $\sim 2$ (see Fig.~\ref{fig:rad.pressure.coupling} and \S~\ref{sec:where} below). This is similar to the effect of metallicity on the momentum flux in early (O/B) stellar winds (discussed in \S~\ref{sec:early}). Meanwhile, for $Q\sim1$ disks, the typical velocity dispersions in large GMC complexes scale with the dispersion in the disk \citep[see e.g.][]{hopkins:fb.ism.prop}; so in MW-mass systems, HII regions expanding at the ionized gas sound speed $\sim 8-10\,{\rm km\,s^{-1}}$ are only marginally able to unbind clouds with similar dispersions, while in dwarfs the lower densities and cloud internal velocity dispersions allow them to be more easily un-bound (for a detailed demonstration of this in cloud simulations, see \citealt{grudic:sfe.cluster.form.surface.density}).

In our LEBRON RHD method, we can also control separately the short-range (kernel-scale surrounding each star) and long-range (propagated via the gravity tree) components of the radiative feedback. For photo-ionization heating, Fig.~\ref{fig:rad.fx.m10q} demonstrates that the long-range component is actually the most important in our dwarf galaxy. In other words, heating diffuse gas (e.g.\ the extended gas disks which form when there are inflows, and diffuse gas within the halo/CGM surrounding the galaxy) is physically most important for the slowing/regulating star formation we see in the dwarf galaxies. The short-range term primarily manifests as local HII regions around the youngest star particles -- these help destroy GMCs before SNe explode, so have some significant effects, but (because the gas is dense) involve proportionally little gas mass. In contrast, photo-heating a diffuse disk to $\sim 10^{4}\,$K in a galaxy with $V_{c}\sim 30\,{\rm km\,s^{-1}}$ raises the local Toomre $Q\gg1$, suppressing star formation significantly (for a more detailed study using ray-tracing in comparison to ``local'' ionization treatments in non-cosmological simulations, see \citealt{emerick:rad.fb.important.stromgren.ok}).

\vspace{-0.5cm}
\subsubsection{Photo-Electric Heating}
\label{sec:feedback:radiation:tests:effects:photoelectric.heating}

At both dwarf and MW masses, we see weak galaxy-scale effects from photo-electric heating. This is consistent with nearly all previous studies of (non-cosmological) galaxy and star formation simulations \citep[see e.g.][]{tasker:2008.gas.turb.vs.gal.prop,tasker:2011.photoion.heating.gmc.evol,dobbs:2008.gmc.behavior.insensitive.to.photoelectric.heating,hopkins:fb.ism.prop,su:2016.weak.mhd.cond.visc.turbdiff.fx,richings:2016.chemistry.uvb.photoelec.fx,hu:photoelectric.heating,hu:2017.rad.fb.model.photoelectric}. Essentially all of these studies (some of which reach mass resolution $\sim 1\,\msun$) conclude that while the details of cold gas phase structure ($\ll 10^{4}\,$K) and fragmentation down to stellar mass scales (i.e.\ the scales where the thermal Jeans mass in GMCs and cold gas becomes relevant) are sensitive to photo-electric heating (and indeed we do see effects in the temperature-density distribution in Figs.~\ref{fig:phase}-\ref{fig:phase.2D}), the behavior we resolve in our studies here is all well in the regime where super-sonic turbulence dominates the dynamics. Given that \paperone\ showed there were very weak effects on most properties studied here if one simply turned off {\em all} cooling below $\sim 10^{4}\,$K, and that perhaps the most important effects of radiation in dense/cold gas gas (namely regulating the IMF; see \citealt{offner:2009.rhd.lowmass.stars,offner:2013.imf.review,bate:2012.rmhd.sims,hansen:2012.lowmass.sf.radsims,guszejnov.2015:feedback.imf.invariance,guszejnov:protostellar.feedback.stellar.clustering.multiplicity}) are implicitly sidestepped by assuming a fixed stellar IMF, it should not be surprising that explicit treatment of photo-electric heating is a generally sub-dominant effect here.

\vspace{-0.5cm}
\subsubsection{Single-Scattering Radiation Pressure (UV/Optical/NIR)}
\label{sec:feedback:radiation:tests:effects:radiation.pressure}

As discussed above in \S~\ref{sec:feedback:radiation:tests:effects:photoionization.heating}, in our dwarf {\bf m10q}, both radiative heating and radiation pressure produce similar effects, and the two in concert produce a smoother (less bursty) star formation history that continues down to $z=0$ (Figs.~\ref{fig:rad.fx.m10q}-\ref{fig:rad.fx.m10q.m1}). Fig.~\ref{fig:rad.fx.m10q} does confirm that the effects of radiation pressure are dominated by the local/short-range (i.e.\ kernel-scale) coupling: specifically, we disable the momentum coupling from the locally absorbed photons in the LEBRON method (those absorbed within the ``short-range'' component calculated within a single kernel around the stellar source, as described in \paperone), but retain the radiation pressure from the long-range component propagated through the gravity tree, and find this is very similar to simply disabling all radiation pressure. This is expected, given the arguments below, since most of the coupling occurs around young stars embedded in GMCs. But it is also reassuring: in our default RHD method, only the long-range component is non-photon-conserving -- this and the agreement between our default method and the M1 method suggest this is not a significant source of error.\footnote{In previous work, e.g.\ \citet{hopkins:stellar.fb.winds}, we argued that the long-range component might be key to re-accelerating or ``lofting'' winds, in particular in massive starbursts, but we did not explicitly attempt to separate the various components as we do here. While this certainly might still occur, our results here suggest this is not critical for {\em most} of the outflows in dwarfs, which are predominantly driven by SNe explosions.} In our MW-mass {\bf m12i} run in Fig.~\ref{fig:rad.fx.z0.m12i.restart}, it seems radiative feedback is important in regulating the SFR into better agreement with the Schmidt-Kennicutt law\footnote{Specifically, the ``Default'' run in Fig.~\ref{fig:rad.fx.z0.m12i.restart} agrees well with the observed Schmidt-Kennicutt relation, as shown in \citet{orr:ks.law}, while the ``No Radiative FB'' run has the same gas surface density (by construction in these restarts, but a factor $\sim3-4\times$ larger SFR.} (similar to our conclusions in \citealt{hopkins:rad.pressure.sf.fb}), but in the absence of just radiation pressure, a similar effect can be ``made up for'' by radiative heating. Although not shown, we again find the radiation pressure terms are dominated by local coupling.

The radiation pressure effects we see in Figs.~\ref{fig:rad.fx.m10q}-\ref{fig:rad.fx.z0.m12i.restart} are almost entirely {\em single-scattering} effects. To see this, Fig.~\ref{fig:rad.pressure.coupling} quantifies the total radiation pressure which has coupled to gas in the galaxy, in a subset of our simulations (using our LEBRON scheme). Specifically, we record the total momentum imparted from photons to gas $p_{\rm coupled} \equiv \sum |\Delta {\bf p}_{a}|$ (with the sum over all particles and timesteps in the simulations, every time a radiation pressure term is calculated), and compare this to the integrated photon momentum from all photons emitted by all stars in the simulation, $p_{\rm available} \equiv c^{-1}\,\sum L_{\rm bol}^{a}(t - t_{\rm form})\,\Delta t_{a}$ (the sum over the bolometric luminosity of all star particles, integrated over all times, since each particle forms). We define $\langle \tilde{\tau} \rangle \equiv p_{\rm coupled} / p_{\rm available}$, and see this is typically $\sim 0.5-0.7$, i.e.\ slightly less than complete single-scattering. As discussed below, that is because some of the emitted optical from older stars escapes, while most of the UV/ionizing radiation is absorbed.

If we directly quantify the multiply-scattered component (here, the radiation pressure from the IR bands), we see it is totally negligible in dwarfs, and rises to just $p_{\rm multiple} \sim 0.1\,p_{\rm available}$ in MW-mass galaxies. This can be important for e.g.\ the dusty nuclei of massive galaxies (during bursts of star formation) and/or individual star cluster formation episodes; but it is not a dominant effect for most star formation.

\begin{figure*}
    \plotsidesize{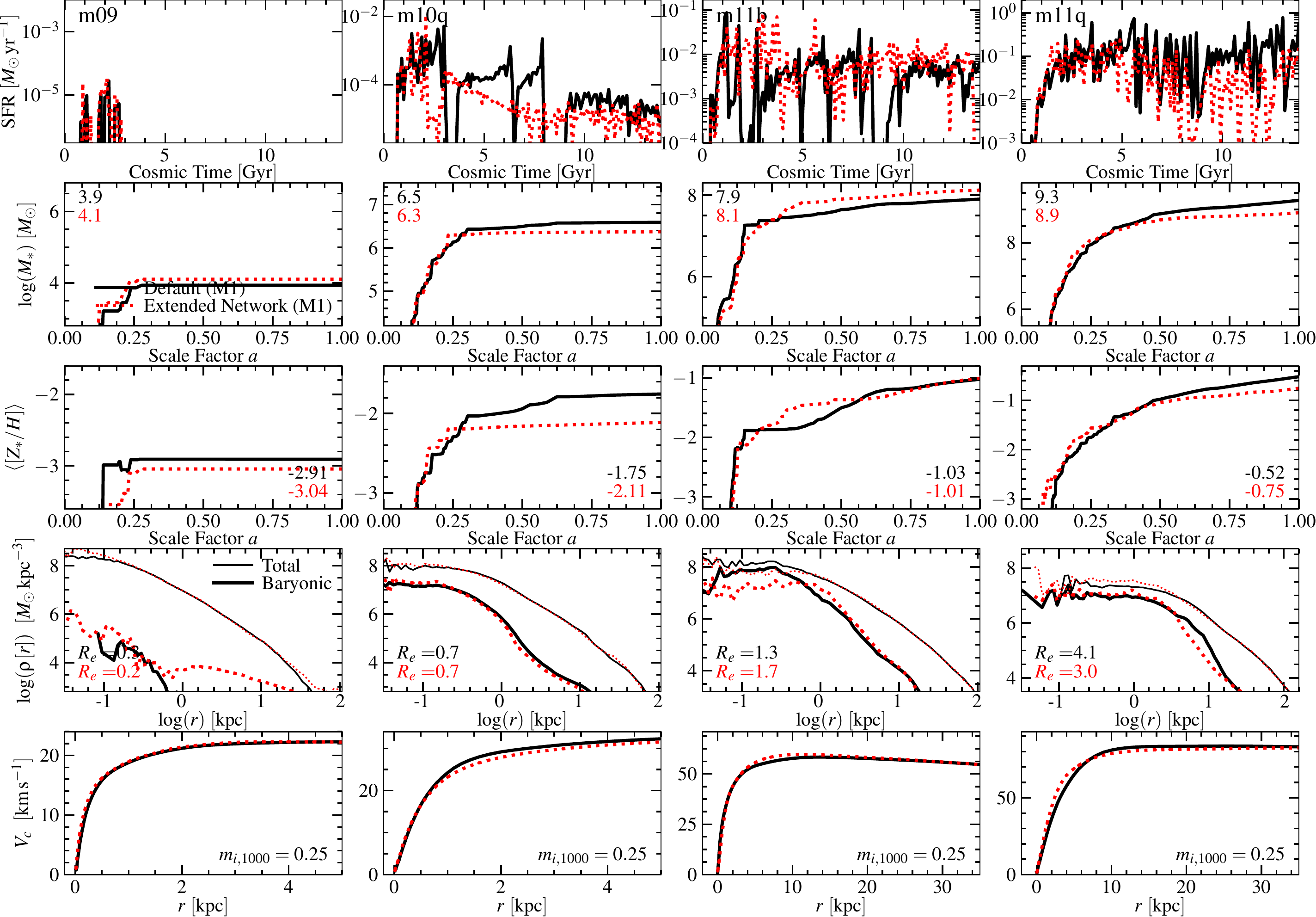}{0.99}
    \vspace{-0.25cm}
    \caption{Effects of extended, additional radiative feedback channels on the simulations, as Fig.~\ref{fig:rad.fx.m10q.m1}. We compare ``Default (M1)'' and runs using the ``Extended Network'' in M1, described in \S~\ref{sec:other.channels}. The latter includes all the identical ``Default (M1)'' physics, but adds several bands including: (1) separate soft and hard X-rays from X-ray binaries, and associated Compton heating, (2) dividing our single-band ionizing spectrum (which assumes a universal spectral shape for ionizing photons) into 4 separate sub-bands (separately tracking He I and He II ionizing photons), (3) treating the IR not as a grey, single-band ``bin'' but self-consistently evolving the dust and IR radiation temperature fields and using a self-consistent opacity, with coupled dust-gas thermal heating and exchange, and (4) adding explicit Lyman-Werner band transport with an approximate treatment of its effects on molecular cooling. These are all expected to have small effects at this mass and redshift range, and we confirm this (the effects on the SFRs are largely consistent with stochastic run-to-run variations). 
    \label{fig:rad.fx.m10q.xnetwork}}
\end{figure*}

\vspace{-0.5cm}
\subsubsection{Additional Channels}
\label{sec:other.channels}

With the extended frequency network active (Fig.~\ref{fig:rad.fx.m10q.xnetwork}), we see relatively little systematic change in our dwarf galaxies. Some differences in the detailed SFR vs time are clearly evident, along with $\sim 10-20\%$ changes in mass, but these appear to be essentially random, and are consistent with stochastic run-to-run variations in these simulations \citep[see][]{su:2016.weak.mhd.cond.visc.turbdiff.fx}. This is not particularly surprising, because the additional mechanisms in this network are not expected to dominate the channels included in our ``default'' simulations, but we briefly discuss these channels in turn to note why this is.

\begin{enumerate}

\item{\bf Compton Heating (via XRBs):} If we assume a constant SFR $\dot{M}_{\ast}$, and use this to estimate the soft and hard X-ray luminosity produced by XRBs (Appendix~\ref{sec:luminosity.opacity.descriptions}), and in turn the Compton heating rate $Q_{\rm Compton}$ (see \paperone) for gas with density $n$ at a distance $r$ from the galaxy, and compare this to the normal cooling rate ($\dot{e}_{\rm cool} = \Lambda\,n^{2}$ with $\Lambda \sim 10^{-23}\,{\rm erg\,cm^{3}\,s^{-1}}$), we obtain $Q_{\rm Compton}/\dot{e}_{\rm cool} \sim 10^{-8}\,(\dot{M}_{\ast} / M_{\sun}\,{\rm yr^{-1}})\,(n/0.01\,{\rm cm^{-3}})^{-1}\,(r/10\,{\rm kpc})^{-2}$. In other words, Compton heating is totally negligible.\footnote{\citet{cantalupo:ionization.by.xrays.from.stars} similarly show that the effect of XRBs on the cooling rates of CGM gas via indirect feedback (altering the ionization of the gas) is small unless the SFRs are extremely large ($\gg 100\,M_{\sun}\,{\rm yr^{-1}}$) and the gas lies in a narrow range of temperatures around $\approx 1\times10^{5}\,$K.} This could be important in the near vicinity ($r < 1\,$pc) of a luminous AGN, but should not be important from stars, and we confirm this by including these terms in our extended network. A more interesting X-ray feedback channel is either non-equilibrium metal-line over-ionization (altering the metal-line cooling rate) or IGM ionization by redshifted X-rays, but our simulations do not include the relevant physics or scales to follow these.

\item{\bf (Explicit) He Ionization:} Our extended network includes a 4-band treatment for photo-ionizing radiation, with separate tracking of HeI and HeII (vs.\ HI) ionizing photons. However, at the coarse-grained level here, we find this does not make a large difference to the bulk galaxy properties we see. Recall, the effects of ionization from local stars alone (if we still include a fixed UVB, radiation pressure, and O/B mass-loss) are relatively subtle, so it is not surprising that making our ``Default (M1)'' ionization treatment slightly more accurate has small effects. Note that in our default treatment, we still account separately for HeI and HeII ionization in the cooling routines and chemistry calculations (see \paperone\ for details); we simply assume a spatially-uniform spectral shape for the ionizing photons (fixed to the UVB spectral shape from \citealt{faucher-giguere:2009.ion.background}, which can evolve in time). The spatial variations occur only on small spatial and timescales in the vicinity of very young, hot stellar populations -- these could easily be important to dynamics in individual HII regions and corresponding emission-line diagnostics, but are second-order on galactic scales. 

\item{\bf Ly-Werner Radiation:} The extended network also includes a very approximate treatment of the effect of H$_{2}$ dissociating (Lyman-Werner or LW) radiation on molecular cooling (specifically calculating an equilibrium molecular fraction depending on the incident flux, and reducing the metal-free contribution to the cooling rate below $<10^{4}$\,K accordingly). However, as we showed and discussed at length in \paperone, {\em completely} removing this plus all metal-line and atomic cooling below a few thousand Kelvin effectively has no influence on our large-scale conclusions. Many other chemical studies have reached the same conclusions, specifically that molecular cooling produces essentially no appreciable dynamical effects on star or galaxy formation above metallicities [Z/H]$\gtrsim -3$ (or $-5$, if dust cooling is included; see \citealt{glover:2011.molecules.not.needed.for.sf,hopkins:fb.ism.prop,dopke.2013:fragmentation.all.dust.levels.but.enhanced.with.crit.dust,ji:2014.silicate.dust.cooling.for.metal.poor.star.criterion.and.tests,moran:2018.metallicity.cooling.direct.collapse}). Recently, \citet{lupi:2018.h2.sfr.rhd} used the M1 implementation in {\small GIZMO}, and broadly similar stellar feedback models, coupled to the non-equilibrium {\small KROME} chemistry module \citep{krome:2014.chemistry}, to explore the non-equilibrium effects of H$_{2}$ dissociating radiation on evolved galaxies in idealized and cosmological simulations. While they concluded the LW transport is, of course, important for correctly modeling the H$_{2}$ and therefore associated diagnostics, it again had no significant dynamical effects on SFRs or other galaxy properties, in the mass and redshift range studied here.

\item{\bf IR Thermal Dust Heating:} Our extended network allows for the dust temperature to come into appropriate equilibrium with the IR radiation field, and thermally couple to the gas via dust-gas collisions. This tends to raise or lower the gas temperature to be in equilibrium with the dust temperature around temperatures below $\sim 30-100\,$K, above densities $\gtrsim 10^{6}\,{\rm cm^{-3}}$ (where it becomes dominant). Unsurprisingly, for the reasons above, and because these densities are well above those resolved here, this has negligible effects here. As discussed above, this might in nature determine the IMF turnover (regulating the thermal Jeans mass around $\gtrsim 0.1\,M_{\sun}$), but we {\em assume} an IMF and do not resolve this, and confirm it has no large-scale {\em direct} dynamical effects (though of course if the IMF changed, this could be important for feedback).

\item{\bf Non-Grey IR:} In the extended network, the IR is no longer treated as a single bin with a single opacity, but rather as a pseudo black-body where the radiation, dust, and gas temperatures are all evolved independently and explicitly, with opacities that depend on these temperatures (and metallicity). It has been argued that this more sophisticated treatment could substantially alter the ability of IR radiation to multiply-scatter (because it will be down-graded to longer wavelengths and lower opacities as it does so). However, we show below (\S~\ref{sec:ir}) that such multiple-scattering accounts for a very small fraction of the total radiative FB, so (unsurprisingly) these higher-order corrections to it make little galaxy-scale difference. 

\item{\bf Ly-$\alpha$ Resonant Scattering:} Our RHD methods do not allow us to consider multiple Lyman $\alpha$ scattering. However, we briefly discuss it here. Recently, \citet{kimm:lyman.alpha.rad.pressure} considered a detailed study of simulations including a sub-grid model for Ly-$\alpha$ RHD (together with a similar multi-band treatment of ionizing, photoelectric, single and multiple scattering in UV/optical/IR to our ``Default (M1)'' runs) in idealized (non-cosmological) simulations of a dwarf galaxy (similar in mass to our {\bf m11b}). Although the details of their numerical hydrodynamic method, treatment of SNe and stellar mass loss differ substantially, they reach remarkably similar conclusions to our study here about the role of radiation and early feedback in dwarf galaxies. In particular, they argue that although the Ly-$\alpha$ luminosity is only a small fraction of the continuum, it can be multiply-scattered giving a net momentum flux $\dot{p}_{{\rm Ly}\alpha} \sim (10-300)\,L_{{\rm Ly}\alpha}/c$, larger by several than the continuum.\footnote{As they note, this effect will likely diminish rapidly in more massive, dust-rich galaxies, as dust destroys Ly-$\alpha$ photons.} This amplifies all the effects studied here: star formation is less bursty, more warm gas is supported, and cluster formation is suppressed. Studies in more idealized galactic wind environments, but using explicit Ly-$\alpha$ RHD, have reached qualitatively similar conclusions \citep[see][]{2017MNRAS.464.2963S,2018MNRAS.479.2065S}.

\end{enumerate}

\vspace{-0.5cm}
\subsubsection{Infrared Radiation \&\ Photon-Trapping or Multiple-Scattering}
\label{sec:ir}

As shown in Fig.~\ref{fig:rad.pressure.coupling}-\ref{fig:rad.pressure.tau.distribution}, the momentum $L/c$ contributed by multiply-scattered IR radiation is completely negligible in dwarf galaxies (reaching values $\ll 0.01$), rising with galaxy mass until it reaches just $\sim 0.1$ in MW-mass galaxies. This rising importance is expected, given the increasing metallicities (hence dust opacities) and surface densities in more massive galaxies (e.g.\ {\bf m10q}, with median gas surface density $\Sigma_{\rm gas} \sim 10\,\msun\,{\rm pc^{-2}}$ and metallicity $Z\sim 0.02\,Z_{\sun}$, has median IR optical depth $\sim 0.0004$). As discussed below, even in MW-mass galaxies, most absorption occurs in ``typical'' GMCs with IR optical depths $\sim 0.1$. Over the limited range of resolution we explicitly probe here (e.g.\ decreasing the mass resolution by a factor $\sim 8-64$), this conclusion appears robust, and (as noted above) our previous resolution studies \citep{hopkins:fire2.methods,guszejnov:imf.var.mw,guszejnov:fire.gmc.props.vs.z} have shown that this characteristic $\Sigma_{\rm gas}$ and (corresponding) IR optical depth is robust over factor $>100$ changes in resolution. However, the ``tail'' of absorption at very high column densities in Fig.~\ref{fig:rad.pressure.tau.distribution} is more prominent (as expected) at higher resolution where we can follow smaller, denser structures (clumps, cores) which typically have higher column densities.

We stress that there is no artificial large ``boost factor'' or ``added optical depth'' applied to radiative feedback in any of our simulations. IR photons, in principle, can be trapped and multiply-scatter: if one has a source of luminosity $L_{\rm bol}$, surrounded by a sphere of gas with flux-mean optical depth $\tau_{\rm single}$ to initial single-scattering (optical/UV) and appropriately-weighted $\tau_{\rm eff,\,IR}$ to the re-emitted IR photons, then the momentum flux imparted to the gas is $\dot{p} = \tilde{\tau}\,L_{\rm bol}/c$, with $\tilde{\tau} = (1 - \exp{(-\tau_{\rm single})} )\,(1 + \tau_{\rm eff,\,IR})$.\footnote{To derive $\tilde{\tau}$, begin by noting that locally at some position ${\bf x}$, the acceleration/momentum flux from photons at frequency $\nu$ is exactly $\partial (\rho\,{\bf v}[{\bf x}])_{\nu}/\partial t = \rho({\bf x})\,\kappa_{\nu}\,{\bf F}_{\nu}({\bf x}) / c$. If we integrate over both volume and frequency to obtain the total radiative force (momentum flux), we obtain $\partial {\bf p}/\partial t = \int d^{3}{\bf x}\,\rho({\bf x})\,{\bf F}({\bf x})\,\kappa_{{\rm F}}({\bf x})$, where ${\bf F}$ is the total flux and $\kappa_{\rm F}$ the flux-mean opacity at ${\bf x}$. If we simplify by assuming spherical symmetry, then we can trivially solve this integral and obtain $\partial {\bf p}/\partial t = \tau_{\rm eff,\,IR}\,L_{\rm IR}/c\,\hat{\bf r}$, where $\tau_{\rm eff,\,IR} \equiv \int dr\,\rho(r)\,\kappa_{\rm F}(r)$ and $L_{\rm IR}$ is some central source luminosity (a similar expression can be written without the symmetry assumption, using an appropriately angle-weighted average $\tau_{\rm eff,\,IR}$). The IR luminosity $L_{\rm IR}$ comes from single-scattered photons absorbed by dust, with $L_{\rm IR} \approx L_{\rm abs} = (1-\exp{(-\tau_{\rm single})})\,L_{\rm bol}$ (where $\tau_{\rm single}$ is the flux-mean opacity for the input spectrum), so including their momentum we have  $\partial {\bf p}/\partial t = \tilde{\tau}\,L_{\rm bol}/c\,\hat{\bf r}$ where $\tilde{\tau} \equiv (1 - \exp{(-\tau_{\rm single})} )\,(1 + \tau_{\rm eff,\,IR})$.} This $\tilde{\tau}$ term is sometimes referred to as a ``boost factor''; it ranges from $\approx \tau_{\rm single} \ll 1$ in the optically-thin limit to $\approx 1+\tau_{\rm eff,\,IR} \gtrsim 1$ in the IR optically-thick limit. In our simulations (FIRE-1 and FIRE-2), we explicitly calculate the radiative acceleration, based only on the local opacity and incident flux at every {\em resolved} gas element position. This means that our simulations will, if anything, tend to {\em under}-estimate the true momentum flux from radiation pressure -- i.e.\ it is likely that if we massively improved our resolution, we might see a more important role for IR multiple-scattering in the cold, dense ISM at densities well above our current star formation threshold \citep[see the discussion in][]{rosdahl:2015.galaxies.shine.rad.hydro}. For example, $\tau_{\rm eff,\,IR} \gg 1$ should in reality occur on sufficiently small scales around individual proto-stellar cores, but since these are not resolved in our simulations this contribution is not included (only the explicitly-resolved contributions to $\tau$ are accounted for). However, the lifetime of this deeply-buried phase is short ($\ll 10^{6}\,{\rm yr}$), so the expectation is that the integrated momentum ``missed'' is therefore small. 

Unfortunately some confusion on this point owes to our older (pre-FIRE) work, specifically in \citet{hopkins:rad.pressure.sf.fb}, so we wish to clarify it here. The \citet{hopkins:rad.pressure.sf.fb} simulations pre-dated the FIRE and {\small GIZMO} codes by several years, were non-cosmological, did not include any feedback other than radiation pressure, and used a sub-grid model to treat radiation which was fundamentally numerically different from either the LEBRON or M1 RHD methods. In that particular study, $\tilde{\tau}$ was multiplied by arbitrary factors to explore its effects; however the conclusion was that for most typical dwarf, dwarf starburst, or MW-like galaxies $\tilde{\tau} \sim 1-2$ in a time-averaged sense (and even for an intentionally extreme dense starburst-nucleus model with disk surface density $>1000\,M_{\sun}\,{\rm pc^{-2}}$, $\tilde{\tau}$ did not exceed $\sim 5-10$). The confusion on this point largely owes to Fig.~5 in \citet{hopkins:rad.pressure.sf.fb}, where we showed that for the starburst disk, the {\em instantaneous}, momentum-coupling-weighted $\tau_{\rm eff,\,IR}$ reached $\sim 30-50$ -- but this is just the statement that, at the moment a star particle ``turns on,'' the resolved surface densities of the star-forming cores reached $\Sigma_{\rm gas} \sim 10^{4}\,\msun\,{\rm pc^{-2}}$ (similar to those observed). As we noted therein, such high $\Sigma_{\rm gas}$ and $\tau_{\rm eff,\,IR}$ means that the core is almost immediately turned into stars or disrupted (the gas is exhausted or pushed away on a dynamical time $\sim 10^{5}\,$yr), such that the ``effective'' $\tilde{\tau}$ (time-averaged) is an order of magnitude lower \citep[for more detailed discussion, see e.g.][]{hopkins:rhd.momentum.optically.thick.issues,grudic:max.surface.density,grudic:sfe.cluster.form.surface.density}. This is essentially identical to the conclusions in many subsequent radiative-transfer studies \citep[e.g.][]{kuiper:2012.rad.pressure.outflow.vs.rt.method,krumholz:2012.rad.pressure.rt.instab,davis:2014.rad.pressure.outflows,tsang:monte.carlo.rhd.dusty.wind}.

\vspace{-0.5cm}
\subsection{Where Does Radiation Couple?}
\label{sec:where}

\begin{figure}
\plotonesize{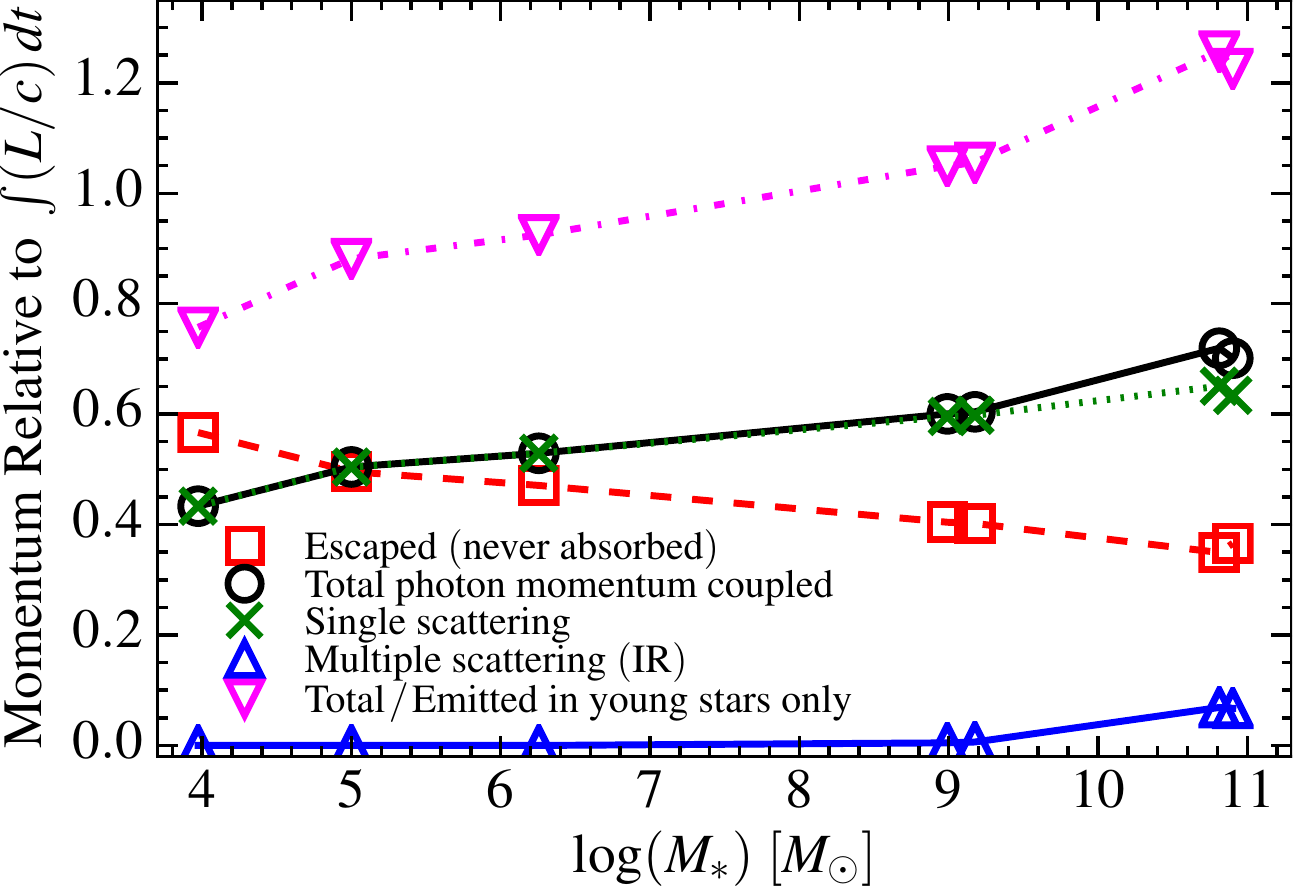}{1.0}
    \vspace{-0.25cm}
    \caption{Diagnostics of photon absorption. We study a subset of our highest-resolution ``Default (LEBRON)'' simulations from Table~\ref{tbl:sims} and \paperone: {\bf m09}, {\bf m10v}, {\bf m10q}, {\bf m11q}, {\bf m11v}, {\bf m12i}, {\bf m12f} (points, left-to-right). We compare the $z=0$ stellar mass of the primary galaxy, and total luminosity/momentum coupled to gas via various channels (labeled), integrated over all cosmic time (including all simulation stars, but this is dominated by the primary galaxy). We compare this to the total produced ($E = \int L\,dt$, or momentum $=E/c$). The fraction coupled rises slowly from $\sim 0.4$ in ultra-faints to $\sim 0.8$ in MW-mass galaxies. The dust opacity (proportional to $Z/Z_{\odot}$) is much larger at MW mass, but most of the ionizing continuum is absorbed by neutral gas in dwarfs (the absorbed fraction is closer to $\sim 100\%$ of the light emitted by young, hot stars $<50\,$Myr old). Roughly $\sim1/2$ of emitted radiation escapes without absorption, primarily optical/NIR from older ($\gtrsim 100\,$Myr) populations. Single-scattering dominates: {\em resolved} multiple-scattering in the IR (the only IR term included in the FIRE simulations) accounts for just $\sim 0.1\,E/c$ in the most massive systems, and much less in dwarfs (with lower dust-to-gas ratios).\label{fig:rad.pressure.coupling}}
\end{figure}

\begin{figure}
\plotonesize{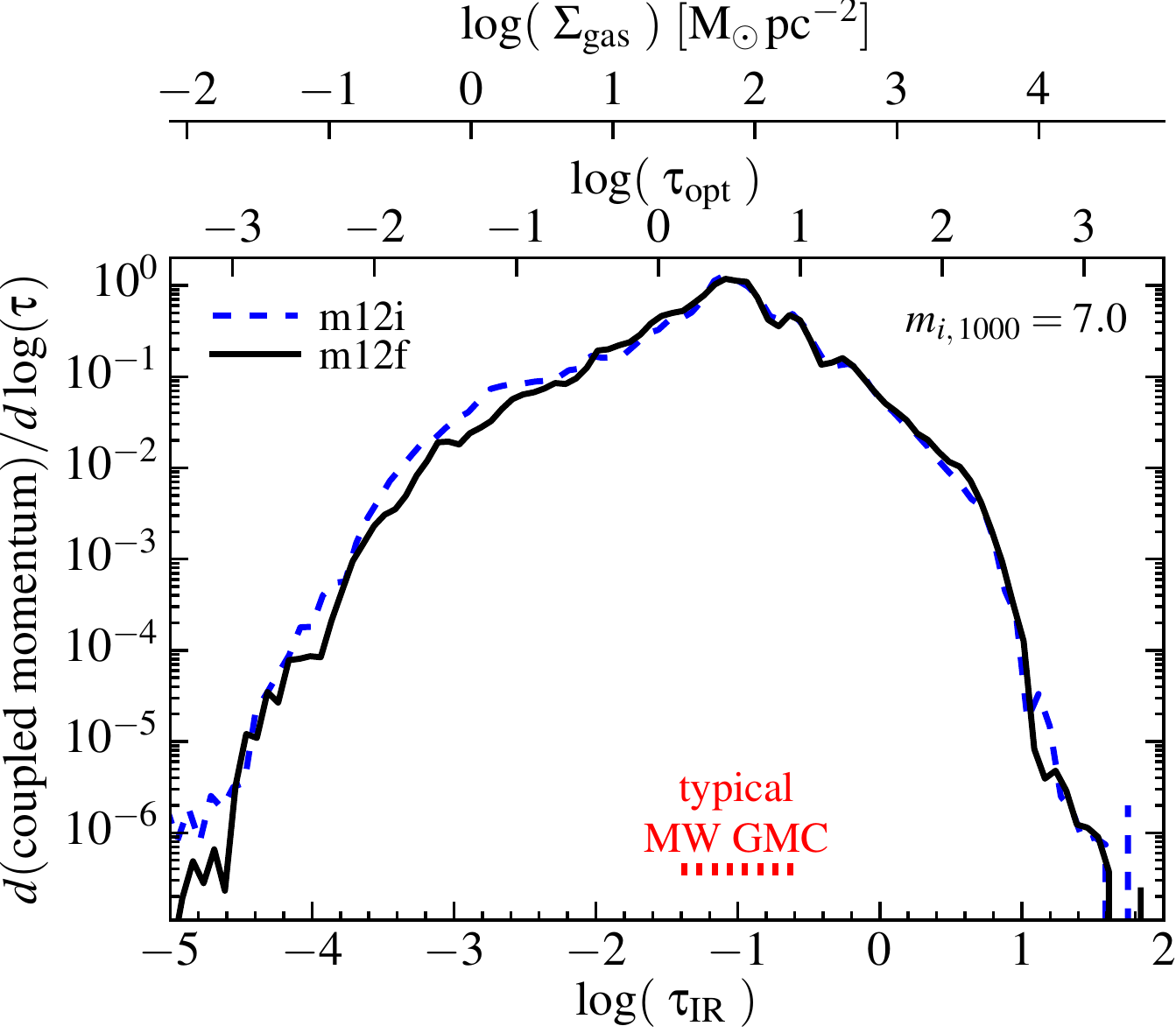}{1.0}
    \vspace{-0.25cm}
    \caption{Distribution of column densities (and approximate corresponding IR and optical depths) at which photons are absorbed, weighted by fraction of total photon momentum coupled to gas. Each time photons are absorbed we estimate the local $\Sigma_{\rm gas}$ from our Sobolev-type approximation and record it, for the highest-resolution ($m_{i,\,1000}=7$) MW-mass ({\bf m12i}, {\bf m12f}) simulations with the ``Default (LEBRON)'' FIRE treatment from \paperone. The distribution has a sharply peaked ``core'' (with most of the absorption) with disperson of just $\sim 0.15\,$dex around $\Sigma_{\rm gas}\sim 100\,\msun\,{\rm pc}^{-2}$ ($\tau_{\rm IR}\sim 0.1$, $\tau_{\rm optical}\sim $\,couple) -- but broad tails of the form $dP/d\log{\tau} \propto \tau^{\pm1}$, similar to what is expected for a quasi-fractal or log-normal ISM density distribution \citep{hopkins:rad.pressure.sf.fb}. Note that just $\sim 5\%$ ($\sim 0.01\%$) of the momentum comes from regions with resolved $\tau_{\rm IR}>1$ ($>10$). 
    We compare the $\pm1\,\sigma$ range of surface densities through to the center of a GMC in the MW ($=(1/2)\,M_{\rm GMC}/\pi\,R_{\rm GMC}^{2}$) from the observed compilation in \citet{bolatto:2008.gmc.properties} -- this is almost exactly the optical depth we see dominating absorption. In other words, most of radiative feedback comes from single-scattering/absorption from embedded stars in ``normal'' GMCs.
    \label{fig:rad.pressure.tau.distribution}}
\end{figure}

As noted above, Fig.~\ref{fig:rad.pressure.coupling} plots the total photon momentum absorbed -- essentially, the fraction of the galaxy-lifetime-integrated stellar luminosity $L$ which is absorbed by gas and dust in the system, and (correspondingly) the fraction which escapes without absorption or scattering. We also plot this total relative to the light emitted (primarily in UV) by young stellar populations (this can be $>1$ if some light from ``old stars'' is also absorbed), and the light re-emitted and then absorbed in the IR (i.e.\ effectively ``multiply scattered''). Fig.~\ref{fig:rad.pressure.tau.distribution} shows the distribution of column densities, in our MW-mass galaxies, at which the absorption occurs (in our ``Default (LEBRON)'' simulations). 

The effective ``coupled fraction'' of the emitted bolometric luminosity, $\langle \tilde{\tau} \rangle \sim 0.5$, ranging from $\sim 0.4$ in the smallest dwarfs to $\sim 0.7$ in MW-mass systems. A fraction $\sim 0.6$ (dwarfs) to $\sim 0.4$ (MW-mass systems) escapes without ever being absorbed. The order-unity coupled and escaped fractions are remarkably weakly dependent on galaxy mass. This is because a comparable fraction of the time-integrated bolometric output, {\em integrated over a Hubble time}, comes from (1) ionizing luminosity from very young stars, almost all of which is absorbed (the opacities are extremely high, and do not require dust, and many of the stars are buried in large columns), and (2) longer-wavelength (optical/NIR) from older ($\gtrsim 50-100\,$Myr), less luminous populations (which given the lower opacities and low dust content in the galaxy outskirts, tend to escape). It is also the case that, because galaxies are super-sonically turbulent, there is a broad distribution of column densities through the disk to a random star at any time -- thus an order-unity fraction of sightlines are always optically thin in the optical. Indeed, if we compare the coupled photon momentum to that integrated over stellar populations only up to an age of $\sim100\,$Myr, we obtain $p_{\rm coupled} \sim p_{\rm available}(t < 100\,{\rm Myr})$. 

Fig.~\ref{fig:rad.pressure.tau.distribution} examines further where absorption occurs. We focus on the MW-mass systems, as this is (a) where the best observational constraints exist, and (b) the only case where any significant IR multiple-scattering or dust absorption occurs. Since we will use our ``Default (LEBRON)'' simulations, we use the highest-resolution versions available, namely those with $m_{i,\,1000}=7$, with all properties shown in detail in \paperone. In the simulations, every time radiation (from stars in the simulation) is coupled to gas in the simulation via the LEBRON algorithm, we calculate a local Sobolev-type estimate of the column density seen by those photons (specifically, $\langle \Sigma \rangle \approx \rho\,[h + \rho/|\nabla \rho|]$, where $h = (m_{i}/\rho)^{1/3}$ is the local resolution element size, and $\nabla\rho$ is the density gradient, to account both for the column density within a single ``cell'' and approximate it integrated out to infinity). We record this and the amount of luminosity $\Delta L$ absorbed (equivalently, the momentum $\Delta L/c$ deposited). We then construct the $z=0$ distribution of ``column densities'' (or approximate optical depths at different wavelengths) weighted by the absorbed luminosity. 

We clearly see that the majority of the coupled radiation pressure occurs around optical depths of order a couple in the optical, or $\tau_{\rm IR}\sim 0.1$ -- which corresponds neatly to the typical surface density of GMCs both in our simulations \citep[see e.g.][]{hopkins:fb.ism.prop} and observed \citep{bolatto:2008.gmc.properties}. In other words, most of the imparted radiation pressure comes from single-scattering of light from massive, still-embedded stars. A few percent of the momentum comes from regions with $\tau_{\rm IR}>1$ -- exactly consistent with the ratio of photon momentum from multiple-scattering to the total imparted (for MW-mass systems) in Fig.~\ref{fig:rad.pressure.coupling}. Given our resolution, this does not come from protostellar cores (which are totally un-resolved), but from periods where the galactic nucleus (or massive ``clump complexes'' at high redshift) experiences tidal compression and rapid gas inflow in a starburst on resolved scales of $\sim 100\,$pc. In these rare phases, the IR terms may dominate -- and in future work we will explore how this does or does not matter for extreme starburst environments. But clearly, the multiple-scattering effects are minimal for most of the galactic star formation on the scales resolved here.

\vspace{-0.5cm}
\subsection{Numerical Methods}
\label{sec:numerics}

Here we briefly summarize the impact of numerical RHD methods on the simulation results. 

For {\em validation} tests of the numerical methods studied here (e.g.\ confirmation that the implementations recover the correct answer in the limits under which their fundamental assumptions are valid), we refer to several previous studies \citep[e.g.][]{hopkins:fb.ism.prop,hopkins:2013.fire,hopkins:fire2.methods,hopkins:rhd.momentum.optically.thick.issues,roth:2012.rad.transfer.agn,ma:2015.fire.escape.fractions,grudic:sfe.cluster.form.surface.density,lupi:2018.h2.sfr.rhd} as well as parallel studies using the same classes of methods in different codes \citep{rosdahl:2013.m1.ramses,rosdahl:m1.method.ramses,rosdahl:2015.galaxies.shine.rad.hydro,kannan:photoion.feedback.sims,hu:photoelectric.heating,hu:2017.rad.fb.model.photoelectric,emerick:rad.fb.important.stromgren.ok}. It is not our intention to repeat these studies here. 

Rather, in the sections above, we considered the {\em physical} consequences of variations in the numerical methods. First, we note the variations of LEBRON or M1 which we have studied. The LEBRON method allows us to formally turn on or off different physical components of the radiative FB: we can specifically disable local (kernel-scale) or long-range (tree-based) radiation pressure and/or photo-heating terms, or we can remove the ``local extinction'' operation which attenuates the spectrum before propagation to long-range distances. These variations are compared in Fig.~\ref{fig:rad.fx.m10q}, and have been discussed above: specifically they allow us to show that most of the RP comes from absorption in the vicinity of stars (consistent with the study of where radiation is absorbed, also above), and that local extinction of non-ionizing radiation is un-important in small dwarfs where the optical depths are relatively small (but important in massive galaxies -- there, failing to account for extinction in the vicinity of stars would lead to a significant over-estimate of the importance of radiation, since one would assume all emergent flux is in the UV/optical). 

Within the M1 models, in addition to the physical variations (turning on and off radiation pressure and different wavebands) discussed above, we have also studied the role of the numerical treatment of the IR (as a single bin with grey opacity versus explicitly-evolved radiation temperature fields with complicated opacities), which has little effect. In our M1 runs, we have also considered variations in a limited subset of runs of {\bf m10q} of the numerical method used to ``deposit'' radiation in the neighboring cells (weighting by solid angle, as in \citealt{hopkins:sne.methods}, or with a simpler kernel weight) and the timestep for star particles (how frequently this is done, as described in \citealt{hopkins:fire2.methods}, varying between our default stellar-evolution timestep and 10x shorter or 10x longer). None of these variations has a significant effect. We have also varied the ``reduced speed of light'' in both {\bf m10q} and {\bf m12i} halos, from $\tilde{c} \sim 300-5000\,{\rm km\,s^{-1}}$; the effects are small for $\tilde{c} \gtrsim 500\,{\rm km\,s^{-1}}$ (generally smaller, for example, than the differences between M1 and LEBRON methods), consistent with the well-known result that this should be converged so long as $\tilde{c}$ is faster than other, explicitly-resolved speeds in the simulations. However it is not completely negligible: the general sense is that increasing $\tilde{c}$ from $\sim 300-1000\,{\rm km\,s^{-1}}$ gives slightly stronger radiative feedback effects (while increasing it beyond this point has little impact), suggesting that at too-small $\tilde{c} \ll 1000\,{\rm km\,s^{-1}}$ photons emitted in dense regions (where massive stars form) may spend ``too long'' streaming out, making them less efficient on large scales. As shown in \citet{rosdahl:2015.galaxies.shine.rad.hydro} and \citet{hopkins:rhd.momentum.optically.thick.issues}, because the mean-free-path of ionizing/UV photons is not explicitly resolved in the simulations (it is many orders-of-magnitude beyond state-of-the-art resolution), naive implementations of M1 that fail to account for two closely-related potential errors in the coupling between photon momentum and gas will under-estimate the radiation pressure forces by orders of magnitude. Our default M1 implementation includes the relevant fixes demonstrated in those papers to resolve this issue: but we have considered one test (not shown) of {\bf m10q} removing these fixes (i.e.\ using the ``M1 (Cell-centred)'' implementation described in \citealt{hopkins:rhd.momentum.optically.thick.issues}). As expected, the results from the incorrect method are essentially identical to our runs with M1 removing radiation pressure entirely. 

{\small GIZMO} also includes two other moments-based methods for RHD, the ``flux-limited diffusion'' (FLD; \citealt{levermore:1984.FLD.M1}) and ``optically-thin, variable Eddington tensor'' (OTVET; \citealt{gnedin.abel.2001:otvet}) methods. These are just the zeroth-moment expansions of the photon transport equation, where one closes the equations at zeroth order by assuming pure diffusion with a ``flux limiter'' as opposed to explicitly evolving the flux vector as in the first-moment ``M1'' method (the primary difference between FLD and OTVET is whether one assumes an isotropic Eddington tensor in FLD, or the Eddington tensor which would be calculated if all sources were optically-thin, in OTVET). Unlike M1, these cannot capture phenomena such as ``shadowing'' by optically thick structures; moreover they are actually more computationally expensive owing to a stricter timestep criterion. We therefore did not consider them primarily here; however we have run both {\bf m10q} and {\bf m12i} at intermediate resolution with both methods ({\bf m12i} only run down to $z\sim1$). The only difference with our ``Default (M1)'' runs is the exact form of the photon-propagation step. We find these give very similar results to M1. 

More interesting is the difference between LEBRON and M1, which are fundamentally distinct methods. Recall, {\em neither} of these methods is exact in general cases, even at infinite resolution. LEBRON converges to exact solutions in the optically thin regime, independent of the number and distribution of sources, but will only converge to approximate solutions in the optically-thick, multiple-scattering regime. Conversely, M1 converges to exact solutions in the optically-thick multiple-scattering regime, but will incorrectly merge photons and reduce to the diffusion limit in the optically-thin regime if there are multiple sources. So we should not regard either of these methods as ``correct.'' But since they are exact in essentially opposite limits, it is plausible to suppose they bracket the reasonable range of behaviors. In detail, if we compute the radiative flux at a given frequency, at any given specific point ${\bf x}$ in the simulations, it is possible for LEBRON and M1 to diverge by orders-of-magnitude from each other (if, e.g.\ the point is shadowed by complex, very optically-thick structures, where most of the photons are absorbed). However, if we are only concerned with galaxy-scale properties, our comparisons show that the two give broadly similar results. This is because in {\em either} method, most of the short-wavelength (UV) photons are absorbed in the ISM near massive stars (while most of the longer-wavelength IR photons escape), so the average heating rates and radiation pressure forces, their characteristic spatial and timescales, are broadly similar. The exact spatial locations where absorption occurs are of secondary importance, and the difference in the {\em dynamics} of a region illuminated by a ``modestly attenuated'' spectrum (say, $\tau \sim 1-10$) versus ``heavily attenuated'' ($\tau \sim 10-100$) is not important (even though the flux differs by orders of magnitude) because in both cases most of the light is blocked and so the resulting radiation effects are weak.\footnote{Because of the Lagrangian nature of the code here, HII regions are always comparably-resolved in gas and stars. Briefly, a Stromgren sphere sourced by $N_{\ast}$ young star particles of mass $m_{i}$ will fully-ionize $\sim 2\,N_{\ast}\,(n_{\rm gas}/100\,{\rm cm^{-3}})^{-1}$ gas resolution elements. Since the maximum densities reached here are comparable to our density threshold for star formation ($\sim 1000\,{\rm cm^{-3}}$), this is at least marginally-resolved for any resolved star cluster. For our highest-resolution simulations ($m_{i}=250\,\msun$), HII regions from individual O-stars can be marginally resolved \citep[see discussion in][]{ma:2015.fire.escape.fractions,su:discrete.imf.fx.fire,wheeler:ultra.highres.dwarfs}.}

We do tend to find that the effects of radiation are slightly {\em stronger} in M1, as compared to LEBRON. In Figs.~\ref{fig:ov}-\ref{fig:ov.2}, the ``Default (M1)'' dwarfs ({\bf m10q}, {\bf m11b}, {\bf m11q}) have slightly smoother SFHs (with slightly larger SFRs owing to less-bursty/violent SF), while the ``Default (M1)'' MW-mass systems ({\bf m12i}, {\bf m12m}) have slightly lower late-time SFRs and central circular velocity curve ``spikes'' (compared to ``Default (LEBRON)''). Our experiments turning on and off different components of the radiative FB imply this is dominated by the effects of the far-UV/ionizing bands. In fact, the dominant source of the difference appears to be lie in how the ``Default (LEBRON)'' method calculates absorption of ionizing photons in the immediate vicinity of the emitting star: as described in \paperone, this uses a Stromgren-type approximation. Moving spherically outwards from the star particle, each time a gas element is encountered, the code calculates the number of ionizing photons needed to fully-ionize it over the timestep (consuming them), until the ``long range'' escape is reached or the photons are exhausted. But imagine a star  surrounded by (mostly) low-density gas with one extremely dense (optically-thick) ``clump.'' In reality, the clump subtends a small area on the sky, so should receive and destroy a small fraction of the ionizing photons; but if it is within the local kernel in LEBRON, it will be encountered in the radial search, and since the number of ionizing photons needed to ionize some volume scales as $\propto n^{2}$, it can essentially ``use up'' the full photon budget. Taking an identical snapshot of {\bf m10q} and {\bf m12i}, at $z=0.05$, and running it for a very short amount of time with both M1 and LEBRON methods in turn (with no UV background), we have confirmed that in M1 a larger total mass of gas is ionized by the {\em same} number of photons emitted from the stars. So ironically, even though LEBRON is formally a non-photon-conserving scheme, it actually tends to artificially {\em reduce} the number of viable photons for feedback. This also suggests improvements to the short-range terms in the LEBRON scheme, based on e.g.\ HEALPIX or other angular tesselation rather than a spherically-symmetric assumption, might reduce the discrepancy.

In future work, we will explore simulations using the RHD scheme from \citet{jiang:2014.rhd.solver.local} implemented in {\small GIZMO}, which is exact in {\em both} optically thin and thick regimes. However it is much more expensive, especially for multi-band transport. %Preliminary comparisons of single-band (ionizing) RHD, running {\bf m10q} at intermediate resolution ($m_{i,\,1000}=2$) for a short time with this scheme, show results somewhere in-between LEBRON and M1 (as we might expect), but more study is needed.

Extensive numerical tests of almost every other aspect of these simulations (resolution, force softening, hydrodynamic solvers, etc.) are presented in \citet{hopkins:fire2.methods} and \citet{hopkins:sne.methods}. These all use the ``Default (LEBRON)'' method, except where otherwise specified. So extensive resolution tests of this particular method are presented there. As shown in \citet{hopkins:fire2.methods}, the simulations do have many predictions which depend on resolution with this default prescription. However, this is not necessarily because the radiation transport {\em for a fixed physical mass configuration} is not converged (in fact \citealt{hopkins:rhd.momentum.optically.thick.issues} show quantities like the radiation pressure coupled are reasonably well-converged with these methods and this setup). Rather, the gas and stellar distributions exhibit more and more complex sub-structure at higher resolution, so (naturally) quantities like where and when and how the radiation couples can, in turn, scale as well. Although, as noted above, quantities such as the GMC mass function at the largest masses (which can we resolved), surface densities/optical depths, linewidth-size relation, and related properties do appear to be particularly robust to resolution (see \citealt{hopkins:fire2.methods,guszejnov:imf.var.mw,guszejnov:fire.gmc.props.vs.z}, likely owing to how these properties are self-regulated by extremely simple bulk galaxy properties like the Toomre mass for a $Q\sim 1$ disk in a turbulent fragmentation cascade \citep[see][]{hopkins:excursion.ism,hopkins:excursion.imf,hopkins:frag.theory,guszejnov:cmf.imf,guszejnov:universal.scalings}. So the most important question, perhaps, is  whether and how new (currently un-resolved) scales like proto-stellar cores will depend on radiative physics beyond the scope explored here.

\vspace{-0.5cm}
\subsection{A Note on ``Sub-Grid'' Models for Radiation}
\label{sec:subgrid}

A variety of ``sub-grid'' models for stellar feedback do not attempt to explicitly model the salient physical processes, but rather to capture their ``net effects.'' These are common (indeed, necessary) in large-volume simulations which cannot resolve the ISM. The most obvious examples are models like \citet{springel:multiphase,dave.2016:mufasa.fire.inspired.cosmo.boxes}, which simply eject mass from a galaxy with some scaling proportional to the star formation rate (and add pressure to dense gas, attributed to un-resolved phase structure). Whether one attributes these scalings to SNe or radiation or some combination of these and other physics, they are obviously fundamentally distinct from the models here. Essentially, we are trying to {\em predict} these effects, using stellar evolution theory (for calculating e.g.\ SNe rates and energetics, radiative luminosities and spectra) as the ``input.'' 

We note this because several such models have been used for radiative FB. For example, \citet{agertz:2013.new.stellar.fb.model} and \citet{agertz:sf.feedback.multiple.mechanisms} add an outward momentum flux in cells immediately adjacent to a star particle scaled to a multiple of $L/c$; \citet{ceverino:2013.rad.fb} add a pressure $P \sim L/(c\,A_{\rm cell})$ to the hydrodynamic pressure (where $A_{\rm cell}\sim \Delta x^{2}$ is the cell area) in cells containing star particles $<5\,$Myr old; \citet{stinson:2013.new.early.stellar.fb.models} add a heating term $\dot{E} \sim L$ to the gas heating/cooling subroutine for the gas containing star particles $<10\,$Myr old. Although these might represent some consequences of radiative FB, none of these models attempts to actually follow radiation (transport or RHD) explicitly. Our comparisons here indicate that an approximation like that in \citet{agertz:sf.feedback.multiple.mechanisms} might be reasonable for the single-scattering RP (with $\sim 1/2\,L/c$ or $\sim 1\,L_{\rm UV}/c$ absorbed), in simulations which do not resolve GMCs (since we find most of the single-scattering RP is imparted in the GMCs in which massive stars are born). In fact, \citet{lupi:2018.h2.sfr.rhd} compare several RHD methods including our LEBRON and M1 algorithms in {\small GIZMO}, to such a ``local momentum flux'' estimator, and do argue that it is able to capture many of the most important effects (and describe a similar comparison for local photo-ionization heating). It is unlikely that the approximation in \citet{ceverino:2013.rad.fb} resembles the radiative FB here, since it is only representative of ``radiation pressure'' in the infinite optical depth (perfect-trapping), multiple-scattering, grey-opacity limit. And the approximation in \citet{stinson:2013.new.early.stellar.fb.models} might capture some effects of photo-ionization heating if HII regions are un-resolved, but physically in these cases the heating should be restricted to photo-heating (i.e.\ not allowed to heat at $T\gg 10^{4}$\,K) from ionizing radiation. 

In any case, it is interesting, but beyond the scope of our study, to explore whether one can define a better sub-grid model for use in lower-resolution simulations, or whether capturing the key effects ultimately requires explicitly tracking multi-band radiation transport as we do here. For one such study, we refer readers to \citet{lupi:2018.h2.sfr.rhd}.

\vspace{-0.5cm}
\section{Summary \&\ Conclusions}
\label{sec:discussion}

We use a survey of $\sim 100$ high-resolution radiation-hydrodynamical cosmological zoom-in simulations of galaxies to study the nature and effects of radiative feedback from stars on galaxy formation. Our simulations span masses from ultra-faint to Milky Way ($M_{\ast} \sim 10^{4}-10^{11}\,M_{\odot}$), and include the FIRE-2 physical models for ISM microphysics (cooling, chemistry), star formation, and stellar feedback (from supernovae and stellar winds, in addition to radiation). We extensively survey several different radiative feedback ``channels'' and wavelength ranges, and consider two fundamentally distinct numerical radiation-hydrodynamics methods, in order to identify the most important and robust results. We note that this is a companion paper to \paperone\ and \papertwo, where more general numerical (e.g.\ resolution, hydro solvers) and mechanical feedback (SNe \&\ stellar mass-loss) methods are explored in detail.

\vspace{-0.5cm}
\subsection{Overview: Different Radiative Feedback Mechanisms in Galaxy Formation}

\begin{itemize}
\item{Averaged over the entire life of a galaxy, most of the emitted far-UV/ionizing radiation ($\sim 1/2$ the total bolometric) is absorbed. Relatively little optical/NIR/FIR is absorbed. Total absorption increases with galaxy mass (as dust masses and densities increase), but the effect is weak because of efficient neutral gas absorption of ionizing photons in even metal-free galaxies.}

\item{As a result, the most important feedback mechanisms, in a galaxy-lifetime-averaged sense, are photo-ionization heating, and single-scattering radiation pressure from UV \&\ ionizing photons. Although we did not study it here, it is possible that resonant Ly-$\alpha$ scattering in metal-poor dwarfs could produce similar effects.}

\item{Photo-electric and IR thermal dust/collisional heating, while important for phase structure in dense, cold ($T \ll 10^{4}$\,K) gas, have weak effects on galactic scales (thermal pressure is always weak in the cold gas, compared to e.g.\ turbulence). Likewise, since we ignore Pop III (metal-free) star formation, Lyman-Werner radiation also plays a minor role (as molecular cooling has essentially no effect on star formation in the presence of even trace metals; see \citealt{glover:2011.molecules.not.needed.for.sf}). We showed in \paperone\ that even much more radical changes to cooling physics in $T<10^{4}\,$K gas have negligible effects. Compton heating from soft/hard X-rays emitted by LMXBs/HMXBs also plays a minor role: the flux is too low to compete with cooling rates in hot gas.}

\item{A more detailed ``breakdown'' of ionizing radiation into a multi-band treatment, e.g.\ separately following HeI, HeII, and HI, makes relatively small differences compared to following ionizing photons in a single-band approximation, with a mean SED calculated for young stellar populations. This does not mean HeII ionization has no effects, but simply that they can be captured (to leading order) by a mean local+UVB SED treatment.}

\item{Multiple-scattering of IR photons produces weak effects, in a galaxy-lifetime-averaged sense, at achievable cosmological resolutions. In metal-poor dwarfs, IR optical depths are almost always small. In MW-mass systems, most photons are absorbed in ``typical'' GMCs which are optically thin in the IR ($\tau_{\rm IR} \sim 0.1$) on average. Only a small fraction of the light, in e.g.\ galaxy nuclei in starbursts or dense, high-redshift clouds, is emitted in regions with $\tau_{\rm IR} \gg 1$ where multiple-scattering is potentially important. However this could change if we resolved extremely small, dense structures in cores, although this is likely not where most radiation is emitted.}

\end{itemize}

\vspace{-0.5cm}
\subsection{Important Dynamical Effects of Radiation on Galaxies}

We summarize the dynamical effects of the key radiative feedback channels identified above. In our study, we focused on bulk galaxy properties including: SFRs; stellar masses; metallicities; stellar, baryonic and dark matter density profiles; rotation curves; and morphology. We also briefly discussed, but did not study in detail, quantities such as outflow rates and ISM phase structure.

\subsubsection{In Dwarf Galaxies}

\begin{itemize}

\item{The most important radiative feedback effect in dwarfs is photo-ionization heating by the meta-galactic background (UVB). Even in galaxies with $V_{\rm max}\sim 50\,{\rm km\,s^{-1}}$, much more massive than the threshold where the UVB ``quenches'' ($\sim 10-20\,{\rm km\,s^{-1}}$), removing the UVB leads to order-of-magnitude enhanced SFRs (much larger than observed) owing to additional late-time cooling and accretion that is otherwise suppressed.}

\item{Radiation pressure and local photo-heating play a similar role to one another: both ``smooth out'' star formation by providing a ``gentle'' form of feedback that can support warm or cool gas which would otherwise lose its pressure support and collapse under self-gravity. If we remove these, the star formation becomes substantially more ``violent'' and ``bursty'' because cool/warm gas more rapidly fragments into GMCs, whose collapse is only halted by SNe after they turn much of their mass into stars (similar conclusions have been reached using idealized, non-cosmological simulations with entirely different numerical methods and treatments of mechanical feedback; \citealt{kannan:photoion.feedback.sims,rosdahl:2015.galaxies.shine.rad.hydro,kimm:lyman.alpha.rad.pressure,emerick:rad.fb.important.stromgren.ok}). Within the galaxy, GMC and ``star forming clump'' lifetimes and star formation efficiencies are obviously strongly modified -- the ``no radiative feedback'' prediction is already ruled out by resolved GMC observations  \citep[compare][]{lee:gmc.sfe,grudic:sfe.gmcs.vs.obs}. But also, if we remove radiative feedback, the more violent SNe feedback makes the galaxies more metal poor (metallicities are suppressed by $\sim 0.5\,$dex, in conflict with the observed mass-metallicity relation; \citealt{ma:2015.fire.mass.metallicity}) and more baryon-poor (lower stellar mass by a factor $\sim 1.5-2$, and substantially more gas-poor) with almost no residual cool gas. These effects are notably exacerbated (with excessive early-time star formation followed by ``self-quenching'' from explosive SNe feedback) if we {\em also} remove ``early'' stellar mass-loss (e.g.\ fast O/B winds), which carry a momentum flux similar to radiation.}

\end{itemize}

\subsubsection{In Massive (Milky Way-mass) Galaxies}

\begin{itemize}

\item{Photo-heating by the UVB plays a negligible role in the evolution of the primary galaxy (though it is important for the dwarf satellites as described above), once it is massive ($V_{\rm max}\gtrsim 100\,{\rm km\,s^{-1}}$).}

\item{Radiation pressure and local photo-heating again play a similar role to each other: by pre-processing GMCs and dense star-forming gas, they allow SNe to more easily able to escape, over-lap and generate super-bubbles, especially in the denser central regions of galaxies. Removing them leads to substantially more compact, dense bulges and steep central rotation curves, in conflict with observations. Radiative FB is important within galaxies on short timescales regulating the rate of conversion of dense gas into stars (e.g.\ the position of the galaxy on the Schmidt-Kennicutt relation), as previous studies have shown \citep{wise:2008.first.star.fb,hopkins:rad.pressure.sf.fb,krumholz:2011.rhd.starcluster.sim,trujillo-gomez:2013.rad.fb.dwarfs,federrath:2014.low.sfe}. More dramatically, without this or some other strong ``early feedback'' acting before SNe explode, massive GMCs turn almost all their mass into stars leaving hyper-dense bound relics, predicting galaxies dominated entirely by dense star clusters \citep[consistent with many previous GMC-scale studies, e.g.][]{fall:2010.sf.eff.vs.surfacedensity,colin:2013.star.cluster.rhd.cloud.destruction,grudic:sfe.cluster.form.surface.density,howard:gmc.rad.fx,rosen:massive.sf.rhd,kim:2017.art.uv.starclusters}.}

\item{Of course, without local photo-heating, one also cannot correctly predict the distribution of ISM phases (the warm medium in particular), which is maintained by these processes.}

\end{itemize}

\vspace{-0.5cm}
\subsection{Numerical Methods \&\ Caveats}

RHD remains a numerical frontier: almost all RHD methods which are efficient enough to use in high-resolution cosmological simulations are approximate in some manner, and uncertainty remains regarding the role this plays in radiative feedback. We cannot definitively resolve this since the methods here are among those approximate classes; however we have compared fundamentally distinct methods and choices within those methods.

\begin{itemize}

\item{Our results are qualitatively similar using either the LEBRON or M1 numerical RHD methods. LEBRON is a ray-based algorithm which is exact in the optically-thin limit (independent of source number), but fails to capture shadowing and exact photon-conservation in the optically-thick, multiple-scattering limits. M1 is a moments-based algorithm which is photon-conserving and exact in the optically-thick, multiple-scattering limit but cannot capture the optically-thin limit with multiple sources (intersecting rays ``collide'' and diffuse out). Both are approximate, but valid in essentially opposite limits. To first order, we find they give similar results. In detail, M1 shows slightly stronger radiative FB effects: this appears to owe (primarily) to the fact that LEBRON artificially allows dense ``clumps'' near massive stars to consume too many ionizing/UV photons. We have also considered limited comparisons of other, less-accurate moments-based methods (e.g.\ FLD) which are similar to M1.}

\item{Reassuringly, our results are also consistent with a growing number of multi-frequency RHD studies using a range of different numerical RHD and hydro methods to treat the same radiative FB mechanisms \citep[see e.g.][]{kannan:photoion.feedback.sims,rosdahl:2015.galaxies.shine.rad.hydro,kimm:lyman.alpha.rad.pressure,emerick:rad.fb.important.stromgren.ok}.}

\item{We stress that the M1 implementation here involves the ``face-centered'' formulation which resolves the numerical errors identified in \citet{rosdahl:2015.galaxies.shine.rad.hydro} and \citet{hopkins:rhd.momentum.optically.thick.issues}. Without these, the radiation pressure is strongly artificially suppressed.}

\item{While \paperone\ (and this paper to a lesser extent) show our conclusions are robust over factors of $\sim 100$ in mass resolution, we stress that this does not mean they are formally ``converged.'' At much higher resolution, new physics, including resolving smaller-scale substructure in gas (e.g.\ small ``holes'' in compact HII regions) and the spatial distribution of massive stars within star clusters may have important effects on how radiative FB acts upon natal GMCs.}

\item{We show that some ``early feedback'' (feedback from massive stars before SNe explode) is critical for regulating collapse of GMCs, bursty/violent SF, and galaxy morphologies. However, for the default simulations here this support can be provided by a mix of O/B stellar mass-loss, single-scattering radiation pressure, and warm photo-ionized gas in HII regions. From a numerical point of view, these are somewhat degenerate if we only consider galaxy-scale properties (all act on similar small time and spatial scales with similar momentum fluxes). Their relative importance, if the coupling is occurring near the resolution limit, can be sensitive to numerical choices (e.g.\ whether one accounts for un-resolved multiple-scattering or leakage, or for ``trapping'' and pressure-driven work done by stellar wind bubbles) -- for this reason we find somewhat different results in FIRE-2 versus FIRE-1. Simulations and observations of smaller (GMC and star cluster) scales are clearly needed to robustly address the relative roles played here.}

\end{itemize}

\vspace{-0.5cm}
\subsection{Additional Caveats, Missing Physics, \&\ Future Work}

In addition to the numerical caveats above, we stress that our simulation set is necessarily limited. 

\begin{itemize}

\item{As we noted above, it is likely that some physics here (e.g.\ infrared multiple-scattering or Lyman-Werner radiation) could be very important in special environments and/or times in the life of galaxies (e.g.\ nuclear starbursts or circum-AGN environments, or in the first metal-free stars), even if it does not alter {\em global} properties of the ``typical'' galaxy at $z=0$ \citep[see e.g.][]{thompson:rad.pressure,costa:qso.outflows.multiple.ir.scattering}.} 

\item{Our treatment of the meta-galactic UVB is not self-consistent because actually predicting the UVB requires volumes $\gg (100\,{\rm Mpc})^{3}$, impossible to achieve at our resolution: one possible approach is to use large-scale (low-resolution) studies to model e.g.\ fluctuations in the local UVB field in e.g.\ QSO proximity zones (or patches surrounding different structures during the process of reionization, where the UVB may be highly inhomogeneous), then use those models for zoom-in simulations to self-consistently predict e.g.\ UV escape fractions.} 

\item{We have also neglected some potentially important radiative feedback channels from \S~\ref{sec:rad.overview}, e.g.\ over-ionization of metal-species in the CGM, or multiple-scattering in resonance lines. The former requires non-equilibrium chemistry for the metals, which will be studied in future work (Richings et al., in prep.), but preliminary studies suggest it may only be significant in the near vicinity of AGN or extreme starbursts \citep{richings:2016.chemistry.uvb.photoelec.fx,oppenheimer:2018.flickering.heating.cgm}. The latter is generally not believed to be important on galaxy scales (though it is probably critical for wind-launching in stellar photospheres and AGN accretion disks), {\em except} perhaps from resonant Lyman-$\alpha$ scattering; but following this requires Lyman-$\alpha$ radiative transfer which cannot be handled by any of the default RHD methods here (it will be studied in future work, Ma et al., in prep.; but see e.g.\ \citealt{2009MNRAS.396..377D,2017MNRAS.464.2963S,kimm:lyman.alpha.rad.pressure}).}

%\item{Some un-related physics (e.g.\ cosmic rays) could also be important and have similar effects -- perhaps falling into the ``early feedback'' category -- but is neglected here.}

\item{We neglect a potentially critically important radiative FB channel, in AGN. AGN are generally sub-dominant in luminosity in low-mass (sub-MW-mass) galaxies, at almost all times and redshifts, relative to stars -- so this is probably a reasonable approximation at for the galaxies we study here. But for more massive galaxies, the radiation pressure, photo-heating, and Compton heating from AGN can easily dominate that from stars by orders of magnitude. It is likely that in these regimes, bright AGN or quasars have very important radiative feedback effects not captured here \citep[see, e.g.][]{proga:disk.winds.2000,sazonov:radiative.feedback,murray:momentum.winds,kurosawa:outflows.w.rad.in.precession.w.proga,choi:2014.mechanical.feedback.model.2,costa:dusty.wind.driving,hopkins:qso.stellar.fb.together,brennan:radiative.bh.fb}.}

\item{We assume perfect momentum re-distribution from dust to gas, when photons are absorbed by dust (i.e.\ assume dust and gas move together), in addition to a constant dust-to-metals ratio. While unimportant for ionizing photons if neutral gas dominates the opacity, in more massive galaxies at longer wavelengths the dust opacities dominate. But \citet{squire.hopkins:RDI,squire:rdi.ppd,hopkins:2017.acoustic.RDI,hopkins:2018.mhd.rdi} recently showed that this scenario (accelerating one of either gas or dust, and relying on drag or Lorentz forces to ``pull'' the other along) is violently unstable, with the ensuing instabilities driving strong turbulence and segregation of the dust and gas. Whether it is in fact possible to have ``dust-driven'' outflows requires further investigation in light of these previously-unrecognized instabilities.}

\end{itemize}

We also stress, once more, that our conclusions here apply to {\em global, galaxy-scale} properties. The radiative physics controlling e.g.\ proto-stellar evolution and the initial mass function, or star cluster formation, or AGN accretion, will be distinct, as the characteristic spatial scales, timescales relative to stellar evolution, opacities, densities, and wavelengths where most of the light is emitted differ enormously. 

Likewise, it should be obvious that the detailed chemical state of a galaxy, and observational diagnostics of this state and the stellar emission itself, depend directly on the radiation from stars (e.g.\ the radiative environment determines quantities like the H$\alpha$, OIII, CO, CII luminosities and excitation, the UV/optical/IR continuum, etc.). Our goal here was not to explore these observables as ``tracers,'' but rather to ask whether and how stellar radiation alters the formation, evolution, and dynamics of bulk galaxy properties.

\vspace{-0.7cm}
\acknowledgments 
We thank Eliot Quataert, Alexander Richings and Alexander Gurvich, with whom we have had a number of useful discussions on topics here. Support for PFH and co-authors was provided by an Alfred P. Sloan Research Fellowship, NSF Collaborative Research Grant \#1715847 and CAREER grant \#1455342, and NASA grants NNX15AT06G, JPL 1589742, 17-ATP17-0214.
AW was supported by NASA, through ATP grant 80NSSC18K1097, and HST grants GO-14734 and AR-15057 from STScI. DK was supported by NSF grant AST-1715101 and the Cottrell Scholar Award from the Research Corporation for Science Advancement.
Numerical calculations were run on the Caltech compute cluster ``Wheeler,'' allocations from XSEDE TG-AST130039 and PRAC NSF.1713353 supported by the NSF, and NASA HEC SMD-16-7592.\\

\vspace{-0.2cm}
%\bibliography{/Users/phopkins/Dropbox/Public/ms}
\bibliography{ms_extracted}

\begin{thebibliography}{181}
\expandafter\ifx\csname natexlab\endcsname\relax\def\natexlab#1{#1}\fi

\bibitem[{{Agertz} \&
  {Kravtsov}(2015)}]{agertz:sf.feedback.multiple.mechanisms}
{Agertz}, O., \& {Kravtsov}, A.~V. 2015, \apj, 804, 18

\bibitem[{{Agertz} {et~al.}(2013){Agertz}, {Kravtsov}, {Leitner}, \&
  {Gnedin}}]{agertz:2013.new.stellar.fb.model}
{Agertz}, O., {Kravtsov}, A.~V., {Leitner}, S.~N., \& {Gnedin}, N.~Y. 2013,
  \apj, 770, 25

\bibitem[{Barkana \& Loeb(2001)}]{barkana:reionization.review}
Barkana, R., \& Loeb, A. 2001, Physics Reports, 349, 125, review reionization

\bibitem[{{Bate}(2012)}]{bate:2012.rmhd.sims}
{Bate}, M.~R. 2012, \mnras, 419, 3115

\bibitem[{{Bate} {et~al.}(2014){Bate}, {Tricco}, \&
  {Price}}]{bate:2014.core.collapse.rad.mhd.sph}
{Bate}, M.~R., {Tricco}, T.~S., \& {Price}, D.~J. 2014, \mnras, 437, 77

\bibitem[{{Behroozi} {et~al.}(2010){Behroozi}, {Conroy}, \&
  {Wechsler}}]{behroozi:mgal.mhalo.uncertainties}
{Behroozi}, P.~S., {Conroy}, C., \& {Wechsler}, R.~H. 2010, \apj, 717, 379

\bibitem[{{Bieri} {et~al.}(2017){Bieri}, {Dubois}, {Rosdahl}, {Wagner}, {Silk},
  \& {Mamon}}]{bieri:qso.rhd.fb}
{Bieri}, R., {Dubois}, Y., {Rosdahl}, J., {Wagner}, A., {Silk}, J., \& {Mamon},
  G.~A. 2017, \mnras, 464, 1854

\bibitem[{{Blitz}(1993)}]{blitz:gmc.properties}
{Blitz}, L. 1993, in Protostars and Planets III (University of Arizona Press:
  Tucson), ed. E.~H. {Levy} \& J.~I. {Lunine}, 125--161

\bibitem[{{Bolatto} {et~al.}(2008){Bolatto}, {Leroy}, {Rosolowsky}, {Walter},
  \& {Blitz}}]{bolatto:2008.gmc.properties}
{Bolatto}, A.~D., {Leroy}, A.~K., {Rosolowsky}, E., {Walter}, F., \& {Blitz},
  L. 2008, \apj, 686, 948

\bibitem[{{Bournaud} {et~al.}(2010){Bournaud}, {Elmegreen}, {Teyssier},
  {Block}, \& {Puerari}}]{bournaud:2010.grav.turbulence.lmc}
{Bournaud}, F., {Elmegreen}, B.~G., {Teyssier}, R., {Block}, D.~L., \&
  {Puerari}, I. 2010, \mnras, 409, 1088

\bibitem[{{Brennan} {et~al.}(2018){Brennan}, {Choi}, {Somerville},
  {Hirschmann}, {Naab}, \& {Ostriker}}]{brennan:radiative.bh.fb}
{Brennan}, R., {Choi}, E., {Somerville}, R.~S., {Hirschmann}, M., {Naab}, T.,
  \& {Ostriker}, J.~P. 2018, \apj, 860, 14

\bibitem[{{Bryan} \& {Norman}(1998)}]{bryan.norman:1998.mvir.definition}
{Bryan}, G.~L., \& {Norman}, M.~L. 1998, \apj, 495, 80

\bibitem[{{Cantalupo}(2010)}]{cantalupo:ionization.by.xrays.from.stars}
{Cantalupo}, S. 2010, \mnras, 403, L16

\bibitem[{{Ceverino} {et~al.}(2014){Ceverino}, {Klypin}, {Klimek},
  {Trujillo-Gomez}, {Churchill}, {Primack}, \& {Dekel}}]{ceverino:2013.rad.fb}
{Ceverino}, D., {Klypin}, A., {Klimek}, E.~S., {Trujillo-Gomez}, S.,
  {Churchill}, C.~W., {Primack}, J., \& {Dekel}, A. 2014, \mnras, 442, 1545

\bibitem[{{Chan} {et~al.}(2015){Chan}, {Kere{\v s}}, {O{\~n}orbe}, {Hopkins},
  {Muratov}, {Faucher-Gigu{\`e}re}, \& {Quataert}}]{chan:fire.dwarf.cusps}
{Chan}, T.~K., {Kere{\v s}}, D., {O{\~n}orbe}, J., {Hopkins}, P.~F., {Muratov},
  A.~L., {Faucher-Gigu{\`e}re}, C.-A., \& {Quataert}, E. 2015, \mnras, 454,
  2981

\bibitem[{{Chan} {et~al.}(2018){Chan}, {Kere{\v s}}, {Wetzel}, {Hopkins},
  {Faucher-Gigu{\`e}re}, {El-Badry}, {Garrison-Kimmel}, \&
  {Boylan-Kolchin}}]{chan:fire.udgs}
{Chan}, T.~K., {Kere{\v s}}, D., {Wetzel}, A., {Hopkins}, P.~F.,
  {Faucher-Gigu{\`e}re}, C.-A., {El-Badry}, K., {Garrison-Kimmel}, S., \&
  {Boylan-Kolchin}, M. 2018, \mnras, 478, 906

\bibitem[{{Choi} {et~al.}(2014){Choi}, {Naab}, {Ostriker}, {Johansson}, \&
  {Moster}}]{choi:2014.mechanical.feedback.model.2}
{Choi}, E., {Naab}, T., {Ostriker}, J.~P., {Johansson}, P.~H., \& {Moster},
  B.~P. 2014, \mnras, 442, 440

\bibitem[{Cole {et~al.}(2000)Cole, Lacey, Baugh, \&
  Frenk}]{cole:durham.sam.initial}
Cole, S., Lacey, C.~G., Baugh, C.~M., \& Frenk, C.~S. 2000, \mnras, 319, 168

\bibitem[{{Col{\'{\i}}n} {et~al.}(2013){Col{\'{\i}}n}, {V{\'a}zquez-Semadeni},
  \& {G{\'o}mez}}]{colin:2013.star.cluster.rhd.cloud.destruction}
{Col{\'{\i}}n}, P., {V{\'a}zquez-Semadeni}, E., \& {G{\'o}mez}, G.~C. 2013,
  \mnras, 435, 1701

\bibitem[{{Conroy} {et~al.}(2006){Conroy}, {Wechsler}, \&
  {Kravtsov}}]{conroy:monotonic.hod}
{Conroy}, C., {Wechsler}, R.~H., \& {Kravtsov}, A.~V. 2006, \apj, 647, 201

\bibitem[{{Corbett Moran} {et~al.}(2018){Corbett Moran}, {Grudi{\'c}}, \&
  {Hopkins}}]{moran:2018.metallicity.cooling.direct.collapse}
{Corbett Moran}, C., {Grudi{\'c}}, M.~Y., \& {Hopkins}, P.~F. 2018, \mnras, in
  press, arXiv:1803.06430

\bibitem[{{Costa} {et~al.}(2018{\natexlab{a}}){Costa}, {Rosdahl}, {Sijacki}, \&
  {Haehnelt}}]{costa:dusty.wind.driving}
{Costa}, T., {Rosdahl}, J., {Sijacki}, D., \& {Haehnelt}, M.~G.
  2018{\natexlab{a}}, \mnras, 473, 4197

\bibitem[{{Costa} {et~al.}(2018{\natexlab{b}}){Costa}, {Rosdahl}, {Sijacki}, \&
  {Haehnelt}}]{costa:qso.outflows.multiple.ir.scattering}
---. 2018{\natexlab{b}}, \mnras, 479, 2079

\bibitem[{{Dav{\'e}} {et~al.}(2016){Dav{\'e}}, {Thompson}, \&
  {Hopkins}}]{dave.2016:mufasa.fire.inspired.cosmo.boxes}
{Dav{\'e}}, R., {Thompson}, R., \& {Hopkins}, P.~F. 2016, \mnras, 462, 3265

\bibitem[{{Davis} {et~al.}(2014){Davis}, {Jiang}, {Stone}, \&
  {Murray}}]{davis:2014.rad.pressure.outflows}
{Davis}, S.~W., {Jiang}, Y.-F., {Stone}, J.~M., \& {Murray}, N. 2014, \apj,
  796, 107

\bibitem[{{Dijkstra} \& {Loeb}(2009)}]{2009MNRAS.396..377D}
{Dijkstra}, M., \& {Loeb}, A. 2009, \mnras, 396, 377

\bibitem[{{Dobbs} {et~al.}(2011){Dobbs}, {Burkert}, \&
  {Pringle}}]{dobbs:2011.why.gmcs.unbound}
{Dobbs}, C.~L., {Burkert}, A., \& {Pringle}, J.~E. 2011, \mnras, 413, 528

\bibitem[{{Dobbs} {et~al.}(2008){Dobbs}, {Glover}, {Clark}, \&
  {Klessen}}]{dobbs:2008.gmc.behavior.insensitive.to.photoelectric.heating}
{Dobbs}, C.~L., {Glover}, S.~C.~O., {Clark}, P.~C., \& {Klessen}, R.~S. 2008,
  \mnras, 389, 1097

\bibitem[{{Dopcke} {et~al.}(2013){Dopcke}, {Glover}, {Clark}, \&
  {Klessen}}]{dopke.2013:fragmentation.all.dust.levels.but.enhanced.with.crit.dust}
{Dopcke}, G., {Glover}, S.~C.~O., {Clark}, P.~C., \& {Klessen}, R.~S. 2013,
  \apj, 766, 103

\bibitem[{{Draine}(2011)}]{draine:ism.book}
{Draine}, B.~T. 2011, {Physics of the Interstellar and Intergalactic Medium}
  (Princeton University Press, Princeton, NJ, USA)

\bibitem[{{Emerick} {et~al.}(2018){Emerick}, {Bryan}, \& {Mac
  Low}}]{emerick:rad.fb.important.stromgren.ok}
{Emerick}, A., {Bryan}, G.~L., \& {Mac Low}, M.-M. 2018, \apjl, in press,
  arXiv:1808.00468

\bibitem[{{Evans} {et~al.}(2009)}]{evans:2009.sf.efficiencies.lifetimes}
{Evans}, N.~J., {et~al.} 2009, \apjs, 181, 321

\bibitem[{{Evans}(1999)}]{evans:1999.sf.gmc.review}
{Evans}, II, N.~J. 1999, \araa, 37, 311

\bibitem[{{Fall} {et~al.}(2010){Fall}, {Krumholz}, \&
  {Matzner}}]{fall:2010.sf.eff.vs.surfacedensity}
{Fall}, S.~M., {Krumholz}, M.~R., \& {Matzner}, C.~D. 2010, \apjl, 710, L142

\bibitem[{{Faucher-Gigu{\`e}re} {et~al.}(2010){Faucher-Gigu{\`e}re}, {Kere{\v
  s}}, {Dijkstra}, {Hernquist}, \&
  {Zaldarriaga}}]{faucher-giguere:2010.lya.cooling.selfshield}
{Faucher-Gigu{\`e}re}, C.-A., {Kere{\v s}}, D., {Dijkstra}, M., {Hernquist},
  L., \& {Zaldarriaga}, M. 2010, \apj, 725, 633

\bibitem[{{Faucher-Gigu{\`e}re} {et~al.}(2011){Faucher-Gigu{\`e}re}, {Kere{\v
  s}}, \& {Ma}}]{faucher-giguere:2011.halo.inflow.properties}
{Faucher-Gigu{\`e}re}, C.-A., {Kere{\v s}}, D., \& {Ma}, C.-P. 2011, \mnras,
  417, 2982

\bibitem[{{Faucher-Gigu{\`e}re} {et~al.}(2009){Faucher-Gigu{\`e}re}, {Lidz},
  {Zaldarriaga}, \& {Hernquist}}]{faucher-giguere:2009.ion.background}
{Faucher-Gigu{\`e}re}, C.-A., {Lidz}, A., {Zaldarriaga}, M., \& {Hernquist}, L.
  2009, \apj, 703, 1416

\bibitem[{{Federrath}(2015)}]{federrath:2014.low.sfe}
{Federrath}, C. 2015, \mnras, 450, 4035

\bibitem[{{Federrath} \&
  {Klessen}(2012)}]{federrath:2012.sfr.vs.model.turb.boxes}
{Federrath}, C., \& {Klessen}, R.~S. 2012, \apj, 761, 156

\bibitem[{{Fielding} {et~al.}(2018){Fielding}, {Quataert}, \&
  {Martizzi}}]{2018MNRAS.481.3325F}
{Fielding}, D., {Quataert}, E., \& {Martizzi}, D. 2018, \mnras, 481, 3325

\bibitem[{{Gentry} {et~al.}(2017){Gentry}, {Krumholz}, {Dekel}, \&
  {Madau}}]{gentry:clustered.sne.momentum.enhancement}
{Gentry}, E.~S., {Krumholz}, M.~R., {Dekel}, A., \& {Madau}, P. 2017, \mnras,
  465, 2471

\bibitem[{{Glover} \& {Clark}(2012)}]{glover:2011.molecules.not.needed.for.sf}
{Glover}, S.~C.~O., \& {Clark}, P.~C. 2012, \mnras, 421, 9

\bibitem[{{Gnedin} \& {Abel}(2001)}]{gnedin.abel.2001:otvet}
{Gnedin}, N.~Y., \& {Abel}, T. 2001, New Astronomy, 6, 437

\bibitem[{{Grassi} {et~al.}(2014){Grassi}, {Bovino}, {Schleicher}, {Prieto},
  {Seifried}, {Simoncini}, \& {Gianturco}}]{krome:2014.chemistry}
{Grassi}, T., {Bovino}, S., {Schleicher}, D.~R.~G., {Prieto}, J., {Seifried},
  D., {Simoncini}, E., \& {Gianturco}, F.~A. 2014, \mnras, 439, 2386

\bibitem[{{Grudi{\'c}} {et~al.}(2018{\natexlab{a}}){Grudi{\'c}}, {Guszejnov},
  {Hopkins}, {Lamberts}, {Boylan-Kolchin}, {Murray}, \&
  {Schmitz}}]{grudic:cluster.properties}
{Grudi{\'c}}, M.~Y., {Guszejnov}, D., {Hopkins}, P.~F., {Lamberts}, A.,
  {Boylan-Kolchin}, M., {Murray}, N., \& {Schmitz}, D. 2018{\natexlab{a}},
  \mnras, 481, 688

\bibitem[{{Grudi{\'c}} {et~al.}(2018{\natexlab{b}}){Grudi{\'c}}, {Hopkins},
  {Faucher-Gigu{\`e}re}, {Quataert}, {Murray}, \& {Kere{\v
  s}}}]{grudic:sfe.cluster.form.surface.density}
{Grudi{\'c}}, M.~Y., {Hopkins}, P.~F., {Faucher-Gigu{\`e}re}, C.-A.,
  {Quataert}, E., {Murray}, N., \& {Kere{\v s}}, D. 2018{\natexlab{b}}, \mnras,
  475, 3511

\bibitem[{{Grudi{\'c}} {et~al.}(2018{\natexlab{c}}){Grudi{\'c}}, {Hopkins},
  {Lee}, {Murray}, {Faucher-Gigu{\`e}re}, \&
  {Johnson}}]{grudic:sfe.gmcs.vs.obs}
{Grudi{\'c}}, M.~Y., {Hopkins}, P.~F., {Lee}, E.~J., {Murray}, N.,
  {Faucher-Gigu{\`e}re}, C.-A., \& {Johnson}, L.~C. 2018{\natexlab{c}}, \mnras,
  in press, arXiv:1809.08348

\bibitem[{{Grudi{\'c}} {et~al.}(2018{\natexlab{d}}){Grudi{\'c}}, {Hopkins},
  {Quataert}, \& {Murray}}]{grudic:max.surface.density}
{Grudi{\'c}}, M.~Y., {Hopkins}, P.~F., {Quataert}, E., \& {Murray}, N.
  2018{\natexlab{d}}, \mnras, in press, arxiv:1804.04137

\bibitem[{{Guszejnov} {et~al.}(2019){Guszejnov}, {Grudi{\'c}}, {Offner},
  {Boylan-Kolchin}, {Faucher-Gig{\`e}re}, {Wetzel}, {Benincasa}, \&
  {Loebman}}]{guszejnov:fire.gmc.props.vs.z}
{Guszejnov}, D., {Grudi{\'c}}, M.~Y., {Offner}, S. S.~R., {Boylan-Kolchin}, M.,
  {Faucher-Gig{\`e}re}, C.-A., {Wetzel}, A., {Benincasa}, S.~M., \& {Loebman},
  S. 2019, arXiv e-prints, arXiv:1910.01163

\bibitem[{{Guszejnov} \& {Hopkins}(2015)}]{guszejnov:cmf.imf}
{Guszejnov}, D., \& {Hopkins}, P.~F. 2015, \mnras, 450, 4137

\bibitem[{{Guszejnov} {et~al.}(2018){Guszejnov}, {Hopkins}, \&
  {Grudi{\'c}}}]{guszejnov:universal.scalings}
{Guszejnov}, D., {Hopkins}, P.~F., \& {Grudi{\'c}}, M.~Y. 2018, \mnras, 477,
  5139

\bibitem[{{Guszejnov} {et~al.}(2017{\natexlab{a}}){Guszejnov}, {Hopkins}, \&
  {Krumholz}}]{guszejnov:protostellar.feedback.stellar.clustering.multiplicity}
{Guszejnov}, D., {Hopkins}, P.~F., \& {Krumholz}, M.~R. 2017{\natexlab{a}},
  \mnras, 468, 4093

\bibitem[{{Guszejnov} {et~al.}(2017{\natexlab{b}}){Guszejnov}, {Hopkins}, \&
  {Ma}}]{guszejnov:imf.var.mw}
{Guszejnov}, D., {Hopkins}, P.~F., \& {Ma}, X. 2017{\natexlab{b}}, \mnras, 472,
  2107

\bibitem[{{Guszejnov} {et~al.}(2016){Guszejnov}, {Krumholz}, \&
  {Hopkins}}]{guszejnov.2015:feedback.imf.invariance}
{Guszejnov}, D., {Krumholz}, M.~R., \& {Hopkins}, P.~F. 2016, \mnras, 458, 673

\bibitem[{{Haid} {et~al.}(2016){Haid}, {Walch}, {Naab}, {Seifried}, {Mackey},
  \& {Gatto}}]{haid:snr.in.clumpy.ism}
{Haid}, S., {Walch}, S., {Naab}, T., {Seifried}, D., {Mackey}, J., \& {Gatto},
  A. 2016, \mnras, 460, 2962

\bibitem[{{Hansen} {et~al.}(2012){Hansen}, {Klein}, {McKee}, \&
  {Fisher}}]{hansen:2012.lowmass.sf.radsims}
{Hansen}, C.~E., {Klein}, R.~I., {McKee}, C.~F., \& {Fisher}, R.~T. 2012, \apj,
  747, 22

\bibitem[{{Harper-Clark} \& {Murray}(2011)}]{harper-clark:2011.gmc.sims}
{Harper-Clark}, E., \& {Murray}, N. 2011, in Computational Star Formation;
  Cambridge University Press, ed. {J.~Alves, B.~G.~Elmegreen, J.~M.~Girart \&
  V.~Trimble}, Vol. 270 (Cambridge, UK: Cambridge University Press), 235--238

\bibitem[{{Harris}(1996)}]{harris96:mw.gcs}
{Harris}, W.~E. 1996, \aj, 112, 1487

\bibitem[{{Heckman} {et~al.}(2000){Heckman}, {Lehnert}, {Strickland}, \&
  {Armus}}]{heckman:superwind.abs.kinematics}
{Heckman}, T.~M., {Lehnert}, M.~D., {Strickland}, D.~K., \& {Armus}, L. 2000,
  \apjs, 129, 493

\bibitem[{{Hopkins}(2012{\natexlab{a}})}]{hopkins:excursion.ism}
{Hopkins}, P.~F. 2012{\natexlab{a}}, \mnras, 423, 2016

\bibitem[{{Hopkins}(2012{\natexlab{b}})}]{hopkins:excursion.imf}
---. 2012{\natexlab{b}}, \mnras, 423, 2037

\bibitem[{{Hopkins}(2013{\natexlab{a}})}]{hopkins:lagrangian.pressure.sph}
---. 2013{\natexlab{a}}, \mnras, 428, 2840

\bibitem[{{Hopkins}(2013{\natexlab{b}})}]{hopkins:frag.theory}
---. 2013{\natexlab{b}}, \mnras, 430, 1653

\bibitem[{{Hopkins}(2015)}]{hopkins:gizmo}
---. 2015, \mnras, 450, 53

\bibitem[{{Hopkins}(2016)}]{hopkins:cg.mhd.gizmo}
---. 2016, \mnras, 462, 576

\bibitem[{{Hopkins}(2017)}]{hopkins:gizmo.diffusion}
---. 2017, \mnras, 466, 3387

\bibitem[{{Hopkins} \&
  {Grudic}(2018)}]{hopkins:rhd.momentum.optically.thick.issues}
{Hopkins}, P.~F., \& {Grudic}, M.~Y. 2018, \mnras, in press, arXiv:1803.07573

\bibitem[{{Hopkins} {et~al.}(2014){Hopkins}, {Keres}, {Onorbe},
  {Faucher-Giguere}, {Quataert}, {Murray}, \& {Bullock}}]{hopkins:2013.fire}
{Hopkins}, P.~F., {Keres}, D., {Onorbe}, J., {Faucher-Giguere}, C.-A.,
  {Quataert}, E., {Murray}, N., \& {Bullock}, J.~S. 2014, \mnras, 445, 581

\bibitem[{{Hopkins} {et~al.}(2013{\natexlab{a}}){Hopkins}, {Narayanan}, \&
  {Murray}}]{hopkins:virial.sf}
{Hopkins}, P.~F., {Narayanan}, D., \& {Murray}, N. 2013{\natexlab{a}}, \mnras,
  432, 2647

\bibitem[{{Hopkins} {et~al.}(2013{\natexlab{b}}){Hopkins}, {Narayanan},
  {Murray}, \& {Quataert}}]{hopkins:dense.gas.tracers}
{Hopkins}, P.~F., {Narayanan}, D., {Murray}, N., \& {Quataert}, E.
  2013{\natexlab{b}}, \mnras, 433, 69

\bibitem[{{Hopkins} {et~al.}(2011){Hopkins}, {Quataert}, \&
  {Murray}}]{hopkins:rad.pressure.sf.fb}
{Hopkins}, P.~F., {Quataert}, E., \& {Murray}, N. 2011, \mnras, 417, 950

\bibitem[{{Hopkins} {et~al.}(2012{\natexlab{a}}){Hopkins}, {Quataert}, \&
  {Murray}}]{hopkins:stellar.fb.winds}
---. 2012{\natexlab{a}}, \mnras, 421, 3522

\bibitem[{{Hopkins} {et~al.}(2012{\natexlab{b}}){Hopkins}, {Quataert}, \&
  {Murray}}]{hopkins:fb.ism.prop}
---. 2012{\natexlab{b}}, \mnras, 421, 3488

\bibitem[{{Hopkins} \& {Raives}(2016)}]{hopkins:mhd.gizmo}
{Hopkins}, P.~F., \& {Raives}, M.~J. 2016, \mnras, 455, 51

\bibitem[{{Hopkins} \&
  {Squire}(2018{\natexlab{a}})}]{hopkins:2017.acoustic.RDI}
{Hopkins}, P.~F., \& {Squire}, J. 2018{\natexlab{a}}, \mnras, 480, 2813

\bibitem[{{Hopkins} \& {Squire}(2018{\natexlab{b}})}]{hopkins:2018.mhd.rdi}
---. 2018{\natexlab{b}}, \mnras, 479, 4681

\bibitem[{{Hopkins} {et~al.}(2016){Hopkins}, {Torrey}, {Faucher-Gigu{\`e}re},
  {Quataert}, \& {Murray}}]{hopkins:qso.stellar.fb.together}
{Hopkins}, P.~F., {Torrey}, P., {Faucher-Gigu{\`e}re}, C.-A., {Quataert}, E.,
  \& {Murray}, N. 2016, \mnras, 458, 816

\bibitem[{{Hopkins} {et~al.}(2018{\natexlab{a}}){Hopkins}, {Wetzel}, {Kere{\v
  s}}, {Faucher-Gigu{\`e}re}, {Quataert}, {Boylan-Kolchin}, {Murray},
  {Hayward}, {Garrison-Kimmel}, {Hummels}, {Feldmann}, {Torrey}, {Ma},
  {Angl{\'e}s-Alc{\'a}zar}, {Su}, {Orr}, {Schmitz}, {Escala}, {Sanderson},
  {Grudi{\'c}}, {Hafen}, {Kim}, {Fitts}, {Bullock}, {Wheeler}, {Chan},
  {Elbert}, \& {Narayanan}}]{hopkins:fire2.methods}
{Hopkins}, P.~F., {et~al.} 2018{\natexlab{a}}, \mnras, 480, 800

\bibitem[{{Hopkins} {et~al.}(2018{\natexlab{b}}){Hopkins}, {Wetzel}, {Kere{\v
  s}}, {Faucher-Gigu{\`e}re}, {Quataert}, {Boylan-Kolchin}, {Murray},
  {Hayward}, \& {El-Badry}}]{hopkins:sne.methods}
---. 2018{\natexlab{b}}, \mnras, 477, 1578

\bibitem[{{Hopkins} {et~al.}(2019){Hopkins}, {Chan}, {Garrison-Kimmel}, {Ji},
  {Su}, {Hummels}, {Keres}, {Quataert}, \&
  {Faucher-Giguere}}]{hopkins:cr.mhd.fire2}
---. 2019, \mnras, in press, arXiv:1905.04321, arXiv:1905.04321

\bibitem[{{Howard} {et~al.}(2016){Howard}, {Pudritz}, \&
  {Harris}}]{howard:gmc.rad.fx}
{Howard}, C.~S., {Pudritz}, R.~E., \& {Harris}, W.~E. 2016, \mnras, 461, 2953

\bibitem[{{Hu} {et~al.}(2017){Hu}, {Naab}, {Glover}, {Walch}, \&
  {Clark}}]{hu:2017.rad.fb.model.photoelectric}
{Hu}, C.-Y., {Naab}, T., {Glover}, S.~C.~O., {Walch}, S., \& {Clark}, P.~C.
  2017, \mnras, in press, arXiv:1701.08779

\bibitem[{{Hu} {et~al.}(2016){Hu}, {Naab}, {Walch}, {Glover}, \&
  {Clark}}]{hu:photoelectric.heating}
{Hu}, C.-Y., {Naab}, T., {Walch}, S., {Glover}, S.~C.~O., \& {Clark}, P.~C.
  2016, \mnras, 458, 3528

\bibitem[{{Ji} {et~al.}(2014){Ji}, {Frebel}, \&
  {Bromm}}]{ji:2014.silicate.dust.cooling.for.metal.poor.star.criterion.and.tests}
{Ji}, A.~P., {Frebel}, A., \& {Bromm}, V. 2014, \apj, 782, 95

\bibitem[{{Jiang} {et~al.}(2013){Jiang}, {Davis}, \&
  {Stone}}]{jiang:RT.RHD.instabilities}
{Jiang}, Y.-F., {Davis}, S.~W., \& {Stone}, J.~M. 2013, \apj, 763, 102

\bibitem[{{Jiang} {et~al.}(2014){Jiang}, {Stone}, \&
  {Davis}}]{jiang:2014.rhd.solver.local}
{Jiang}, Y.-F., {Stone}, J.~M., \& {Davis}, S.~W. 2014, \apjs, 213, 7

\bibitem[{{Kannan} {et~al.}(2014{\natexlab{a}}){Kannan}, {Stinson},
  {Macci{\`o}}, {Brook}, {Weinmann}, {Wadsley}, \&
  {Couchman}}]{kannan:2013.early.fb.gives.good.highz.mgal.mhalo}
{Kannan}, R., {Stinson}, G.~S., {Macci{\`o}}, A.~V., {Brook}, C., {Weinmann},
  S.~M., {Wadsley}, J., \& {Couchman}, H.~M.~P. 2014{\natexlab{a}}, \mnras,
  437, 3529

\bibitem[{{Kannan} {et~al.}(2018){Kannan}, {Vogelsberger}, {Marinacci},
  {McKinnon}, {Pakmor}, \& {Springel}}]{kannan:2018.arepo.rhd}
{Kannan}, R., {Vogelsberger}, M., {Marinacci}, F., {McKinnon}, R., {Pakmor},
  R., \& {Springel}, V. 2018, \mnras, in press, arxiv:1804.01987

\bibitem[{{Kannan} {et~al.}(2014{\natexlab{b}}){Kannan}, {Stinson},
  {Macci{\`o}}, {Hennawi}, {Woods}, {Wadsley}, {Shen}, {Robitaille},
  {Cantalupo}, {Quinn}, \& {Christensen}}]{kannan:photoion.feedback.sims}
{Kannan}, R., {et~al.} 2014{\natexlab{b}}, \mnras, 437, 2882

\bibitem[{{Katz} {et~al.}(1996){Katz}, {Weinberg}, \&
  {Hernquist}}]{katz:treesph}
{Katz}, N., {Weinberg}, D.~H., \& {Hernquist}, L. 1996, \apjs, 105, 19

\bibitem[{{Kennicutt}(1998)}]{kennicutt98}
{Kennicutt}, Jr., R.~C. 1998, \apj, 498, 541

\bibitem[{{Kere{\v s}} {et~al.}(2009){Kere{\v s}}, {Katz}, {Dav{\'e}},
  {Fardal}, \& {Weinberg}}]{keres:fb.constraints.from.cosmo.sims}
{Kere{\v s}}, D., {Katz}, N., {Dav{\'e}}, R., {Fardal}, M., \& {Weinberg},
  D.~H. 2009, \mnras, 396, 2332

\bibitem[{{Kim} {et~al.}(2017){Kim}, {Kim}, {Ostriker}, \&
  {Skinner}}]{kim:2017.art.uv.starclusters}
{Kim}, J.-G., {Kim}, W.-T., {Ostriker}, E.~C., \& {Skinner}, M.~A. 2017, \apj,
  851, 93

\bibitem[{{Kim} {et~al.}(2018){Kim}, {Ma}, {Grudi{\'c}}, {Hopkins}, {Hayward},
  {Wetzel}, {Faucher-Gigu{\`e}re}, {Kere{\v s}}, {Garrison-Kimmel}, \&
  {Murray}}]{kim:gc.form.FIRE}
{Kim}, J.-h., {et~al.} 2018, \mnras, 474, 4232

\bibitem[{{Kimm} {et~al.}(2018){Kimm}, {Haehnelt}, {Blaizot}, {Katz},
  {Michel-Dansac}, {Garel}, {Rosdahl}, \&
  {Teyssier}}]{kimm:lyman.alpha.rad.pressure}
{Kimm}, T., {Haehnelt}, M., {Blaizot}, J., {Katz}, H., {Michel-Dansac}, L.,
  {Garel}, T., {Rosdahl}, J., \& {Teyssier}, R. 2018, \mnras, 475, 4617

\bibitem[{{Kroupa}(2002)}]{kroupa:imf}
{Kroupa}, P. 2002, Science, 295, 82

\bibitem[{{Krumholz} \& {Gnedin}(2011)}]{krumholz:2011.molecular.prescription}
{Krumholz}, M.~R., \& {Gnedin}, N.~Y. 2011, \apj, 729, 36

\bibitem[{{Krumholz} {et~al.}(2011){Krumholz}, {Klein}, \&
  {McKee}}]{krumholz:2011.rhd.starcluster.sim}
{Krumholz}, M.~R., {Klein}, R.~I., \& {McKee}, C.~F. 2011, \apj, 740, 74

\bibitem[{{Krumholz} \&
  {Thompson}(2012)}]{krumholz:2012.rad.pressure.rt.instab}
{Krumholz}, M.~R., \& {Thompson}, T.~A. 2012, \apj, 760, 155

\bibitem[{{Kuiper} {et~al.}(2012){Kuiper}, {Klahr}, {Beuther}, \&
  {Henning}}]{kuiper:2012.rad.pressure.outflow.vs.rt.method}
{Kuiper}, R., {Klahr}, H., {Beuther}, H., \& {Henning}, T. 2012, \aap, 537,
  A122

\bibitem[{Kurosawa \&
  Proga(2008)}]{kurosawa:outflows.w.rad.in.precession.w.proga}
Kurosawa, R., \& Proga, D. 2008, The Astrophysical Journal, 674, 97

\bibitem[{{Lamberts} {et~al.}(2016){Lamberts}, {Garrison-Kimmel}, {Clausen}, \&
  {Hopkins}}]{lamberts:bh.mgr.progenitors.ligo}
{Lamberts}, A., {Garrison-Kimmel}, S., {Clausen}, D.~R., \& {Hopkins}, P.~F.
  2016, \mnras, 463, L31

\bibitem[{{Lamberts} {et~al.}(2018){Lamberts}, {Garrison-Kimmel}, {Hopkins},
  {Quataert}, {Bullock}, {Faucher-Gigu{\`e}re}, {Wetzel}, {Kere{\v s}},
  {Drango}, \& {Sanderson}}]{lamberts:lisa.bh.binary.pop.pred}
{Lamberts}, A., {et~al.} 2018, \mnras, 480, 2704

\bibitem[{{Lee} {et~al.}(2016){Lee}, {Miville-Desch{\^e}nes}, \&
  {Murray}}]{lee:gmc.sfe}
{Lee}, E.~J., {Miville-Desch{\^e}nes}, M.-A., \& {Murray}, N.~W. 2016, \apj,
  833, 229

\bibitem[{{Leitherer} {et~al.}(1999)}]{starburst99}
{Leitherer}, C., {et~al.} 1999, \apjs, 123, 3

\bibitem[{{Levermore}(1984)}]{levermore:1984.FLD.M1}
{Levermore}, C.~D. 1984, Journal of Quantitative Spectroscopy and Radiative
  Transfer, 31, 149

\bibitem[{{Lopez} {et~al.}(2011){Lopez}, {Krumholz}, {Bolatto}, {Prochaska}, \&
  {Ramirez-Ruiz}}]{lopez:2010.stellar.fb.30.dor}
{Lopez}, L.~A., {Krumholz}, M.~R., {Bolatto}, A.~D., {Prochaska}, J.~X., \&
  {Ramirez-Ruiz}, E. 2011, \apj, 731, 91

\bibitem[{{Lupi} {et~al.}(2018){Lupi}, {Bovino}, {Capelo}, {Volonteri}, \&
  {Silk}}]{lupi:2018.h2.sfr.rhd}
{Lupi}, A., {Bovino}, S., {Capelo}, P.~R., {Volonteri}, M., \& {Silk}, J. 2018,
  \mnras, 474, 2884

\bibitem[{{Ma} {et~al.}(2016{\natexlab{a}}){Ma}, {Hopkins},
  {Faucher-Gigu{\`e}re}, {Zolman}, {Muratov}, {Kere{\v s}}, \&
  {Quataert}}]{ma:2015.fire.mass.metallicity}
{Ma}, X., {Hopkins}, P.~F., {Faucher-Gigu{\`e}re}, C.-A., {Zolman}, N.,
  {Muratov}, A.~L., {Kere{\v s}}, D., \& {Quataert}, E. 2016{\natexlab{a}},
  \mnras, 456, 2140

\bibitem[{{Ma} {et~al.}(2016{\natexlab{b}}){Ma}, {Hopkins}, {Kasen},
  {Quataert}, {Faucher-Gigu{\`e}re}, {Kere{\v s}}, {Murray}, \&
  {Strom}}]{ma.2016:binary.star.escape.fraction.effects}
{Ma}, X., {Hopkins}, P.~F., {Kasen}, D., {Quataert}, E., {Faucher-Gigu{\`e}re},
  C.-A., {Kere{\v s}}, D., {Murray}, N., \& {Strom}, A. 2016{\natexlab{b}},
  \mnras, 459, 3614

\bibitem[{{Ma} {et~al.}(2015){Ma}, {Kasen}, {Hopkins}, {Faucher-Gigu{\`e}re},
  {Quataert}, {Kere{\v s}}, \& {Murray}}]{ma:2015.fire.escape.fractions}
{Ma}, X., {Kasen}, D., {Hopkins}, P.~F., {Faucher-Gigu{\`e}re}, C.-A.,
  {Quataert}, E., {Kere{\v s}}, D., \& {Murray}, N. 2015, \mnras, 453, 960

\bibitem[{{Ma} {et~al.}(2018){Ma}, {Hopkins}, {Garrison-Kimmel},
  {Faucher-Gigu{\`e}re}, {Quataert}, {Boylan-Kolchin}, {Hayward}, {Feldmann},
  \& {Kere{\v s}}}]{ma:fire2.reion.gal.lfs}
{Ma}, X., {et~al.} 2018, \mnras, 478, 1694

\bibitem[{{Martin}(1999)}]{martin99:outflow.vs.m}
{Martin}, C.~L. 1999, \apj, 513, 156

\bibitem[{{Martin} {et~al.}(2010){Martin}, {Scannapieco}, {Ellison}, {Hennawi},
  {Djorgovski}, \& {Fournier}}]{martin:2010.metal.enriched.regions}
{Martin}, C.~L., {Scannapieco}, E., {Ellison}, S.~L., {Hennawi}, J.~F.,
  {Djorgovski}, S.~G., \& {Fournier}, A.~P. 2010, \apj, 721, 174

\bibitem[{{Morrison} \&
  {McCammon}(1983)}]{morrison.mccammon.83:photoelectric.absorption}
{Morrison}, R., \& {McCammon}, D. 1983, \apj, 270, 119

\bibitem[{{Moster} {et~al.}(2010){Moster}, {Somerville}, {Maulbetsch}, {van den
  Bosch}, {Macci{\`o}}, {Naab}, \& {Oser}}]{moster:stellar.vs.halo.mass.to.z1}
{Moster}, B.~P., {Somerville}, R.~S., {Maulbetsch}, C., {van den Bosch}, F.~C.,
  {Macci{\`o}}, A.~V., {Naab}, T., \& {Oser}, L. 2010, \apj, 710, 903

\bibitem[{{Muratov} {et~al.}(2015){Muratov}, {Kere{\v s}},
  {Faucher-Gigu{\`e}re}, {Hopkins}, {Quataert}, \&
  {Murray}}]{muratov:2015.fire.winds}
{Muratov}, A.~L., {Kere{\v s}}, D., {Faucher-Gigu{\`e}re}, C.-A., {Hopkins},
  P.~F., {Quataert}, E., \& {Murray}, N. 2015, \mnras, 454, 2691

\bibitem[{{Murray}(2011)}]{murray:2010.sfe.mw.gmc}
{Murray}, N. 2011, \apj, 729, 133

\bibitem[{{Murray} {et~al.}(2005){Murray}, {Quataert}, \&
  {Thompson}}]{murray:momentum.winds}
{Murray}, N., {Quataert}, E., \& {Thompson}, T.~A. 2005, \apj, 618, 569

\bibitem[{{Murray} {et~al.}(2010){Murray}, {Quataert}, \&
  {Thompson}}]{murray:molcloud.disrupt.by.rad.pressure}
---. 2010, \apj, 709, 191

\bibitem[{{O{\~n}orbe} {et~al.}(2015){O{\~n}orbe}, {Boylan-Kolchin}, {Bullock},
  {Hopkins}, {Kere{\v s}}, {Faucher-Gigu{\`e}re}, {Quataert}, \&
  {Murray}}]{onorbe:2015.fire.cores}
{O{\~n}orbe}, J., {Boylan-Kolchin}, M., {Bullock}, J.~S., {Hopkins}, P.~F.,
  {Kere{\v s}}, D., {Faucher-Gigu{\`e}re}, C.-A., {Quataert}, E., \& {Murray},
  N. 2015, \mnras, 454, 2092

\bibitem[{{Offner} {et~al.}(2013){Offner}, {Clark}, {Hennebelle}, {Bastian},
  {Bate}, {Hopkins}, {Moraux}, \& {Whitworth}}]{offner:2013.imf.review}
{Offner}, S.~S.~R., {Clark}, P.~C., {Hennebelle}, P., {Bastian}, N., {Bate},
  M.~R., {Hopkins}, P.~F., {Moraux}, E., \& {Whitworth}, A.~P. 2013, Protostars
  and Planets VI, University of Arizona Press (2014), eds. H. Beuther, R. S.
  Klessen, C. P. Dullemond, Th. Henning (arXiv:1312.5326)

\bibitem[{{Offner} {et~al.}(2009){Offner}, {Klein}, {McKee}, \&
  {Krumholz}}]{offner:2009.rhd.lowmass.stars}
{Offner}, S.~S.~R., {Klein}, R.~I., {McKee}, C.~F., \& {Krumholz}, M.~R. 2009,
  \apj, 703, 131

\bibitem[{{Oklop{\v c}i{\'c}} {et~al.}(2017){Oklop{\v c}i{\'c}}, {Hopkins},
  {Feldmann}, {Kere{\v s}}, {Faucher-Gigu{\`e}re}, \&
  {Murray}}]{oklopcic:clumpy.highz.gals.fire.case.study.clumps.not.long.lived}
{Oklop{\v c}i{\'c}}, A., {Hopkins}, P.~F., {Feldmann}, R., {Kere{\v s}}, D.,
  {Faucher-Gigu{\`e}re}, C.-A., \& {Murray}, N. 2017, \mnras, 465, 952

\bibitem[{{Oppenheimer} {et~al.}(2018){Oppenheimer}, {Segers}, {Schaye},
  {Richings}, \& {Crain}}]{oppenheimer:2018.flickering.heating.cgm}
{Oppenheimer}, B.~D., {Segers}, M., {Schaye}, J., {Richings}, A.~J., \&
  {Crain}, R.~A. 2018, \mnras, 474, 4740

\bibitem[{{Orr} {et~al.}(2018{\natexlab{a}}){Orr}, {Hayward}, \&
  {Hopkins}}]{orr:non.eqm.sf.model}
{Orr}, M.~E., {Hayward}, C.~C., \& {Hopkins}, P.~F. 2018{\natexlab{a}}, \mnras,
  submitted, arXiv:1810.09460

\bibitem[{{Orr} {et~al.}(2018{\natexlab{b}}){Orr}, {Hayward}, {Hopkins},
  {Chan}, {Faucher-Gigu{\`e}re}, {Feldmann}, {Kere{\v s}}, {Murray}, \&
  {Quataert}}]{orr:ks.law}
{Orr}, M.~E., {et~al.} 2018{\natexlab{b}}, \mnras, 478, 3653

\bibitem[{{Peng} {et~al.}(2008){Peng}, {Jord{\'a}n}, {C{\^o}t{\'e}},
  {Takamiya}, {West}, {Blakeslee}, {Chen}, {Ferrarese}, {Mei}, {Tonry}, \&
  {West}}]{peng:2008.gc.specific.frequencies}
{Peng}, E.~W., {et~al.} 2008, \apj, 681, 197

\bibitem[{{Pettini} {et~al.}(2003){Pettini}, {Madau}, {Bolte}, {Prochaska},
  {Ellison}, \& {Fan}}]{pettini:2003.igm.metal.evol}
{Pettini}, M., {Madau}, P., {Bolte}, M., {Prochaska}, J.~X., {Ellison}, S.~L.,
  \& {Fan}, X. 2003, \apj, 594, 695

\bibitem[{{Power} {et~al.}(2003){Power}, {Navarro}, {Jenkins}, {Frenk},
  {White}, {Springel}, {Stadel}, \&
  {Quinn}}]{power:2003.nfw.models.convergence}
{Power}, C., {Navarro}, J.~F., {Jenkins}, A., {Frenk}, C.~S., {White},
  S.~D.~M., {Springel}, V., {Stadel}, J., \& {Quinn}, T. 2003, \mnras, 338, 14

\bibitem[{{Proga}(2000)}]{proga:disk.winds.2000}
{Proga}, D. 2000, \apj, 538, 684

\bibitem[{{Raskutti} {et~al.}(2016){Raskutti}, {Ostriker}, \&
  {Skinner}}]{raskutti:2016.m1.cloud.sims}
{Raskutti}, S., {Ostriker}, E.~C., \& {Skinner}, M.~A. 2016, \mnras, submitted,
  arXiv:1608.04469

\bibitem[{{Rice} {et~al.}(2016){Rice}, {Goodman}, {Bergin}, {Beaumont}, \&
  {Dame}}]{rice:2016.gmc.mw.catalogue}
{Rice}, T.~S., {Goodman}, A.~A., {Bergin}, E.~A., {Beaumont}, C., \& {Dame},
  T.~M. 2016, \apj, 822, 52

\bibitem[{{Richings} \&
  {Schaye}(2016)}]{richings:2016.chemistry.uvb.photoelec.fx}
{Richings}, A.~J., \& {Schaye}, J. 2016, \mnras, 458, 270

\bibitem[{{Rosdahl} {et~al.}(2013){Rosdahl}, {Blaizot}, {Aubert}, {Stranex}, \&
  {Teyssier}}]{rosdahl:2013.m1.ramses}
{Rosdahl}, J., {Blaizot}, J., {Aubert}, D., {Stranex}, T., \& {Teyssier}, R.
  2013, \mnras, 436, 2188

\bibitem[{{Rosdahl} {et~al.}(2015){Rosdahl}, {Schaye}, {Teyssier}, \&
  {Agertz}}]{rosdahl:2015.galaxies.shine.rad.hydro}
{Rosdahl}, J., {Schaye}, J., {Teyssier}, R., \& {Agertz}, O. 2015, \mnras, 451,
  34

\bibitem[{{Rosdahl} \& {Teyssier}(2015)}]{rosdahl:m1.method.ramses}
{Rosdahl}, J., \& {Teyssier}, R. 2015, \mnras, 449, 4380

\bibitem[{{Rosen} {et~al.}(2016){Rosen}, {Krumholz}, {McKee}, \&
  {Klein}}]{rosen:massive.sf.rhd}
{Rosen}, A.~L., {Krumholz}, M.~R., {McKee}, C.~F., \& {Klein}, R.~I. 2016,
  \mnras, 463, 2553

\bibitem[{{Roth} {et~al.}(2012){Roth}, {Kasen}, {Hopkins}, \&
  {Quataert}}]{roth:2012.rad.transfer.agn}
{Roth}, N., {Kasen}, D., {Hopkins}, P.~F., \& {Quataert}, E. 2012, \apj, 759,
  36

\bibitem[{{Ro{\v s}kar} {et~al.}(2014){Ro{\v s}kar}, {Teyssier}, {Agertz},
  {Wetzstein}, \& {Moore}}]{roskar:2014.stellar.rad.fx.approx.model}
{Ro{\v s}kar}, R., {Teyssier}, R., {Agertz}, O., {Wetzstein}, M., \& {Moore},
  B. 2014, \mnras, 444, 2837

\bibitem[{{Sales} {et~al.}(2013){Sales}, {Marinacci}, {Springel}, \&
  {Petkova}}]{sales:2013.phototion.fb.strong}
{Sales}, L.~V., {Marinacci}, F., {Springel}, V., \& {Petkova}, M. 2013, \mnras,
  in press, arXiv:1310.7572

\bibitem[{{Sato} {et~al.}(2009){Sato}, {Martin}, {Noeske}, {Koo}, \&
  {Lotz}}]{sato:2009.ulirg.outflows}
{Sato}, T., {Martin}, C.~L., {Noeske}, K.~G., {Koo}, D.~C., \& {Lotz}, J.~M.
  2009, \apj, 696, 214

\bibitem[{{Sazonov} {et~al.}(2005){Sazonov}, {Ostriker}, {Ciotti}, \&
  {Sunyaev}}]{sazonov:radiative.feedback}
{Sazonov}, S.~Y., {Ostriker}, J.~P., {Ciotti}, L., \& {Sunyaev}, R.~A. 2005,
  \mnras, 358, 168

\bibitem[{{Sazonov} {et~al.}(2004){Sazonov}, {Ostriker}, \&
  {Sunyaev}}]{sazonov04:qso.radiative.heating}
{Sazonov}, S.~Y., {Ostriker}, J.~P., \& {Sunyaev}, R.~A. 2004, \mnras, 347, 144

\bibitem[{{Semenov} {et~al.}(2003){Semenov}, {Henning}, {Helling}, {Ilgner}, \&
  {Sedlmayr}}]{semenov:2003.dust.opacities}
{Semenov}, D., {Henning}, T., {Helling}, C., {Ilgner}, M., \& {Sedlmayr}, E.
  2003, \aap, 410, 611

\bibitem[{{Semenov} {et~al.}(2018){Semenov}, {Kravtsov}, \&
  {Gnedin}}]{semenov:local.vs.global.sfe}
{Semenov}, V.~A., {Kravtsov}, A.~V., \& {Gnedin}, N.~Y. 2018, \apj, 861, 4

\bibitem[{{Shetty} \& {Ostriker}(2008)}]{shetty:2008.sf.feedback.model}
{Shetty}, R., \& {Ostriker}, E.~C. 2008, \apj, 684, 978

\bibitem[{{Skinner} \& {Ostriker}(2015)}]{skinner:2015.cloud.sf.frag}
{Skinner}, M.~A., \& {Ostriker}, E.~C. 2015, \apj, 809, 187

\bibitem[{{Smith} {et~al.}(2017){Smith}, {Bromm}, \&
  {Loeb}}]{2017MNRAS.464.2963S}
{Smith}, A., {Bromm}, V., \& {Loeb}, A. 2017, \mnras, 464, 2963

\bibitem[{{Smith} {et~al.}(2018{\natexlab{a}}){Smith}, {Ma}, {Bromm},
  {Finkelstein}, {Hopkins}, {Faucher-Gigu{\`e}re}, \& {Kere{\v
  s}}}]{smith:lyalpha.rt.fire.sim.escape.fraction}
{Smith}, A., {Ma}, X., {Bromm}, V., {Finkelstein}, S.~L., {Hopkins}, P.~F.,
  {Faucher-Gigu{\`e}re}, C.-A., \& {Kere{\v s}}, D. 2018{\natexlab{a}}, \mnras,
  in press, arXiv:1810.08185

\bibitem[{{Smith} {et~al.}(2018{\natexlab{b}}){Smith}, {Tsang}, {Bromm}, \&
  {Milosavljevi{\'c}}}]{2018MNRAS.479.2065S}
{Smith}, A., {Tsang}, B.~T.-H., {Bromm}, V., \& {Milosavljevi{\'c}}, M.
  2018{\natexlab{b}}, \mnras, 479, 2065

\bibitem[{{Solomon} {et~al.}(1987){Solomon}, {Rivolo}, {Barrett}, \&
  {Yahil}}]{solomon:gmc.scalings}
{Solomon}, P.~M., {Rivolo}, A.~R., {Barrett}, J., \& {Yahil}, A. 1987, \apj,
  319, 730

\bibitem[{{Somerville} \& {Primack}(1999)}]{somerville99:sam}
{Somerville}, R.~S., \& {Primack}, J.~R. 1999, \mnras, 310, 1087

\bibitem[{{Songaila}(2005)}]{songaila:2005.igm.metal.evol}
{Songaila}, A. 2005, \aj, 130, 1996

\bibitem[{{Springel} \& {Hernquist}(2003{\natexlab{a}})}]{springel:multiphase}
{Springel}, V., \& {Hernquist}, L. 2003{\natexlab{a}}, \mnras, 339, 289

\bibitem[{{Springel} \& {Hernquist}(2003{\natexlab{b}})}]{springel:lcdm.sfh}
---. 2003{\natexlab{b}}, \mnras, 339, 312

\bibitem[{{Squire} \& {Hopkins}(2018{\natexlab{a}})}]{squire:rdi.ppd}
{Squire}, J., \& {Hopkins}, P.~F. 2018{\natexlab{a}}, \mnras, 477, 5011

\bibitem[{{Squire} \& {Hopkins}(2018{\natexlab{b}})}]{squire.hopkins:RDI}
---. 2018{\natexlab{b}}, \apjl, 856, L15

\bibitem[{{Steidel} {et~al.}(2010){Steidel}, {Erb}, {Shapley}, {Pettini},
  {Reddy}, {Bogosavljevi{\'c}}, {Rudie}, \&
  {Rakic}}]{steidel:2010.outflow.kinematics}
{Steidel}, C.~C., {Erb}, D.~K., {Shapley}, A.~E., {Pettini}, M., {Reddy}, N.,
  {Bogosavljevi{\'c}}, M., {Rudie}, G.~C., \& {Rakic}, O. 2010, \apj, 717, 289

\bibitem[{{Stinson} {et~al.}(2013){Stinson}, {Brook}, {Macci{\`o}}, {Wadsley},
  {Quinn}, \& {Couchman}}]{stinson:2013.new.early.stellar.fb.models}
{Stinson}, G.~S., {Brook}, C., {Macci{\`o}}, A.~V., {Wadsley}, J., {Quinn},
  T.~R., \& {Couchman}, H.~M.~P. 2013, \mnras, 428, 129

\bibitem[{{Su} {et~al.}(2017){Su}, {Hopkins}, {Hayward}, {Faucher-Gigu{\`e}re},
  {Kere{\v s}}, {Ma}, \& {Robles}}]{su:2016.weak.mhd.cond.visc.turbdiff.fx}
{Su}, K.-Y., {Hopkins}, P.~F., {Hayward}, C.~C., {Faucher-Gigu{\`e}re}, C.-A.,
  {Kere{\v s}}, D., {Ma}, X., \& {Robles}, V.~H. 2017, \mnras, 471, 144

\bibitem[{{Su} {et~al.}(2018){Su}, {Hopkins}, {Hayward}, {Ma},
  {Boylan-Kolchin}, {Kasen}, {Kere{\v s}}, {Faucher-Gigu{\`e}re}, {Orr}, \&
  {Wheeler}}]{su:discrete.imf.fx.fire}
{Su}, K.-Y., {et~al.} 2018, \mnras, 480, 1666

\bibitem[{{Takeuchi} {et~al.}(2014){Takeuchi}, {Ohsuga}, \&
  {Mineshige}}]{takeuchi:rhd.mhd.dusty.wind.instability}
{Takeuchi}, S., {Ohsuga}, K., \& {Mineshige}, S. 2014, \pasj, 66, 48

\bibitem[{{Tasker}(2011)}]{tasker:2011.photoion.heating.gmc.evol}
{Tasker}, E.~J. 2011, \apj, 730, 11

\bibitem[{{Tasker} \& {Bryan}(2008)}]{tasker:2008.gas.turb.vs.gal.prop}
{Tasker}, E.~J., \& {Bryan}, G.~L. 2008, \apj, 673, 810

\bibitem[{{Thompson} {et~al.}(2005){Thompson}, {Quataert}, \&
  {Murray}}]{thompson:rad.pressure}
{Thompson}, T.~A., {Quataert}, E., \& {Murray}, N. 2005, \apj, 630, 167

\bibitem[{{Thoul} \& {Weinberg}(1996)}]{thoul.weinberg:uvb.effects.on.dwarfs}
{Thoul}, A.~A., \& {Weinberg}, D.~H. 1996, \apj, 465, 608

\bibitem[{{Tielens}(2005)}]{tielens:2005.book}
{Tielens}, A.~G.~G.~M. 2005, {The Physics and Chemistry of the Interstellar
  Medium} (Cambridge, UK: Cambridge University Press)

\bibitem[{{Torrey} {et~al.}(2017){Torrey}, {Hopkins}, {Faucher-Gigu{\`e}re},
  {Vogelsberger}, {Quataert}, {Kere{\v s}}, \&
  {Murray}}]{torrey.2016:fire.galactic.nuclei.star.formation.instability}
{Torrey}, P., {Hopkins}, P.~F., {Faucher-Gigu{\`e}re}, C.-A., {Vogelsberger},
  M., {Quataert}, E., {Kere{\v s}}, D., \& {Murray}, N. 2017, \mnras, 467, 2301

\bibitem[{{Trujillo-Gomez} {et~al.}(2013){Trujillo-Gomez}, {Klypin}, {Colin},
  {Ceverino}, {Arraki}, \& {Primack}}]{trujillo-gomez:2013.rad.fb.dwarfs}
{Trujillo-Gomez}, S., {Klypin}, A., {Colin}, P., {Ceverino}, D., {Arraki}, K.,
  \& {Primack}, J. 2013, \mnras, in press, arXiv:1311.2910

\bibitem[{{Tsang} \&
  {Milosavljevi{\'c}}(2015)}]{tsang:monte.carlo.rhd.dusty.wind}
{Tsang}, B.~T.-H., \& {Milosavljevi{\'c}}, M. 2015, \mnras, 453, 1108

\bibitem[{{Walch} {et~al.}(2015){Walch}, {Girichidis}, {Naab}, {Gatto},
  {Glover}, {W{\"u}nsch}, {Klessen}, {Clark}, {Peters}, {Derigs}, \&
  {Baczynski}}]{2015MNRAS.454..238W}
{Walch}, S., {et~al.} 2015, \mnras, 454, 238

\bibitem[{{Wheeler} {et~al.}(2017){Wheeler}, {Pace}, {Bullock},
  {Boylan-Kolchin}, {O{\~n}orbe}, {Elbert}, {Fitts}, {Hopkins}, \& {Kere{\v
  s}}}]{wheeler.2015:dwarfs.isolated.not.rotating}
{Wheeler}, C., {et~al.} 2017, \mnras, 465, 2420

\bibitem[{{Wheeler} {et~al.}(2018){Wheeler}, {Hopkins}, {Pace},
  {Garrison-Kimmel}, {Boylan-Kolchin}, {Wetzel}, {Bullock}, {Keres},
  {Faucher-Giguere}, \& {Quataert}}]{wheeler:ultra.highres.dwarfs}
---. 2018, arXiv e-prints, arXiv:1812.02749

\bibitem[{{Williams} \& {McKee}(1997)}]{williams:1997.gmc.prop}
{Williams}, J.~P., \& {McKee}, C.~F. 1997, \apj, 476, 166

\bibitem[{{Wise} \& {Abel}(2008)}]{wise:2008.first.star.fb}
{Wise}, J.~H., \& {Abel}, T. 2008, \apj, 685, 40

\bibitem[{{Wise} {et~al.}(2012){Wise}, {Abel}, {Turk}, {Norman}, \&
  {Smith}}]{wise:2012.rad.pressure.effects}
{Wise}, J.~H., {Abel}, T., {Turk}, M.~J., {Norman}, M.~L., \& {Smith}, B.~D.
  2012, \mnras, 427, 311

\bibitem[{{Wolfire} {et~al.}(1995){Wolfire}, {Hollenbach}, {McKee}, {Tielens},
  \& {Bakes}}]{wolfire:1995.neutral.ism.phases}
{Wolfire}, M.~G., {Hollenbach}, D., {McKee}, C.~F., {Tielens}, A.~G.~G.~M., \&
  {Bakes}, E.~L.~O. 1995, \apj, 443, 152

\bibitem[{{Zhang} \& {Davis}(2017)}]{zhang:2017.rhd.dusty.winds}
{Zhang}, D., \& {Davis}, S.~W. 2017, \apj, 839, 54

\bibitem[{{Zhang} {et~al.}(2018){Zhang}, {Davis}, {Jiang}, \&
  {Stone}}]{zhang:dusty.cloud.acceleration}
{Zhang}, D., {Davis}, S.~W., {Jiang}, Y.-F., \& {Stone}, J.~M. 2018, \apj, 854,
  110

\bibitem[{{Zuckerman} \& {Evans}(1974)}]{zuckerman:1974.gmc.constraints}
{Zuckerman}, B., \& {Evans}, II, N.~J. 1974, \apjl, 192, L149

\end{thebibliography}

\begin{appendix}

\vspace{-0.5cm}
\section{Source Luminosities \&\ Opacities}
\label{sec:luminosity.opacity.descriptions}

\subsection{``Default FIRE''}

Our ``default'' RT network uses the five-band transport described in detail in \paperone. For completeness we provide the adopted source luminosities and opacities here, determined as described in \paperone\ by integrating over the relevant bands after computing detailed spectra from standard stellar evolution models (the same models used to compute all feedback quantities). For sources, define the light-to-mass ratio in a given band $\Psi_{\nu}$, in units of $L_{\sun}/M_{\sun}$. Then the bolometric $\Psi_{\rm bol} =1136.59$ for $t_{\rm Myr}<3.5$, and $\Psi_{\rm bol}=1500\,\exp{[-4.145\,x + 0.691\,x^{2} - 0.0576\,x^{3}]}$ with $x\equiv \log_{10}(t_{\rm Myr}/3.5)$ for $t_{\rm Myr} > 3.5$. In mid/far IR, $\Psi_{\rm IR}=0$. In optical/NIR, $\Psi_{\rm opt}=f_{\rm opt}\,\Psi_{\rm bol}$ with $f_{\rm opt}=0.09$ for $t_{\rm Myr}<2.5$; $f_{\rm Opt}=0.09\,(1 + [(t_{\rm Myr}-2.5)/4]^{2})$ for $2.5 < t_{\rm Myr} < 6$; $f_{\rm Opt}=1-0.841/(1+[(t_{\rm Myr}-6)/300])$ for $t_{\rm Myr}>6$. In FUV (photo-electric band) $\Psi_{\rm FUV} = 271\,[1+(t_{\rm Myr}/3.4)^{2}]$ for $t_{\rm Myr}<3.4$; $\Psi_{\rm FUV}=572\,(t_{\rm Myr}/3.4)^{-1.5}$ for $t_{\rm Myr}>3.4$. In the ionizing band $\Psi_{\rm ion} = 500$ for $t_{\rm Myr}<3.5$;  $\Psi_{\rm ion}=60\,(t_{\rm Myr}/3.5)^{-3.6} + 470\,(t_{\rm Myr}/3.5)^{0.045 - 1.82\,\ln{t_{\rm Myr}}}$ for $3.5<t_{\rm Myr}<25$; $\Psi_{\rm ion}=0$ for $t_{\rm Myr}>25$. In NUV, $\Psi_{\rm NUV} = \Psi_{\rm bol}-(\Psi_{\rm IR}+\Psi_{\rm opt}+\Psi_{\rm FUV}+\Psi_{\rm ion})$. The adopted flux-mean dust opacities are $(\kappa_{\rm FUV},$ $\kappa_{\rm NUV},$ $\kappa_{\rm opt},$ $\kappa_{\rm IR})$ $=(2000,\,1800,\,180,\,10)\,(Z/Z_{\sun})\,{\rm cm^{2}\,g^{-1}}$. 

As described in \paperone, Appendices A-B, the photo-electric and photo-ionization terms are coupled directly to the gas chemistry and radiative heating/cooling subroutines.

%The photo-ionization rate (and corresponding $\kappa_{\rm ion}$) is calculated from the neutral hydrogen density as described in Appendix~\ref{sec:radiative.fb.implementation} and Appendix~\ref{sec:cooling.approximations} below.
%[FIRE-1 photo-ion $L_{\rm ion}/(M_{\ast}\,L_{\sun}/M_{\sun}) = 328$ ($t_{\rm Myr}<2.6$) and $=468\,(t_{\rm Myr}/2.384)^{-1.13 - 1.28\,\ln{t_{\rm Myr}}}$ ($3.5<t_{\rm Myr}<25$), $=0$ ($t_{\rm Myr}>25$).

\vspace{-0.5cm}
\subsection{Extended Network}

As described in the text, we also (briefly) consider a more extended network. This includes the same optical/NIR ($\Psi_{\rm opt}$) and NUV ($\Psi_{\rm NUV}$) bands as above, but a more complex treatment of the photo-electric ($\Psi_{\rm PE}$), ionizing ($\Psi_{\rm ion}$), Lyman-Werner ($\Psi_{\rm LW}$), soft ($0.5-2$\,keV) and hard ($0.5-10$\,keV) X-ray ($\Psi_{\rm SX}$, $\Psi_{\rm HX}$), and IR bands (with $\Psi_{\rm FIR}=0$ again). The source terms for the non-trivial added bands are: 
\begin{align}
\Psi_{\rm PE} =&
\begin{cases}
271\,\left(1 + t_{3.4}^{2}\right) \hfill \ (t_{3.4} \le 1) \\ 
542\,t_{3.4}^{-1.5} \hfill \ (t_{3.4} > 1) 
\end{cases} \\ 
\Psi_{\rm LW} =&\,
\begin{cases}
109\,\left(1 + t_{3.4}^{2}\right) \hfill \ (t_{3.4} \le 1) \\ 
243\,t_{3.4}^{-1.6}\,\exp{(-t_{400})} \hfill \ (t_{3.4} > 1) 
\end{cases} \\ 
\Psi_{\rm SX} =&\,
\begin{cases}
1.6\times10^{-6} \hfill \ (t_{10} \le 1) \\ 
2.1\times10^{-6} + 0.10\,t_{10}^{-2} \hfill \ (t_{10} > 1) 
\end{cases} \\ 
\Psi_{\rm HX} =&\,
\begin{cases}
1.6\times10^{-6} \hfill \ (t_{10} \le 1) \\ 
1.6\times10^{-6} + 0.15\,t_{10}^{-2} \hfill \ (t_{10} > 1) 
\end{cases} 
\end{align}
where $t_{N} \equiv t / (N\,{\rm Myr})$. For the ionizing bands, we calculate $\Psi_{{\rm ion},\,\nu}$ in four sub-bands $\nu$ from 13.6-24.6, 24.6-54.4, 54.4-70, and 70-500 eV in frequency, with $\Psi_{{\rm ion},\,\nu} = f_{\nu}\,\Psi_{\rm ion}$ where $\Psi_{\rm ion}$ is the total ionizing luminosity defined for our simpler default method above, and $f_{\nu}$ is the fraction in each band calculated assuming the emergent spectrum has a constant effective temperature set to $4\times10^{4}$\,K. The optical/NIR and NUV bands use the same $\Psi$ defined above for the ``default'' network. All of the above are calculated from {\small STARBURST99} average spectra, with the soft and hard X-ray bands empirically calibrated to observed X-ray binary populations.

The opacities in each band are given by $(\kappa_{\rm NUV}$, $\kappa_{\rm opt}$, $\kappa_{\rm PE}$, $\kappa_{\rm LW}$, $\kappa_{\rm SX}$, $\kappa_{\rm HX}) = (1800\,Z^{\prime},\,180\,Z^{\prime},\,0.2 + 2000\,Z^{\prime},\,2400\,Z^{\prime},\,127 + 50\,Z^{\prime},\,0.53 + 0.27\,Z^{\prime})$ where $Z^{\prime} = Z / Z_{\odot}$. The X-ray cross opacities come from Thompson scattering and metal absorption following \citet{morrison.mccammon.83:photoelectric.absorption} assuming solar abundance ratios. The LW cross-section accounts for dust shielding, but the self-shielding by molecular hydrogen is treated approximately as described below. The photo-electric opacity accounts approximately for molecular opacity at low temperatures and metallicities plus dust. For the FIR band, we calculate the opacity using the tables from \citet{semenov:2003.dust.opacities} (specifically their Rosseland-mean opacities from their ``porous 5-layered sphere'' dust models, as a bivariate function of the IR radiation temperature $T_{\rm IR}$ and dust temperature $T_{\rm dust}$). For the ionizing bands, the cross sections in each frequency range follow from the usual expressions for photo-ionizing absorption, scaling with the neutral H or neutral and partially-ionized fractions for He (calculated as described in \paperone). Recall, these are used in our ``standard'' scheme as well, the only difference is that we assume in the ``default'' scheme that the ionizing photon spectrum always traces the UV background, while here each band is explicitly evolved independently. 

The photo-ionization and photo-electric terms couple directly to the photo-heating rates calculated in-code, as described in detail in \paperone. The NUV, optical/NIR, and FIR terms do not directly couple to the heating/cooling subroutines (except via the dust temperature, below). The Lyman Werner band is treated approximately as follows: in \paperone\ we describe in detail the cooling function used for the combination of molecular+metal-line (fine structure) cooling in primarily neutral ($\lesssim 10^{4}$\,K) gas, which are fit to a table of {\small CLOUDY} simulations of gas slabs as a function of density, temperature, and metallicity. The terms presented in \paperone\ can be trivially divided into a term which vanishes as $Z\rightarrow 0$ (which represents the contribution from metal cooling) and one which remains constant (which is dominated by $H_{2}$ cooling). In these simulations, we simply multiply the latter (metal-free) term by a function $f_{\rm LW}$ which ranges from $0-1$, determined by re-running the {\small CLOUDY} calculations for primordial $Z=0$ gas, illuminated by the given LW background. The soft and hard X-ray terms are included as Compton-heating (in addition to the other Compton heating/cooling terms described in \paperone), with heating rate per electron given by $dW/dt = u_{\gamma}\,\sigma_{\rm T}\,(\langle E_{\gamma} \rangle - 4\,k_{B}\,T)/(m_{e}\,c)$ where $u_{\gamma}$ is the photon energy density in the soft/hard band, $\sigma_{\rm T}$ the Thompson cross-section, $k_{B}$ the Boltzmann constant, $m_{e}$ the electron mass, $c$ the speed of light, $T$ the gas temperature, and $E_{\gamma}$ the mean photon energy in the band (since the bands are narrow we simply take this to be the photon energy at the band median). 

Finally, for the FIR band, we explicitly evolve the IR radiation field and dust temperature. We ignore PAH's and other very small grains where single-photon effects are important, and assume the dust-gas collision rate is lower than the dust-radiation absorption (a good assumption at densities $\lesssim 10^{6-10}\,{\rm cm^{-3}}$), so the dust is simply taken to be in thermal equilibrium (and we assume geometric absorption here), with $T_{\rm dust}^{4} = u_{\gamma}\,c/4\sigma_{B}$ where $u_{\gamma}$ is the local photon energy density integrated across all bands where dust dominates the opacity. The dust temperature $T_{\rm dust}$ then influences the gas temperature via the dust-gas collisional heating term given in \paperone. The radiation field is updated at each timestep as an effective blackbody, assuming the dust emits radiation with radiation temperature equal to the dust temperature: if the dust re-emits a total energy $\Delta E_{\rm dust}$ in one timestep into the zone, then $T_{\rm IR}^{4}(t+\Delta t) = (E_{\gamma,\,{\rm IR}}(t)\,T_{\rm IR}^{4}(t) + \Delta E_{\rm dust}\,T_{\rm dust}^{4}) / (E_{\gamma,\,{\rm IR}} + \Delta E_{\rm dust})$. Likewise when radiation is exchanged between cells, the radiation temperature is updated accordingly.

\end{appendix}

\end{document}